%% file: manuscript_mnras.tex
\pdfoutput=1
\documentclass[useAMS,fleqn,usenatbib]{mnras2}  

\usepackage[T1]{fontenc}
\usepackage{ae,aecompl}

\usepackage[]{hyperref}

\hypersetup{%
  colorlinks=true,
  linkcolor=blue, 
  citecolor=black, 
  urlcolor=blue,
  bookmarksopen=true,
  bookmarksnumbered=true,
  pdfstartpage={1},  
  pdftitle = {},
  pdfsubject = {},
  pdfauthor = {} ,
  pdfkeywords = {},
  pdfproducer = {LaTeX with hyperref}
}

\usepackage{amsmath}
\usepackage{amsopn}
\usepackage{graphicx}
\usepackage[british]{babel}
\usepackage[varg]{txfonts}
\usepackage{biblio} 
\usepackage{natbib}
\usepackage{color}

\usepackage{graphicx}

\input{defs_mnras}

\newcommand{\myemail}{tepper@physics.usyd.edu.au}
\newcommand{\ramses}{{\sc ramses}}
\newcommand{\dice}{{\sc dice}}

\title[ A high speed transit of the Galactic disc ]{ The Smith Cloud: surviving a high speed transit of the Galactic disc }

\author[T.~Tepper-Garc\'\i{}a and J.~Bland-Hawthorn]{%
Thor Tepper-Garc\'\i{}a\thanks{\myemail} and Joss Bland-Hawthorn
\\
Sydney Institute for Astronomy, School of Physics, The University of Sydney, NSW 2006, Australia\\
}

\date{Accepted 2017 October 10. Received 2017 October 5; in original form 2017 August 6}

\pubyear{\date{year}}

\defcitealias{pla14b}{\textcolor{black}{Planck Collaboration} 2014}

\begin{document}
\label{firstpage}
\pagerange{\pageref{firstpage}--\pageref{lastpage}}
\maketitle

\pdfminorversion=5
\begin{abstract}

The origin and survival of the Smith high-velocity \HI\ cloud has so far defied explanation. This object has several remarkable properties: (i) its {\em prograde} orbit is $\approx100~\kms$ faster than the underlying Galactic rotation;  (ii) its total gas mass ($\gtrsim4\times10^6~\Msun$) exceeds the mass of all other high-velocity clouds (HVC) outside of the Magellanic Stream;  (iii) its head-tail morphology extends to the Galactic \HI\ disc, indicating some sort of interaction. The Smith Cloud's kinetic energy rules out models based on ejection from the disc. We construct a dynamically self-consistent, multi-phase model of the Galaxy with a view to exploring whether the Smith Cloud can be understood in terms of an infalling,  compact HVC that has transited the Galactic disc. We show that while a dark-matter (DM) free HVC of sufficient mass and density can reach the disc, it does not survive the transit. The most important ingredient to survival during a transit is a confining DM subhalo around the cloud; radiative gas cooling and high spatial resolution ($\lesssim10~\pc$) are also essential. In our model, the cloud develops a head-tail morphology within $\sim10~\Myr$ {\it before and after} its first disc crossing; after the event, the tail is left behind and accretes onto the disc within $\sim400~\Myr$. In our interpretation, the Smith Cloud corresponds to a gas `streamer' that detaches, falls back and fades after the DM subhalo, distorted by the disc passage, has moved on. We conclude that subhalos with $M_{\textnormal{\sc dm}}\lesssim10^9~\Msun$ have accreted $\sim10^9~\Msun$ of gas into the Galaxy over cosmic time -- a small fraction of the total baryon budget.

\end{abstract}


\section{Introduction} \label{sec:intro}

The Smith Cloud \citep[][]{smi63a} is enshrouded in mystery. It is a massive object ($M_{\HI} \gtrsim 10^6 ~\Msun$) close to the Galactic plane ($z \approx 3 ~\kpc$, $R \approx 8 ~\kpc$), moving rapidly ($v_{\textnormal{\sc gsr}} \approx 300 ~\kms$) on a prograde orbit \citep[][]{put03b,loc08a,wak08a}. It is virtually devoid of stars down to the detection limit \citep[][]{sta15b}. Despite the cloud's large mass, its density is not high enough to allow for the formation of molecular gas, suggesting it could be intergalactic gas accreted by the Galaxy. But metallicity measurements along its wake reveal a patchy sulphur abundance distribution ([S/H]) with an average value half that of the Sun \citep[][]{fox16a}, arguing against an extragalactic origin. Its high ionisation fraction \citep[$M_{\HII} \gtrsim 3 \times 10^6 ~\Msun$;][]{hil09a}, together with the discovery of enhanced Faraday rotation measures ($\approx 8$ \mG) spatially coincident in projection with the Smith Cloud \citep[][]{hil13a}, indicate a strong interaction with the disc-halo interface.

So what {\em is} the Smith Cloud? Since its initial interpretation as an extension of the Galactic gas layer by its discoverer, several other ideas have been put forward: a possible remnant associated with the Sagittarius dwarf galaxy \citep[][]{bla98a}; a gas ejection from the Galactic disc \citep[][see also \citealt{mar17a}]{sof04a}; or an extragalactic, dark-matter (DM) confined gas cloud \citep[`dark galaxy';][]{nic09a}. The latter possibility was explored by \citet[][]{nic14b} with the outcome that it must be on its first approach towards the disc, unless it is embedded in a DM subhalo. This result holds true even in the presence of an ambient magnetic field \citep[][]{gal16a}. If confined by dark matter, then it is possible that the Smith Cloud may have accreted gas from the interstellar medium (ISM) during an earlier passage through the Galactic disc. This could explain its relatively high content of heavy elements compared to the Magellanic Stream and other halo clouds \citep[][]{hen17a}. The Smith Cloud thus remains an enigma and, like the Magellanic Stream, it has become a benchmark for any model that aims at explaining gas circulation processes in galaxies.

Here we consider the possibility that the Smith Cloud is associated with a DM-confined gas structure that has transited the disc in the past. Earlier simulations \citep[][]{nic14b,gal16a} were contrived in that they did not attempt to explore the interaction with a realistic Galactic ecosystem nor in a cosmological context, as we do now. In essence, we investigate the evolution of a DM-confined gas cloud that collides with the Galactic disc, using for the first time a multi-component, multi-phase, self-consistent, dynamically stable, and {\em cosmologically motivated} Galaxy model. It is important to state from the outset that we do not attempt to match in detail the properties of the Smith Cloud, nor do we intend to model every aspect of the Galaxy. Such an ambitious goal is futile unless all relevant physical processes (detailed micro-physics, ambient radiation field, magnetic fields, turbulence, cosmic rays, thermal and kinetic feedback, etc.) are carefully accounted for, and at a sufficiently high resolution, which is currently beyond reach. Our aim is to introduce a framework that allows for more realistic experiments in the near future. The importance of our work relies mainly on the fact that state-of-the-art, full cosmological simulations are not yet able to fully resolve systems like this with enough detail, despite their high level of complexity \citep[e.g.][]{vog14b,dub14a,sch15a}.

In addition to an upgraded Galaxy model, our work has several improvements over previous studies. With respect to both \citet[][]{nic14b} and \citet[][]{gal16a}, we address the mixing of gas, and the effect of resolution, although rather crudely. Gas mixing has indeed been studied by \citet[][]{hen17a}, but under somewhat simplified conditions. In contrast to \citeauthor[][]{gal16a}, we include radiative cooling but do not include magnetic fields, as they do. \citeauthor[][]{gal16a} find that the presence of even a weak Galactic magnetic field reduces the baryonic matter retained by the dark matter subhalo after the transit. In our related MHD study \citep[][]{gro17a} we find that magnetic fields can delay cloud destruction \citep[see also][]{kon02a} {\em if sufficiently well resolved}, so they may be worth considering in future simulations.

Note that, where relevant, we assume a flat, dark-energy- and matter (baryonic and cold dark-matter; CDM) dominated Universe, and a cosmology defined by the set of parameters $h = 0.7$, $\Omega_m = 0.3$, and $\Omega_{\Lambda} = 0.7$.

\section{A Galaxy surrogate}

\begin{figure*}
\centering
\includegraphics[width=0.33\textwidth]{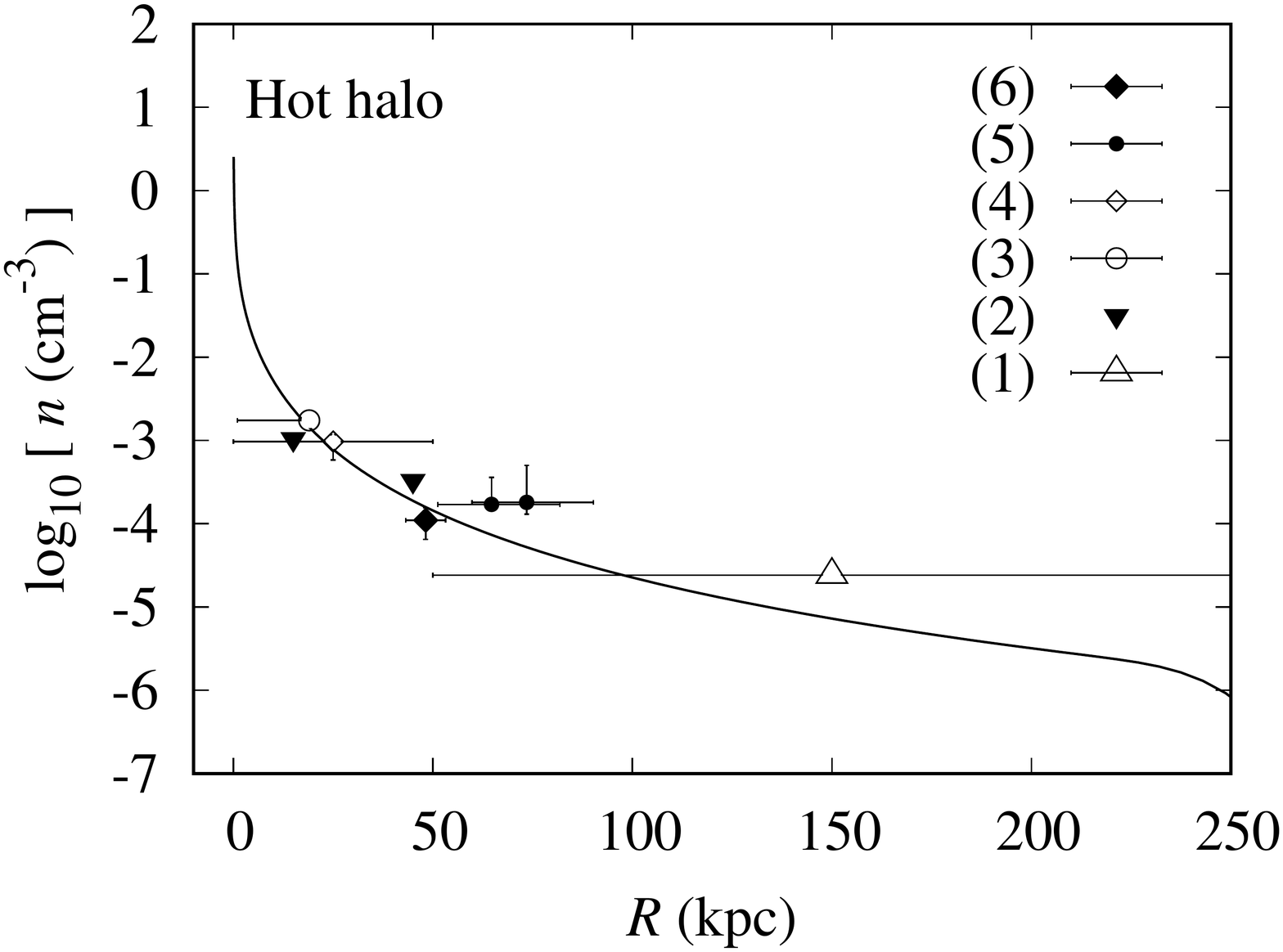}
\hfill
\includegraphics[width=0.33\textwidth]{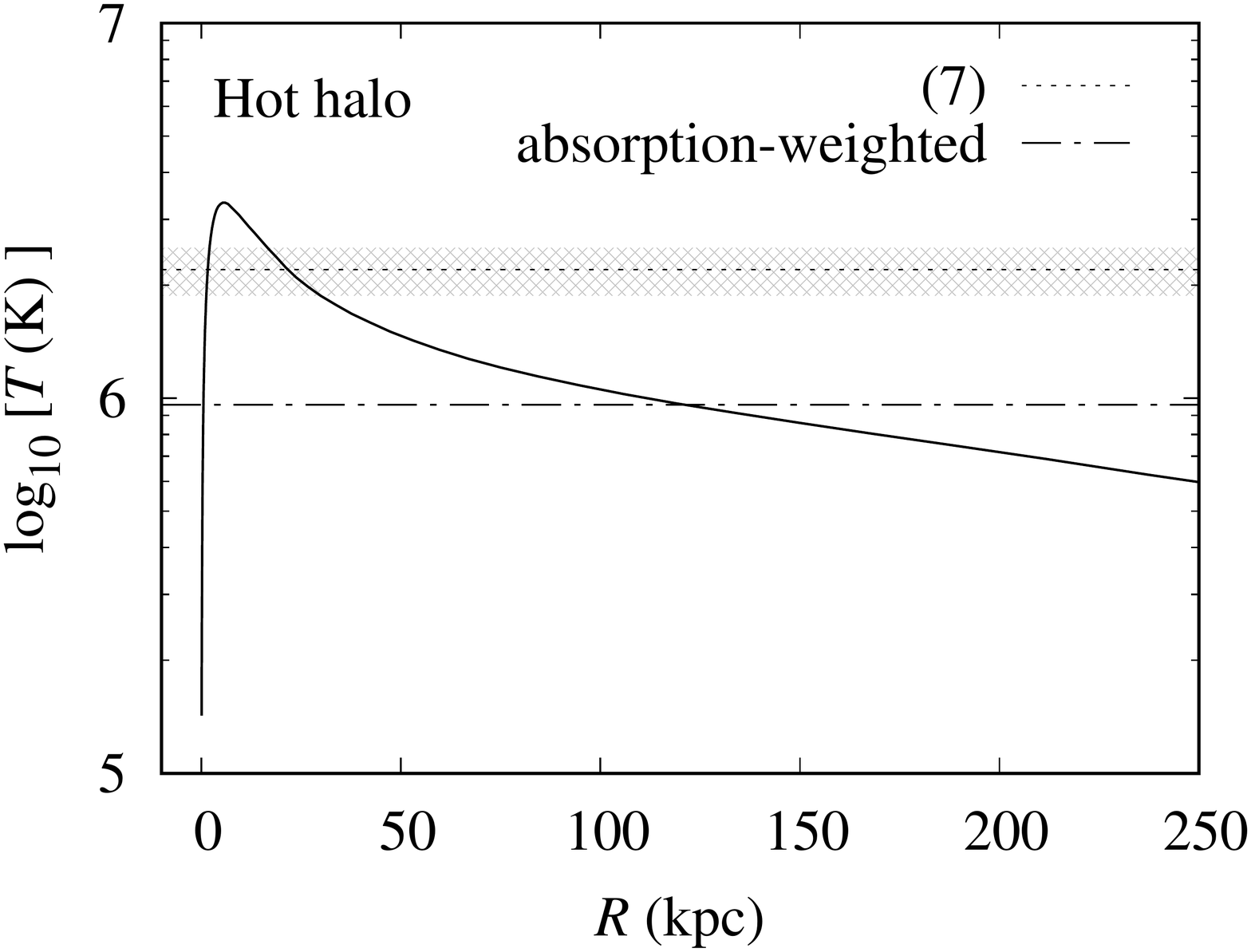}
\hfill
\includegraphics[width=0.33\textwidth]{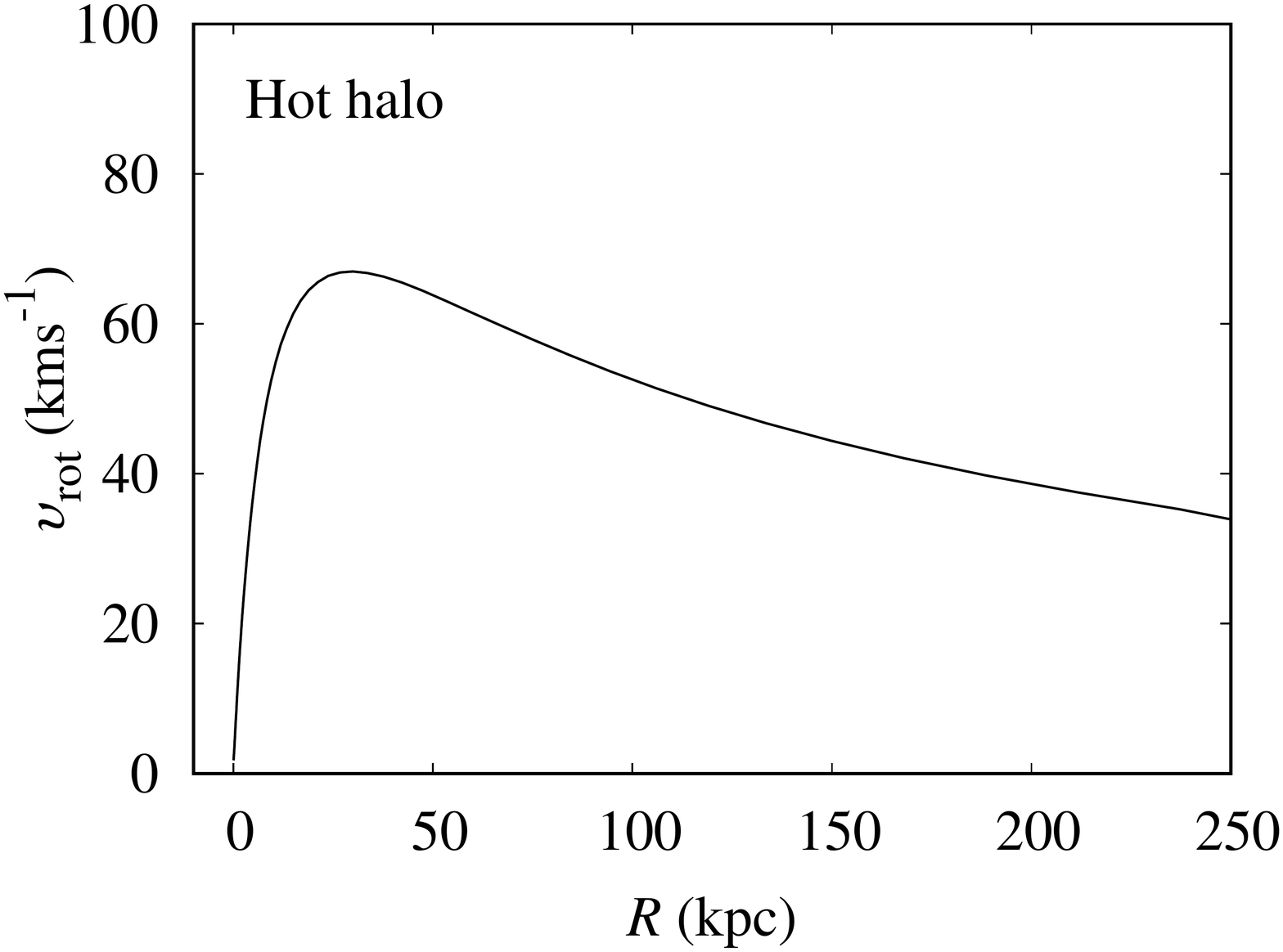}
\caption[ Hot halo ]{ Properties of the hot halo (prior to mapping onto the AMR grid). Total particle density (left); temperature profile (middle); rotation curve (right). The horizontal dot-dashed line in the middle panel indicates the absorption-weighted (i.e. proportional to density) gas temperature across the halo, included for reference. The top dotted line and hatched area indicate the mean temperature and interquartile range inferred from observations, respectively. In the right panel, the maximum rotation speed is $\sim 70 ~\kms$, significantly lower than the value $\sim 180 ~\kms$ measured by \citet[][not shown]{hod16a}. Note that the halo is truncated at a spherical radius $250 ~\kpc$, by construction. Reference key list is as follows. 1: \citet[][]{bli00a}; 2: \citet[][]{sta02a}; 3: \citet[][]{bre07b}; 4: \citet[][]{and10a}; 5: \citet[][]{gat13a}; 6: \citet[][]{sal15a}; 7: \citet[][]{hen13e}.}
\label{fig:ghalo}
\end{figure*}

We are interested only in the global dynamical and structural properties of the Galaxy which are relevant to the event of a DM-confined gas cloud crossing the outer disc. Therefore, we make a number of simplifying assumptions. First, the Galaxy is assumed to be an (initially) perfectly axisymmetric system where substructure such as the spiral arms and the stellar bar are ignored. In addition, we neglect for the moment the details of the Galaxy's stellar population, or the structure and detailed physical and chemical properties of the ISM (e.g., the abundance gradient across the disc). Crucially, non-thermal components (cosmic rays, magnetic fields) that together make a significant contribution to the vertical pressure support of the gas disc \citep[][]{cox05a} are not considered here. We further ignore the stellar halo and bulge components. Our model thus consists of a `live' DM halo, a `live' stellar disc, a warm gas disc, and most notably, a {\em rotating}, non-uniform hot halo. In other words, we adopt a minimal description of the Galaxy which incorporates the dominant large-scale components.

The DM halo is the most dominant gravitational component of any galaxy over cosmic time \citep[][]{fal80a}. A stellar disc increases the potential in the plane of the Galaxy; this stabilises the gas disc and reduces its scaleheight \citep[][]{wan10a}. A rotating, hot halo is expected in our modern picture of galaxy formation \citep[][]{ree77a}, and it has now been observed directly in the Galaxy \citep[q.v.][]{hod16a}. Although its structure and extension are not well understood, the hot halo is now 
believed to constitute a major component of the Galaxy's baryons, from $\sim 2\times 10^{10}\Msun$ (Tepper-Garcia et al 2015) to $\sim 10^{11} ~\Msun$ \citep{fae17a}, and is thus dynamically important. Given its mean temperature $T \sim 10^6 ~\K$ \citep[][]{hen13e} and density in the range $n \sim 10^{-5}$ to $10^{-4} ~\pcc$ within $\sim 200 ~\kpc$ (Figure \ref{fig:ghalo}, left panel), the
Galactic corona represents a high pressure environment that cannot be neglected in the study of gas flows around the Galaxy.

The details of each component that make up our model follow. All of our adopted parameters constitute a plausible set of values taken from a recent review of the Galaxy \citep[][]{bla16a}.

The DM host halo is modelled using a \citet[][NFW]{nav97a} profile, with a mass $\Mvir \approx 10^{12} ~\Msun$ within $\rvir \approx 250 ~\kpc$, and a scale radius of $r_s \approx 17 ~\kpc$ (corresponding to a `concentration' $\xvir \equiv \rvir / r_{s} \approx 15$). Note that we could have used instead an {\em equivalent} \citet[][]{her90a} profile with same mass and a scale radius $r_s \approx 33 ~\kpc$ \citep[][]{spr05c}. In
our model, the DM halo has no net rotation and we
truncate it at a spherical radius of $250 ~\kpc$.

\begin{figure}
\centering
\includegraphics[width=0.42\textwidth]{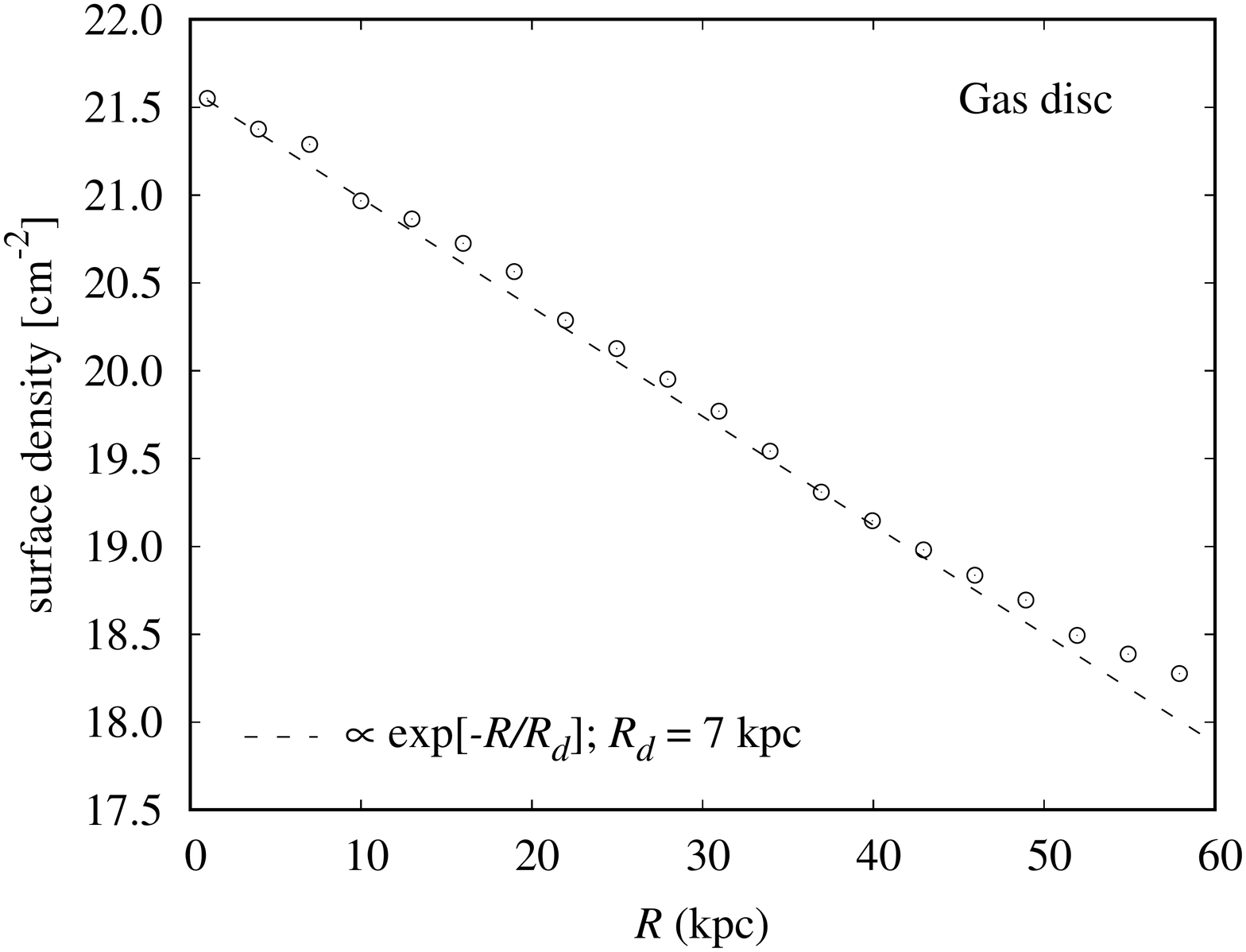}\\
\includegraphics[width=0.42\textwidth]{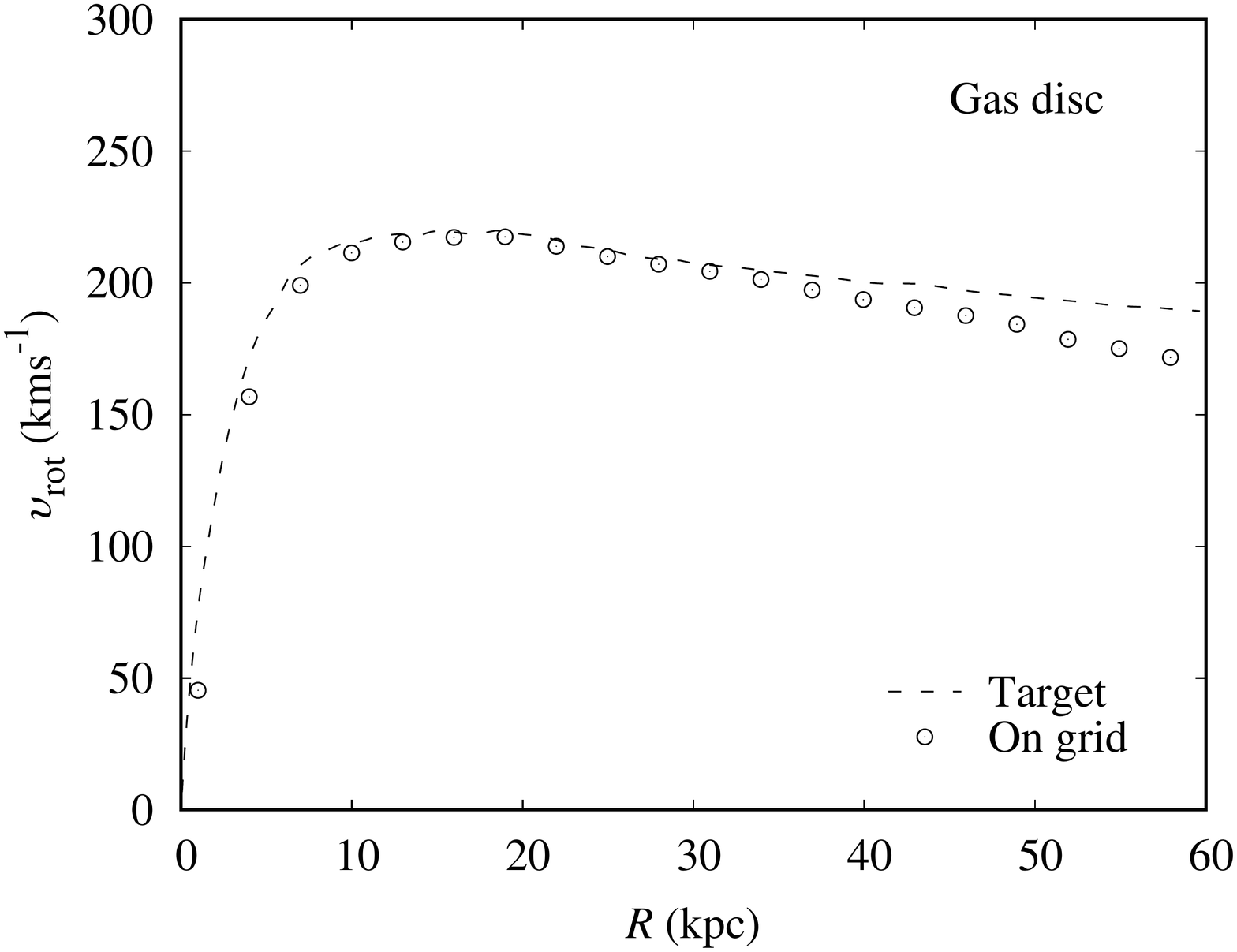}
\caption[ Gas disc ]{ Properties of the gas disc: Surface (particle) density (top); rotation curve (bottom). The target properties are indicated by a dashed line; the actual properties on the AMR grid (at $t = 0$), by the symbols. The latter have been computed by taking the azimuthal average of the gas density and velocity, respectively. The gas surface density is slightly higher than the target density everywhere, and the amplitude of the actual rotation curve departs from the the target rotation curve at $R \gtrsim 30 ~\kpc$, both due to the non-negligible contribution of the hot halo to the gas mass relative to the disc at these radii (see Appendix \ref{sec:sta}). Note that the gas disc is truncated at $R \geq 60 ~\kpc$, by construction.}
\label{fig:gdisc}
\end{figure}

Although there is no clear physical motivation, we assume the density profile of the hot halo follows the same profile as the DM host halo, and with the same structural parameters, i.e. a NFW profile with $\xvir = 15$. In our model, it consists of $\sim 2 \times 10^{10} ~\Msun$ of gas (i.e. a mass fraction $m_h = 0.02$ relative to \Mvir) within a spherical radius of $250 ~\kpc$ in {\em thermal} equilibrium with the composite potential $\Phi$ of the model galaxy (see below), with temperatures in the range $\sim 10^5 ~\K$ to $\sim 10^6 ~\K$ and a density profile consistent with a range of estimates from the literature (Figure \ref{fig:ghalo}). The absorption-weighted temperature of our halo model broadly agrees with the temperature inferred from spectroscopic X-ray observations. Moreover, the temperature profile is remarkably similar to the average profile of halos with a virial mass $\sim 10^{12} ~\Msun$ found in cosmological simulations \citep[][]{nuz14a,nel16a}. A slight temperature gradient in the hot halo of the Galaxy, as well as an adopted gas metallicity of $Z = 0.3 ~\Zsun$ are consistent with observations \citep[][]{mil15a}.

Thermal equilibrium of the halo gas is achieved by calculating the radial velocity dispersion required to maintain dynamic equilibrium with the total potential, derived from the spherically symmetric \citet[][]{jea15a} equation, corrected for the centrifugal potential, and equating the internal energy of the gas to the result thus obtained (which in turn sets the gas temperature). The rotation speed -- or equivalently the centrifugal potential -- is set according to the specific angular momentum, which follows roughly the cumulative mass profile \citep[][]{bul01b}, adopting a value for the spin parameter $\lambda = 0.08$. This value is consistent with the mean value found at $z = 0$ in $N$-body, full hydrodynamic cosmological simulations of structure formation \citep[][but see \citealt{bry13a}]{zju17a}. The value we adopt is optimal to yield a hot, dynamically stable halo with a significant rotation speed. This is forced upon us by recent claims from X-ray spectroscopy of the unexpected rapid rotation in the Galaxy's hot corona \citep[$\sim 180 ~\kms$][]{hod16a}. We find that it is very difficult to construct a fast-rotating corona which is hot and in equilibrium. So we have adopted the fastest halo rotation speed ($70 ~\kms$ at $R \approx 30 ~\kpc$; see Figure \ref{fig:ghalo}) that still affords long-term stability (see Appendix \ref{sec:sta}). This value is much lower than the value inferred by Hodges-Kluck and collaborators. However, it has been shown that quasi-isothermal hot halos feature rotation speeds well below 40 percent of the rotation speed of their embedded gas discs (or $\sim 80 ~\kms$ in our case) except perhaps very close to the centre \citep[][]{pez17a}. Therefore, we believe that the velocity measured by \citeauthor{hod16a} probably arises from the inner halo region close to the disc ($R \lesssim 10 ~\kpc$), contrary to the authors' interpretation.

The stellar disc follows a \citet[][MN]{miy75a} profile, characterised by its total mass $M_{\star} \approx 5 \times 10^{10} ~\Msun$ \citep[][]{bov13a}, and scale parameters $a = 5 ~\kpc$ and $b = 0.5 ~\kpc$. The choice of this profile is motivated by the need for numerical stability. In test $N$-body simulations we have run, discs with an initial exponentially decaying profile (in $R$ and $z$) are found to be unstable due to their significantly higher densities close to the centre. We are aware that a single MN profile is a crude approximation to the observed profile of disc galaxies. In particular, the parameter $a$ may fall short by a factor $\sim 2$ compared to the scalelength of an equivalent exponential disc \citep[][]{fly96a}. Also, the exponential scaleheight can differ from $b$ \cite[][]{smi15a}. To overcome this issue, we take the scale parameter values from \citet{kaf14a} who use stellar kinematic information to measure the mass distribution of the Galaxy, assuming the thin and thick stellar discs are well described by a single MN profile. In this case, the radial scalelength turns out to be intermediate between that of the thin and the thick stellar discs, and the total mass is slightly higher than their combined mass. 

In our model, the stellar (radial) velocity dispersion is adjusted such that \citet[][]{too64a}'s $Q_{\star} \equiv \left( \sigma_r \, \kappa \, / \, 3.36 \, G \, \Sigma_{\star} \right) > 1.5$ everywhere, thus making the stellar disc stable against fragmentation. Here $\sigma_r^2$, $\kappa$ and $\Sigma_{\star}$ are the stellar radial velocity dispersion, the epicyclic frequency and the stellar surface density, respectively. We note that the condition $Q_{\star} >1.5$ is not enough in general to ensure stability \citep[][]{zas17a}, but it is sufficient in our case (Appendix \ref{sec:sta}).

The gas disc has a mass of $\sim 9 \times 10^9 ~\Msun$ (or a mass fraction $m_d = 0.01$ relative to \Mvir) within $R \lesssim 60 ~\kpc$. It is set in vertical hydrostatic equilibrium, following a radial exponential profile $\propto \exp[-R/R_d]$ with scalelength $R_d = 7 ~\kpc$, reasonably consistent with the properties of the combined inner and outer \HI\ distributions of the Galaxy \citep[][]{kal08a}.\footnote{\citet[][]{kal08a} have measured $R_d = 3.75 ~\kpc$ for the Galactic \HI\ disc within $R \lesssim 35 ~\kpc$, and roughly twice that value in the outer region to $R \approx 60 ~\kpc$.} The initial gas temperature is uniform across the disc and is set to $10^4 ~\K$, and the gas metallicity to $Z = 0.3 ~\Zsun$. This value is the lowest acceptable value for the ISM metallicity at radii $5 ~\kpc \lesssim R \lesssim 15 \kpc$ \citep[][]{hou00a}. The gas temperature is chosen to make the disc stable against gravitational fragmentation.\footnote{The stability parameter $Q_g \equiv ( c_s \kappa / \pi ~G \Sigma_g ) > 1$ everywhere. Here, $\Sigma_g$ is the gas surface density. A model where $Q_{\star}$ and $Q_g$ are chosen to give a joint, stable system \citep[][]{rom13a} is left for future work.} In addition, its value is similar to the lowest temperature of other gas components (i.e. the hot halo and the subhalo; see below), which is important given the need of a global temperature floor in our simulations (see Section \ref{sec:sim}). It is worth noting that the value of the disc scalelength is consistent with the values we have adopted for \xvir, $\lambda$, and $m_d$ \citep[][]{mo98a}. The resulting surface density profile and rotation curve of the gas disc are both in broad agreement with those of the Galaxy \citep[Figure \ref{fig:gdisc}; cf.][]{kal09a}. 

\section{The Smith Cloud progenitor} \label{sec:sc}

We adopt the simplest model possible for the Smith Cloud progenitor, which consists of a DM subhalo filled with gas in {\em thermostatic} (i.e., with no net rotation) equilibrium with their combined potential. The relevant parameters in this case are the total DM mass, the subhalo extension, the baryon : total mass fraction, and the gas metallicity.

With an estimated size of $D$, a mass $M$, and an hydrogen fraction (by mass) \XH, the mean hydrogen density of a (spherical) gas cloud is
\begin{equation} \label{eq:npart}
	\nH \approx 0.06 ~\pcc \left( \frac{ \XH }{ 0.76} \right) \left( \frac{ M }{ 10^6 ~\Msun } \right) \left( \frac{ D }{ 1 ~\kpc } \right)^{-3} \, .
\end{equation}
The projected size of the Smith Cloud is $D \gtrsim 1 ~\kpc$ \citep{loc08a} and its total mass is $ \gtrsim 4 \times 10^6$ M$_\odot$ \citep[with at least half of it in ionised form; ][]{hil09a}. Thus we estimate its mean hydrogen density at $\nH \sim 0.2 ~\pcc$ for an assumed $\XH = 0.76$. This is a crude estimate because the Smith Cloud has a non-spherical structure. If we instead take its maximum observed \HI\ column density \citep[$\NHI \sim 5 \times 10^{20} ~\psc$;][]{loc08a} and assume its linear size along the sightline is $D \approx 1 ~\kpc$, then correcting a mean ionisation fraction of 0.5 we arrive at $\nH \approx 0.3 ~\pcc$. Thus, the mean {\em particle} density of the Smith Cloud to order of magnitude accuracy should be $n \sim 0.1 - 1 ~\pcc$ for any reasonable metallicity and ionisation state.

Presumably, both the mass and the density of the Smith Cloud are lower than those of its progenitor. Previous work \citep[][]{nic09a,nic14b} has shown that, in order for a DM confined cloud to survive a disc passage, DM masses in the range $\sim 10^8 - 10^9 ~\Msun$ are required. We take a subhalo in the high-mass limit ($10^9 ~\Msun$) with a profile initially well described by the NFW fitting function adopting a concentration $\xvir = 25$ appropriate for halos of this mass \citep[][]{ste02b}, truncated at a spherical radius of $2 ~\kpc$. For a cosmic baryon : total mass ratio $f_b \approx 0.16$ \citepalias[][]{pla14b}, we get a total gas mass of $1.5 \times 10^8 ~\Msun$ and a mean hydrogen density $\nH \sim 1 ~\pcc$ within a 1 kpc radius. The subhalo's gas metallicity is set to $Z = 0.1 ~\Zsun$ consistent with the metallicity floor at the present epoch.

For our comparative study, we also adopt a DM subhalo at the low-mass end, $\sim10^8 ~\Msun$, resulting in a gas mass and a particle density roughly an order of magnitude lower compared to the high-mass case. The assumption of thermostatic equilibrium leads to the gas having a temperature in the range of $\sim 10^4 - 10^5 ~\K$ (increasing towards the edge) for the high-mass subhalo, and $\sim 10^4 ~\K$ for the low-mass case. In addition, we consider the case of a pure gas cloud with a total gas mass of $1.5 \times 10^8 ~\Msun$, i.e. roughly identical to the gas content of the high-mass subhalo (see Table \ref{tab:runs}).

It is worth mentioning that our model subhalo composed of DM and gas maybe regarded as a dwarf galaxy whose stellar component may be too dispersed to be detected (or may have been lost altogether) due to past interactions prior to falling into the Galaxy.

\section{Simulating an interaction with the galaxy} \label{sec:sim}

Our approach is to set up a multi-component stable system comparable to the Galaxy in terms of mass and overall structure, and to `drop' a DM subhalo from an initial point far from the system's centre. The choice of orbit parameters is discussed further below.

\begin{table*}
\begin{center}
\caption{Relevant galaxy model parameters (initial values).}
\label{tab:comp}
\begin{tabular}{lcccccc}
\hline
\hline
Component 			& Profile	& Mass ($10^{9} ~\Msun$)& Scalelength (kpc)	& Particle mass$^{\,g}$ ($10^{3} ~\Msun$)  & Particle number ($10^{5}$) & Metallicity ($\Zsun$)~\\
\hline\\
DM halo				& NFW	& $10^3$				& 13.6$^{\,c}$		& 	2000			&	5$^{\,h}$		&	--	~\\
Stellar disc$^{\,k}$		& MN	& 46					& 5.0	$^{\,d}$		&	100			&	5$^{\,h}$	&	--	~\\
Gas halo$^{\,a}$ 		& NFW	& 19					& 13.6			&	4$^{\,i}$		&	50			&	0.3	~\\
Gas disc$^{\,b}$		& Exp	& 9					& 7.0	$^{\,e}$		&	1$^{\,i}$		&	100			&	0.3	~\\
~\\
DM subhalo$^{\,j}$		& NFW	& 1, 0.1				& 1.9	$^{\,f}$		&	100			&	1			&	--	~\\
Gas subhalo$^{\,j,l}$		& NFW	& 0.16, 0.02			& 1.9				&	1$^{\,i}$		&	1			&	0.1	~\\
\end{tabular}
\end{center}
\begin{list}{}{}
\item Notes. NFW: \citet{nav97a} profile. MN: \citet{miy75a} profile. $^{a}$In thermal equilibrium. $^{b}$Radial exponential profile in vertical hydrostatic equilibrium initially at constant temperature $T = 10^4 ~\K$. $^{c}$ For a concentration of $\xvir = 15$. $^{d}$ Scaleheight set to 0.5 kpc. $^{e}$ Scaleheight set by vertical hydrostatic equilibrium (giving rise to a `flaring' disc). $^{f}$ For a concentration of $\xvir = 25$. $^{g}$ Adopted when discretising the corresponding density field. $^{h}$ A total of $10^6$ test particles have been used to compute the potential. $^{i}$The minimum gas mass in a cell at maximum refinement is $\sim 2.5 \times 10^6 ~\Msun$. $^{j}$ We adopt both a low-mass, and a high-mass subhalo. $^{k}$ The stellar metallicity is ignored as it is of no relevance for our study. $^{l}$In thermostatic equilibrium. See text for further details. 
 \end{list}
\end{table*}

\subsection{Initial conditions} \label{sec:ics}

The galaxy model is realised by randomly sampling each of the component density fields using a number of `particles' with a fixed mass (different for each component), and adjusting the particles' velocities so that each component is in dynamic equilibrium with the total potential. This is achieved following the approach developed by \citet[][]{spr05c}, as implemented in the \dice\ code \citep[][]{per14c}.\footnote{See footnote \ref{foo:code}.} The particle number and mass for each component are listed in Table \ref{tab:comp}. Note that the particle number of the collisionless components (DM and stars) is below what is generally necessary to fully overcome the Poisson noise that results from the discretisation of the density field \citep[q.v.][]{don13a}. But for now we are not interested in the onset of local instabilities so this is of no concern.

The Smith Cloud progenitor, with its collisionless and gaseous components, is initialised in a similar manner as the Galaxy. In addition, we need to consider the cloud's initial position and trajectory with respect to the galaxy's centre. If the Smith Cloud is associated to a DM subhalo, then it must have fallen in from a large distance ($d \gtrsim 100 ~\kpc$). Nevertheless, it has been argued \citep[][]{nic14b} that, due to its well constrained speed of $\sim 300 ~\kms$, the Smith Cloud cannot have fallen from infinity. Indeed, a (DM confined) cloud with an initial velocity $v_0$, falling (radially) under the gravitational pull of the Galaxy's potential from a distance $d_0$ will hit the disc at a speed (neglecting for the moment other forces)
\begin{equation} \label{eq:vesc}
	v \approx 200 ~\kms \left\{ \left( \frac{ M }{ 10^{12} ~\Msun } \right) \left( \frac{ d_0 }{ 250 ~\kpc } \right)^{-1} \!\!\!+ \left( \frac{ v_0 }{ 200 ~\kms } \right)^2 \right\}^{1/2} \, .
\end{equation}
Here, $M$ is the {\em mean} mass enclosed within $d_0$. This is indeed a very rough estimate, as both the actual enclosed mass and the enclosing radius decrease along the cloud's orbit. Nonetheless, the above calculation shows that, for an initial velocity $v_0 \lesssim 200 ~\kms$, a cloud falling in from $d_0 \gtrsim 250 ~\kpc$ would reach the disc at a speed a factor of a few times 100 ~\kms. It turns out that the change of enclosed mass and drag forces along the cloud's orbit do not significantly affect this estimate. In fact, we have run three test simulations adopting $v_0 = 100 ~\kms$ and $d_0 = 200 ~\kpc$ or 100 kpc (the latter case for both a DM confined cloud and a pure gas cloud) and have found that the cloud's speed right before it reaches the disc is $350 \lesssim ( v / \kms) \lesssim 450$ (see also Figure \ref{fig:kin} below). This is somewhat higher than the observed speed of the Smith Cloud. 

We ignore for now this fact and place the Smith Cloud progenitor on an orbit initially tangential (with respect to the disc plane) at a radial distance of $R = 100 ~\kpc$ and an impact parameter of $\sim 65 ~\kpc$ from the GC (i.e. $d \approx 115 ~\kpc$), and initial speed $v_0 = 100 ~\kms$. A non-vanishing, non-radial initial velocity is necessary to deter the cloud from falling radially towards the Galaxy's centre. Note that the cloud in all our models moves on a {\em prograde} orbit with respect to the rotating hot halo (and the disc) on its way to the galaxy, and thus experience a reduced ram pressure compared to a fully stationary medium. The choice of an initially tangential, prograde orbit is motivated by the observed prograde motion of the Smith Cloud and the angle of its inferred orbit with respect to the disc plane.

\subsection{System's evolution} \label{sec:sys}

The time evolution of the composite system representing the Galaxy and the Smith Cloud progenitor with their collisionless and gaseous components is calculated by solving the Vlasov-Poisson and Euler equations with the adaptive mesh refinement (AMR) $N$-body, gravito-hydrodynamics (GHD) code \ramses\ \citep[version 3.0 of the code last described by][]{tey02a}.\footnote{We use a modified version of \dice\ and our own patched version of \ramses\ which includes the \dice\ patch written by V.~Perret. Our setup files and codes are freely available upon request to the corresponding author (TTG). \label{foo:code}} To this end, the initial conditions (essentially particle masses, positions, and velocities; and additionally internal energies for gas components) are mapped onto an initially regular Cartesian grid, which is then locally refined according to a number of criteria (see below). While collisionless components (DM, stars) in principle retain their particle nature, gas particles initially falling within the same volume element are merged using a cloud-in-cell (CIC) interpolation scheme \citep[][]{hoc88a}, thus setting the initial value of the fluid variables (density, pressure, total energy) in each cell across the grid. All gaseous components are assumed to consists of a monoatomic, ideal gas, with an hydrogen fraction (by mass) $\XH = 0.76$. Note that the contribution of all components to the overall gravitational field is taken into account at all times.

Despite the variety of physical processes which nowadays can be included in simulations of galaxies -- albeit via sub-grid `recipes' -- we opt for a simpler, yet more robust, approach. More specifically, we ignore star formation and feedback processes (in the Galaxy or the Smith Cloud progenitor) of any kind as these are still very crude and dependent of poorly understood parameters. In particular, our calculations do not include the internal heating (e.g. via star formation) that has been shown to assist the stripping of gas off DM subhalos through ram pressure or tidal interactions \citep[][]{nic11a}. The only microphysical process we consider is radiative cooling. This is essential to the stability of the system, which -- in addition to the collisionless components -- consists roughly of two gas phases: a `hot' phase (halo), and a `warm' phase (disc, cloud).

We adopt a simple hydrogen (H) and helium (He) network, in addition to cooling by heavy elements, assuming at all times ionisation equilibrium, following \citet{the98b}.\footnote{This includes a number of processes such as collisional ionisation and excitation, (dielectronic) recombination, Bremsstrahlung, and inverse Compton scattering.} Since we are neglecting both the Galactic and extragalactic ionising radiation fields, we do not include the effect of photo-heating. This, together with our ignoring feedback processes, makes it necessary to impose an artificial temperature floor on the gas to avoid overcooling (thus effectively mimicking heating and thermal feedback). We set this at $T = 10^4 ~\K$, equal to the (initial) temperature of the gas disc, the low-mass subhalo, and on the order of the equilibrium temperature that would be obtained from the balance between cooling and heating for our adopted range in metallicities and a reasonable model of the extragalactic ultra-violet background \citep[e.g.][]{wie09a}. Also, this value is consistent with the temperature (or equivalent pressure) floor required to avoid numerical fragmentation at the relevant densities \citep[q.v.][]{tey10b}.\footnote{It should be mentioned that we do not impose such density-dependent floor; as a consequence, the gas disc develops a dense, `cool' (at $T = 10^4 ~\K$) core over time (see Figure \ref{fig:stable}).}

The system is placed into a cubic box with length 500 kpc on each side, with open (i.e. `zero gradient') boundary conditions, allowing for gas inflow at all times. The reason for the choice of box size is to maximise the resolution for the highest adopted refinement level (see below), while fully enclosing the system. Note that both the DM halo and the hot halo fit tightly into this volume. However, this is not an issue since both the DM and gas densities are extremely low beyond $r \sim 200 ~\kpc$. That being said, we do observe some mass loss from the DM halo through the boundary at the percent level in the course of our simulations, but it is negligible and thus dynamically irrelevant.\footnote{An animation of the DM halo's evolution can be found following this \href{http://www.physics.usyd.edu.au/~tepper/proj_smith_paper.html\#dm_gc_run2}{link}.} 

\begin{figure}
\centering
\includegraphics[width=0.45\textwidth]{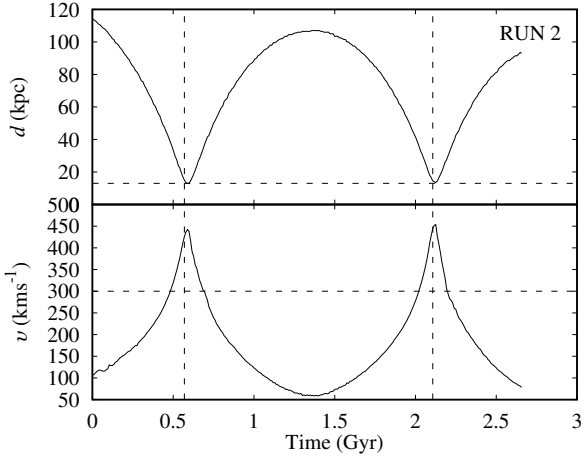}
\caption[ (DM) subhalo kinematics ]{ Subhalo kinematics over a period of roughly $2.5 ~\Gyr$ in RUN 2. Distance to the Galactic Centre (top); speed along its orbit (bottom). The vertical dashed lines indicate the time roughly corresponding to the disc transits at $t \sim 0.6 ~\Gyr$ and $t \sim 2.2 ~\Gyr$. The horizontal dashed line corresponds to the mean distance (top) and mean total speed (bottom) of the Smith Cloud as estimated by \citet{loc08a}, and are only included for reference. Results essentially identical in RUN 1 and qualitatively similar in RUN 3 for $t \lesssim 2 ~\Gyr$ (not shown).}
\label{fig:kin}
\end{figure}

\begin{figure*}
\centering
\includegraphics[width=0.23\textwidth]{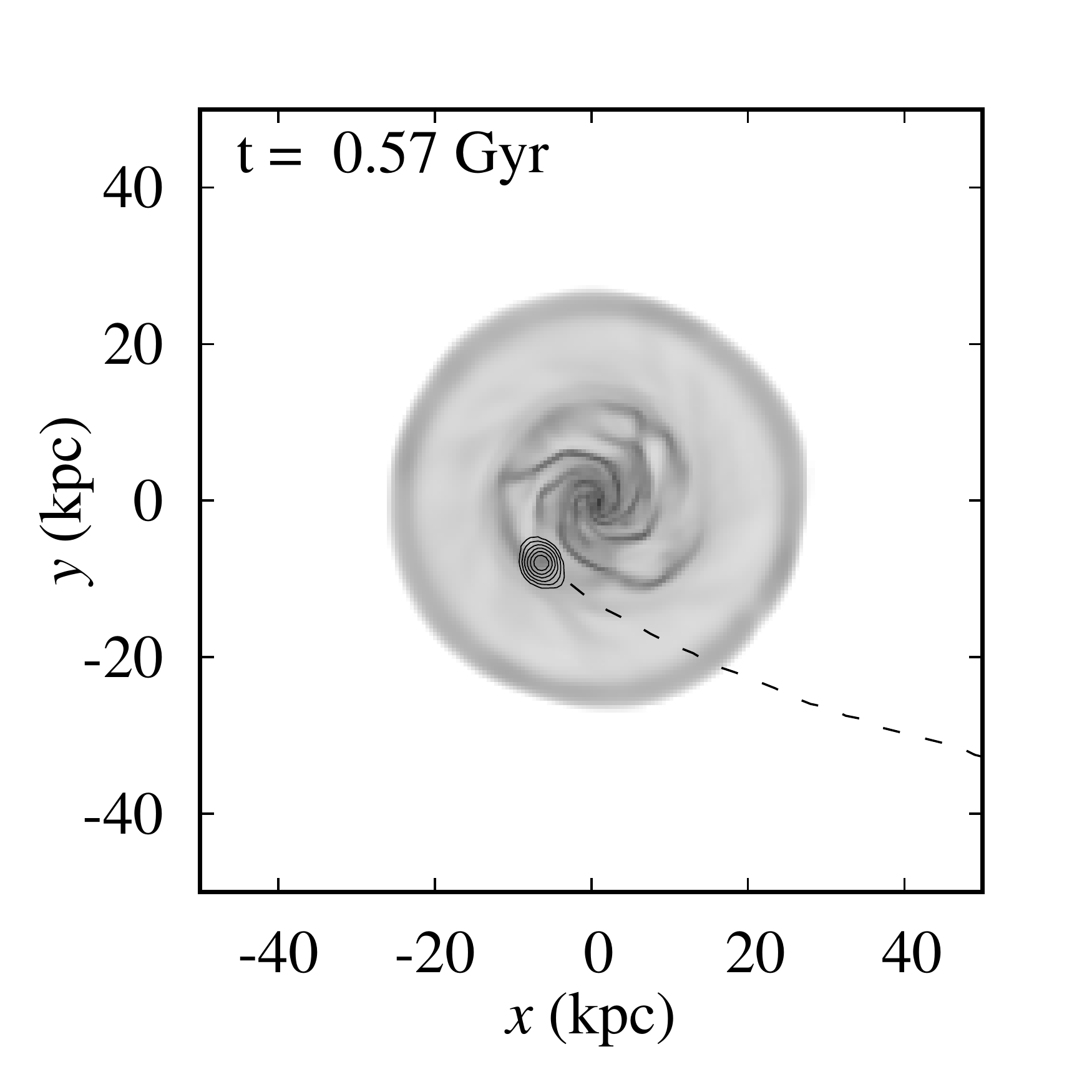}
\includegraphics[width=0.23\textwidth]{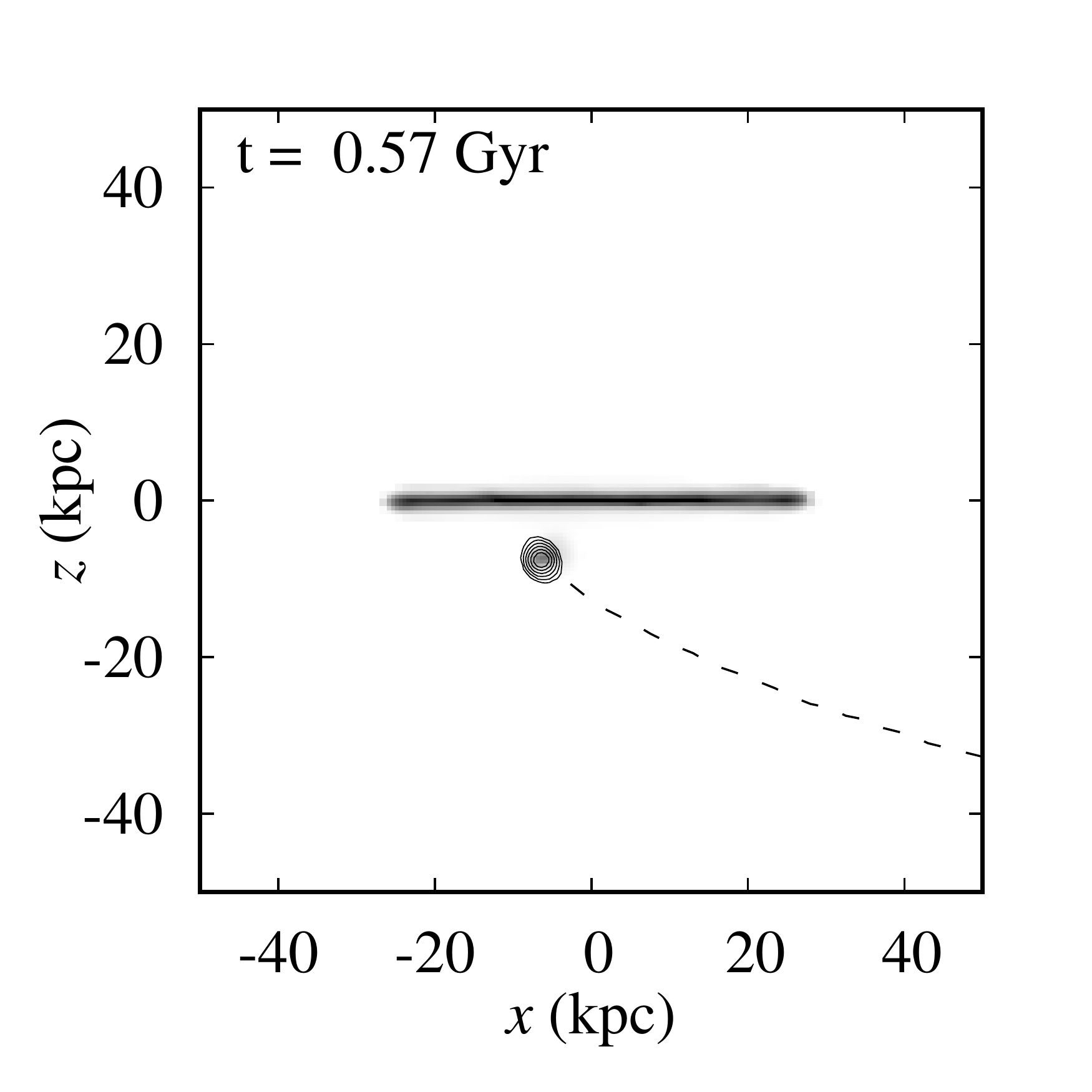}
\includegraphics[width=0.23\textwidth]{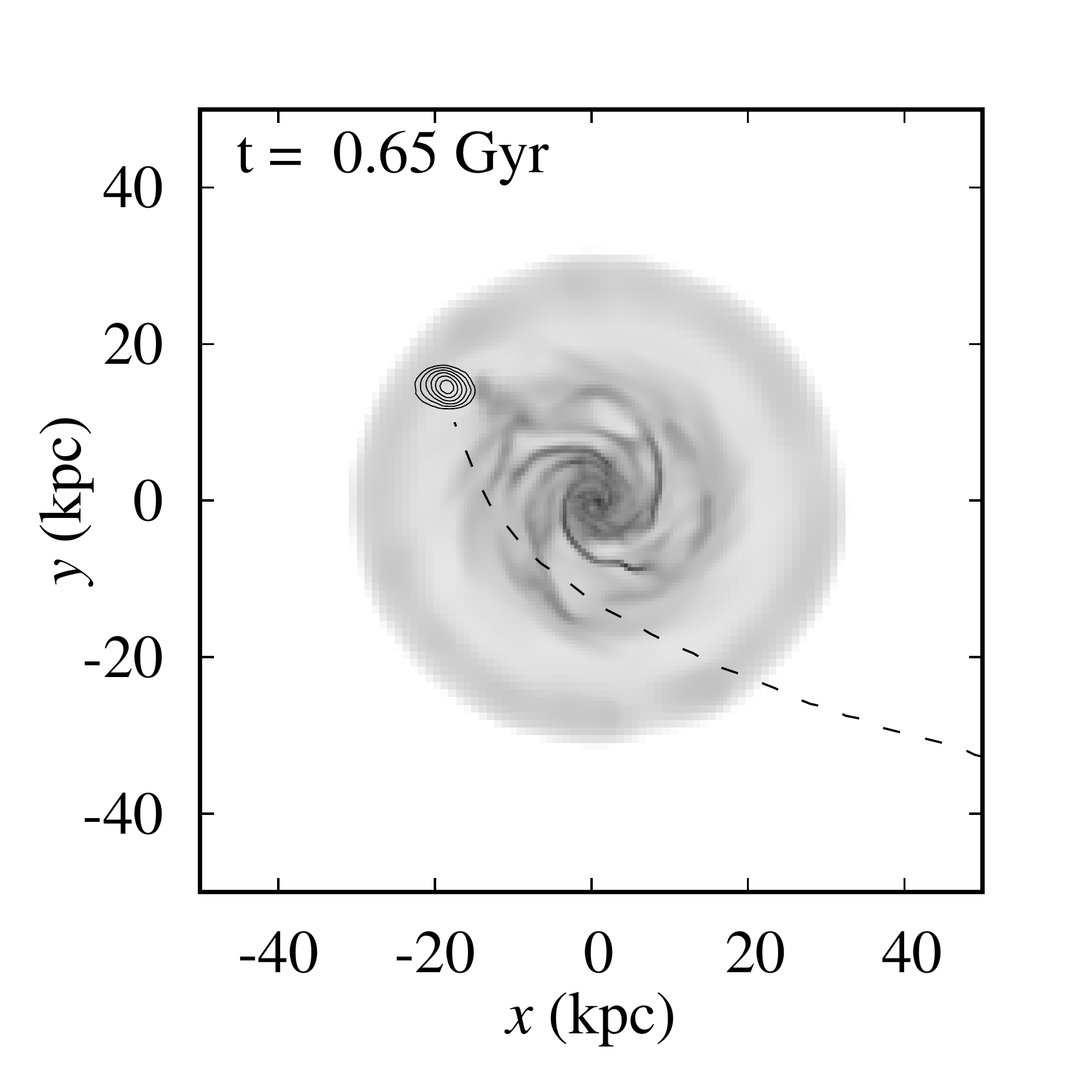}
\includegraphics[width=0.28\textwidth]{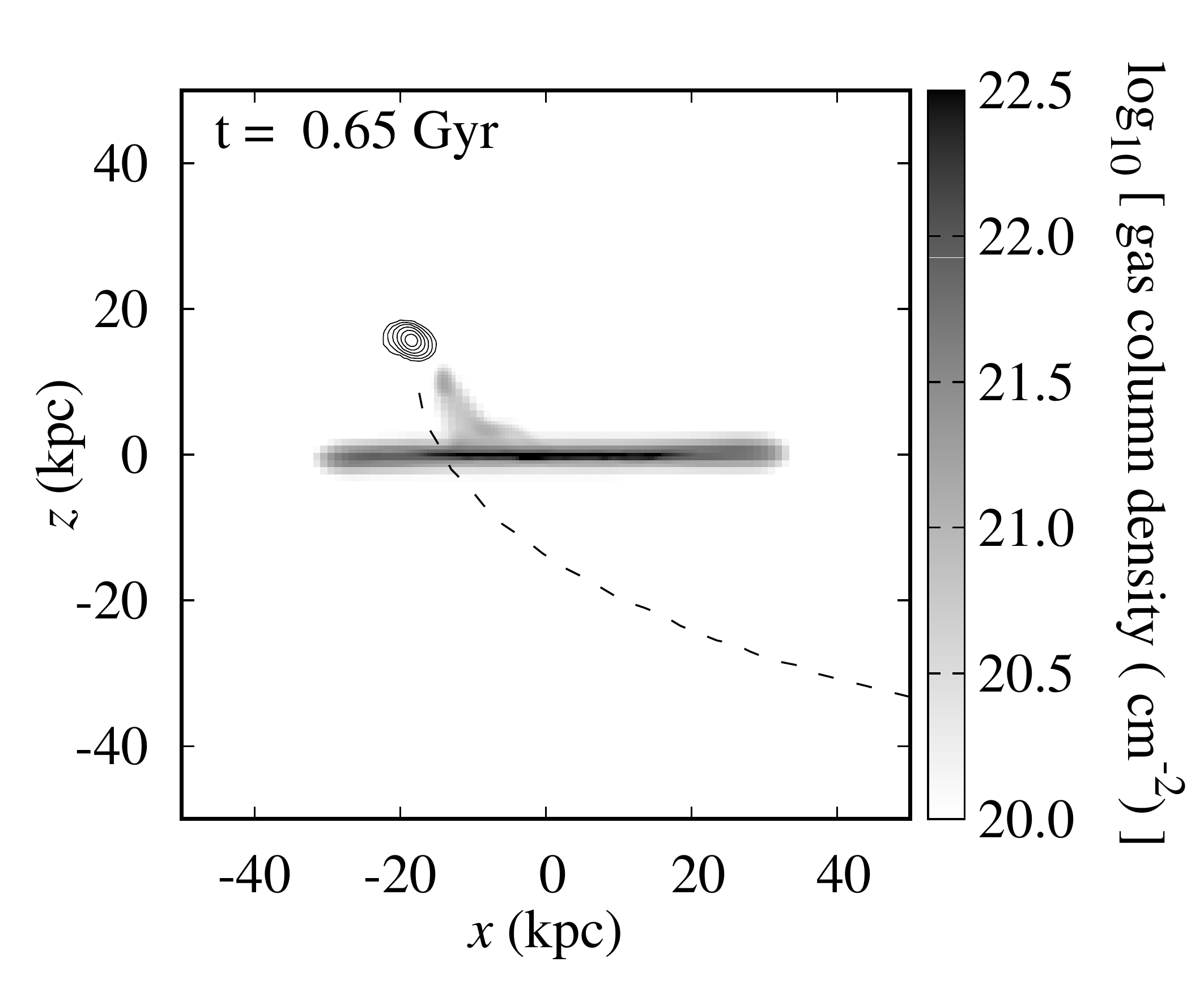}\\
\includegraphics[width=0.23\textwidth]{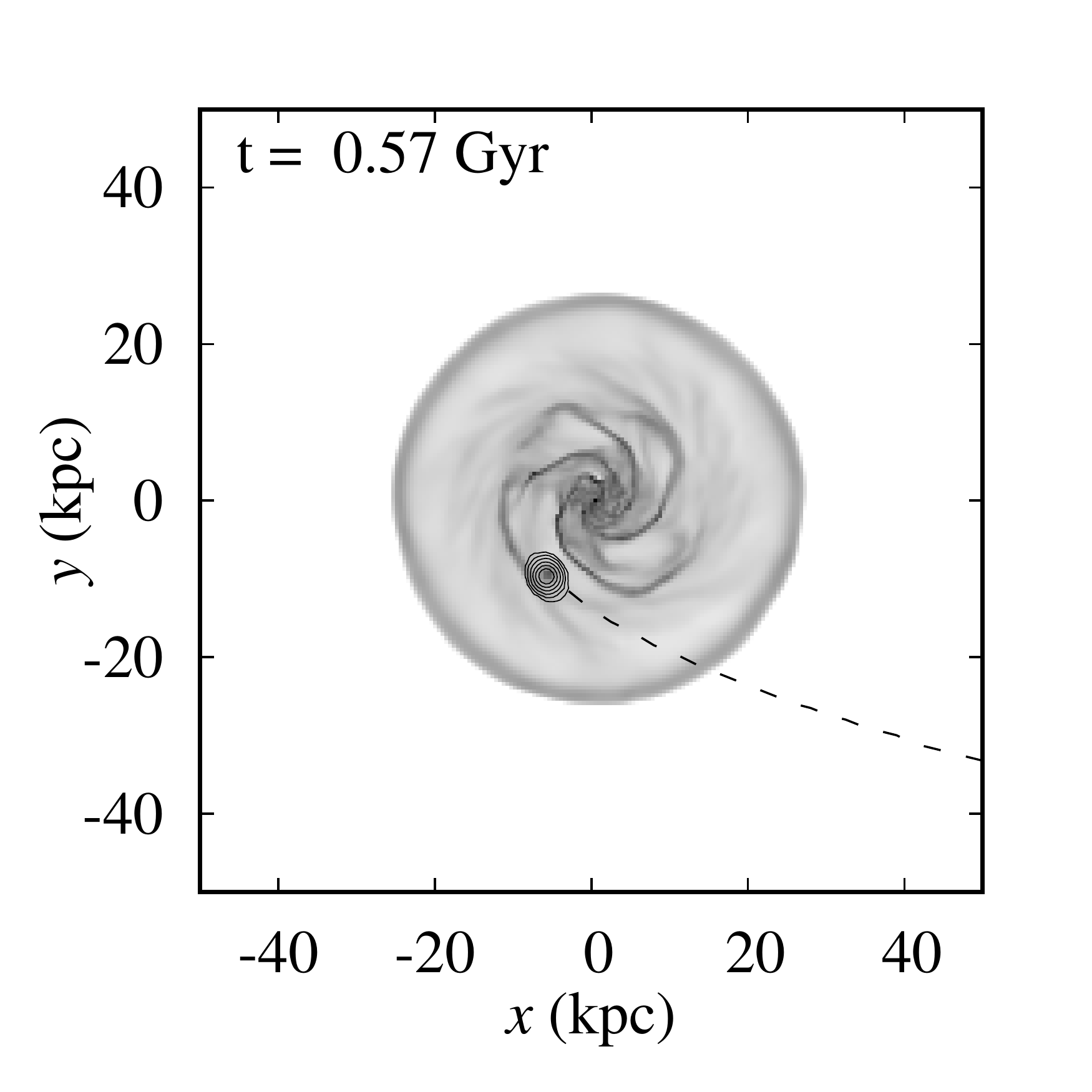}
\includegraphics[width=0.23\textwidth]{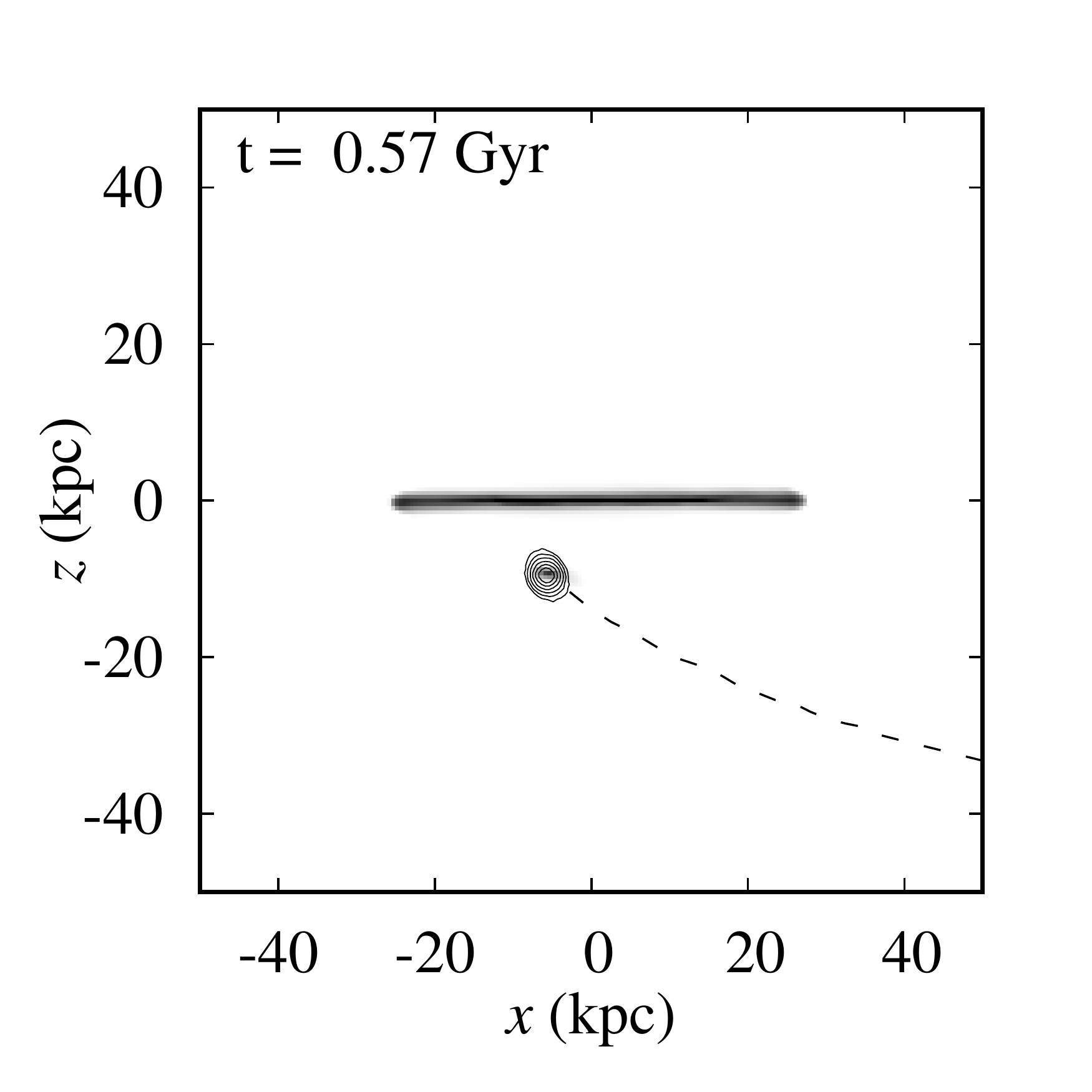}
\includegraphics[width=0.23\textwidth]{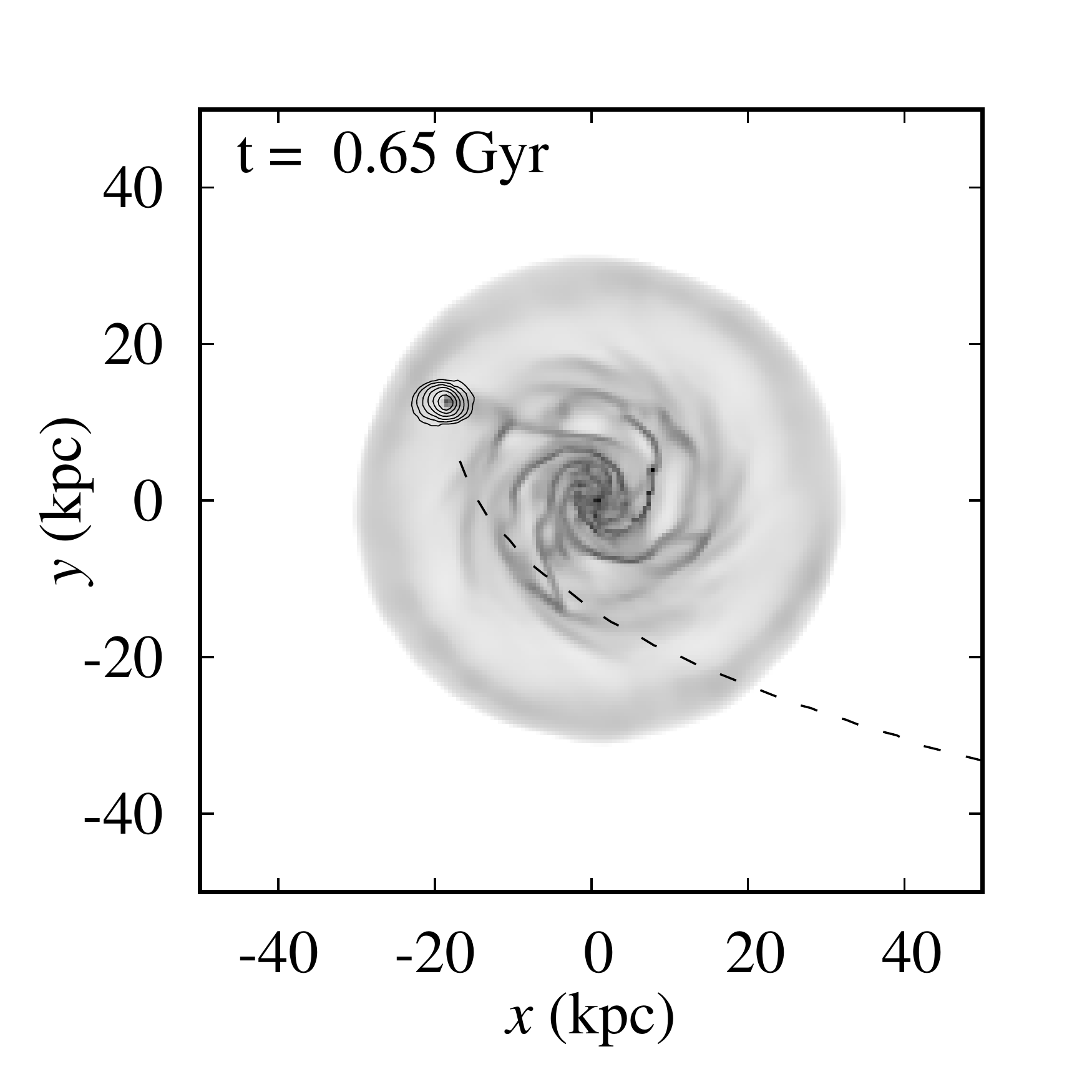}
\includegraphics[width=0.28\textwidth]{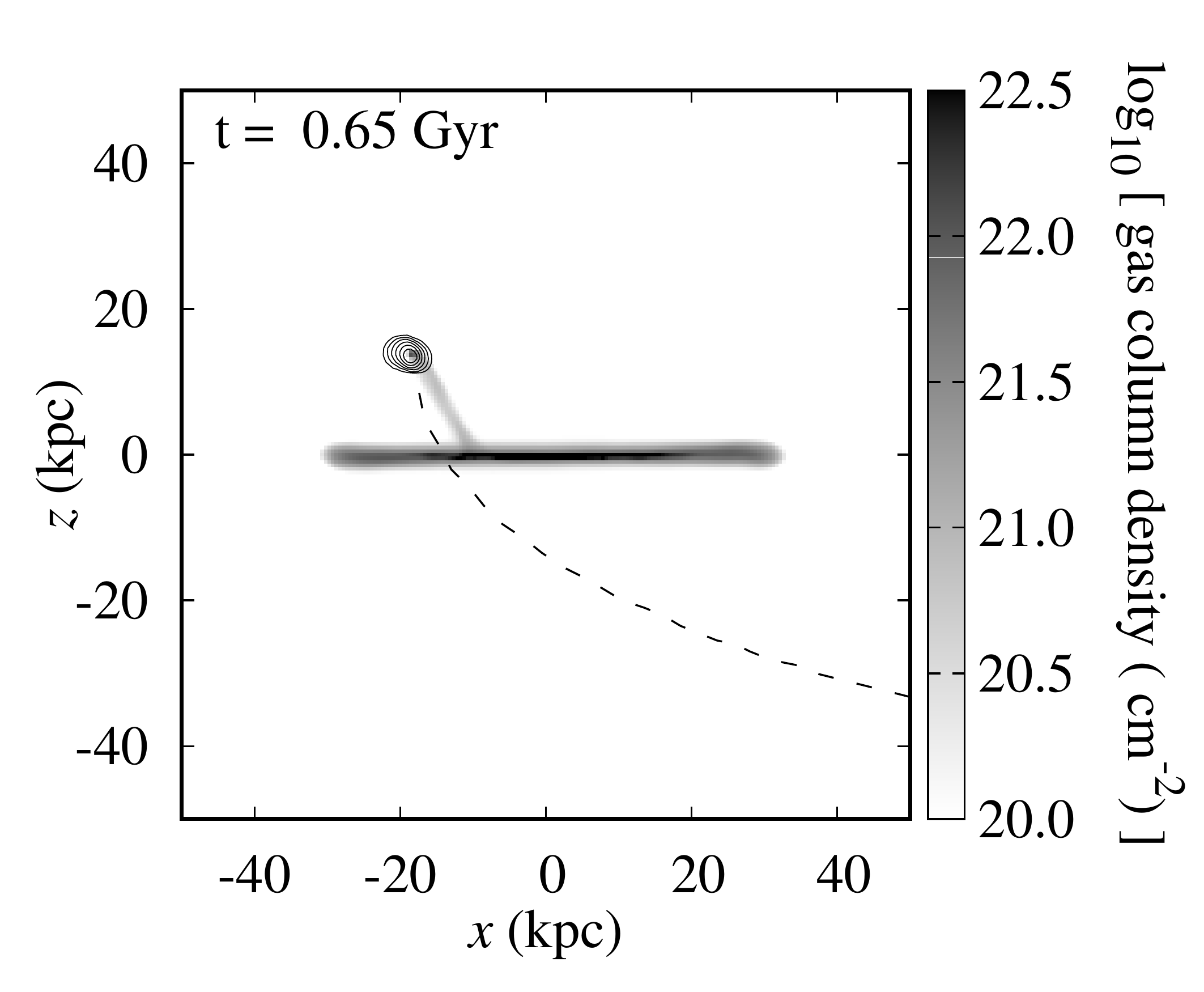}\\
\caption[ Gas / stellar disc (lo /hi res run) ]{ Gas column density (face-on and edge-on projections) prior to (columns 1 - 2) and after (columns 3 - 4) the subhalo's first transit in RUN 1 (top row) and RUN 2 (bottom row). The logarithmic value of the gas column density (in units of $\psc$) is indicated by the grey scale to the right of each row, and is the same for all panels. We have imposed a lower limit at a logarithmic value of 20 for displaying purposes. The contours indicate the projected DM density of the subhalo between $\sim 3 \times 10^6$ and $10^8$ $\Msun ~\kpc^{-2}$ in steps of 0.3 dex. The time indicated in the top left corner of each panel corresponds to the total simulation time. The total speed of the DM subhalo before (after) the transit is $\sim 420 ~\kms$ ($\sim 320 ~\kms$; see Figure \ref{fig:kin}). The dotted line in each panel indicates the subhalo's projected trajectory (up to its  position at each snapshot). The face-on projected gas disc is spinning clockwise; the subhalo is therefore on a {\em prograde} orbit with respect to the disc (and the hot halo; not visible here). The corresponding result in RUN 3 is qualitatively similar to RUN 1, and is therefore omitted. For an animated version of this figures and additional material follow this \href{http://www.physics.usyd.edu.au/~tepper/proj_smith_paper.html\#snpart_gc_run2}{link}.}
\label{fig:trans1_gas}
\end{figure*}

\begin{figure*}
\centering
\includegraphics[width=0.23\textwidth]{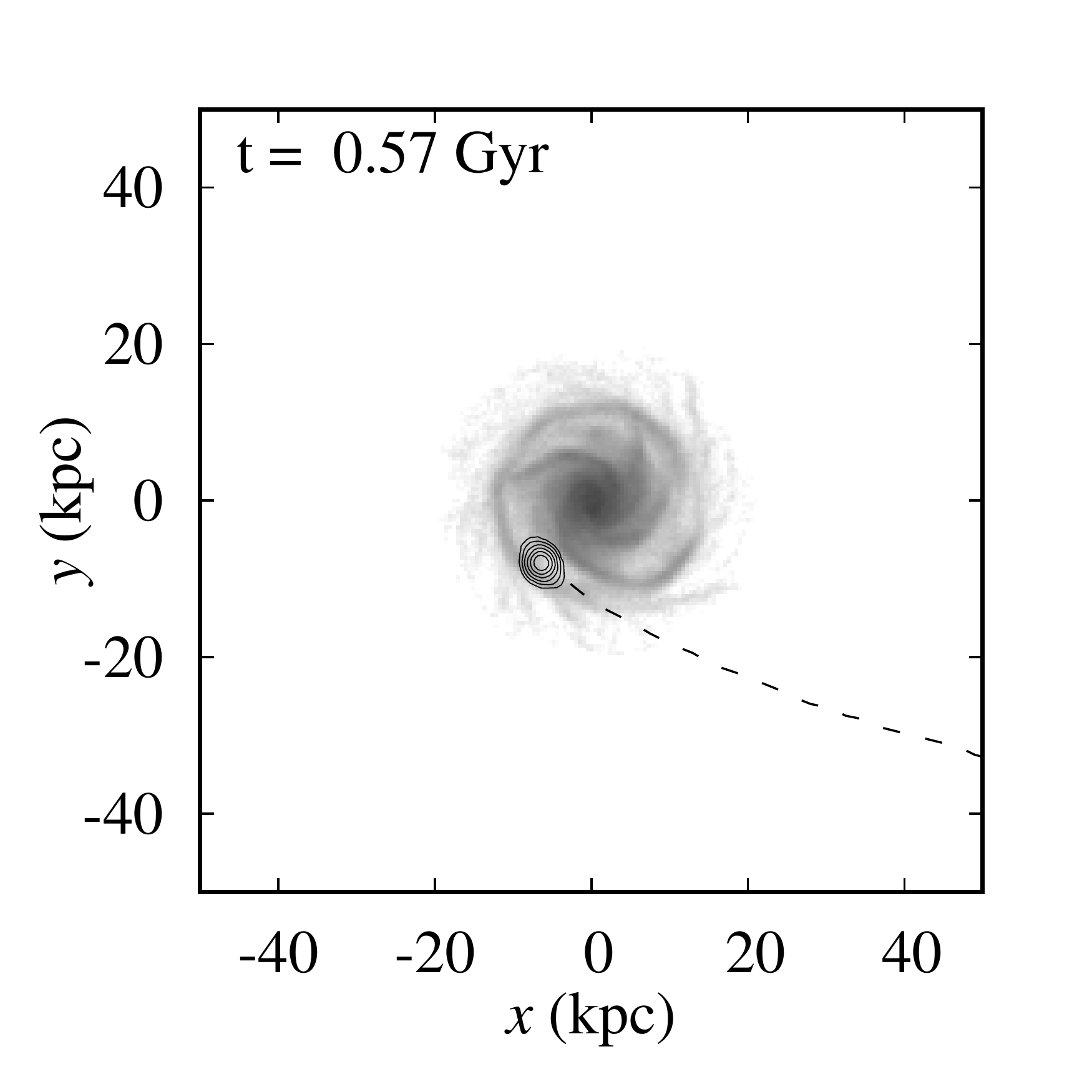}
\includegraphics[width=0.23\textwidth]{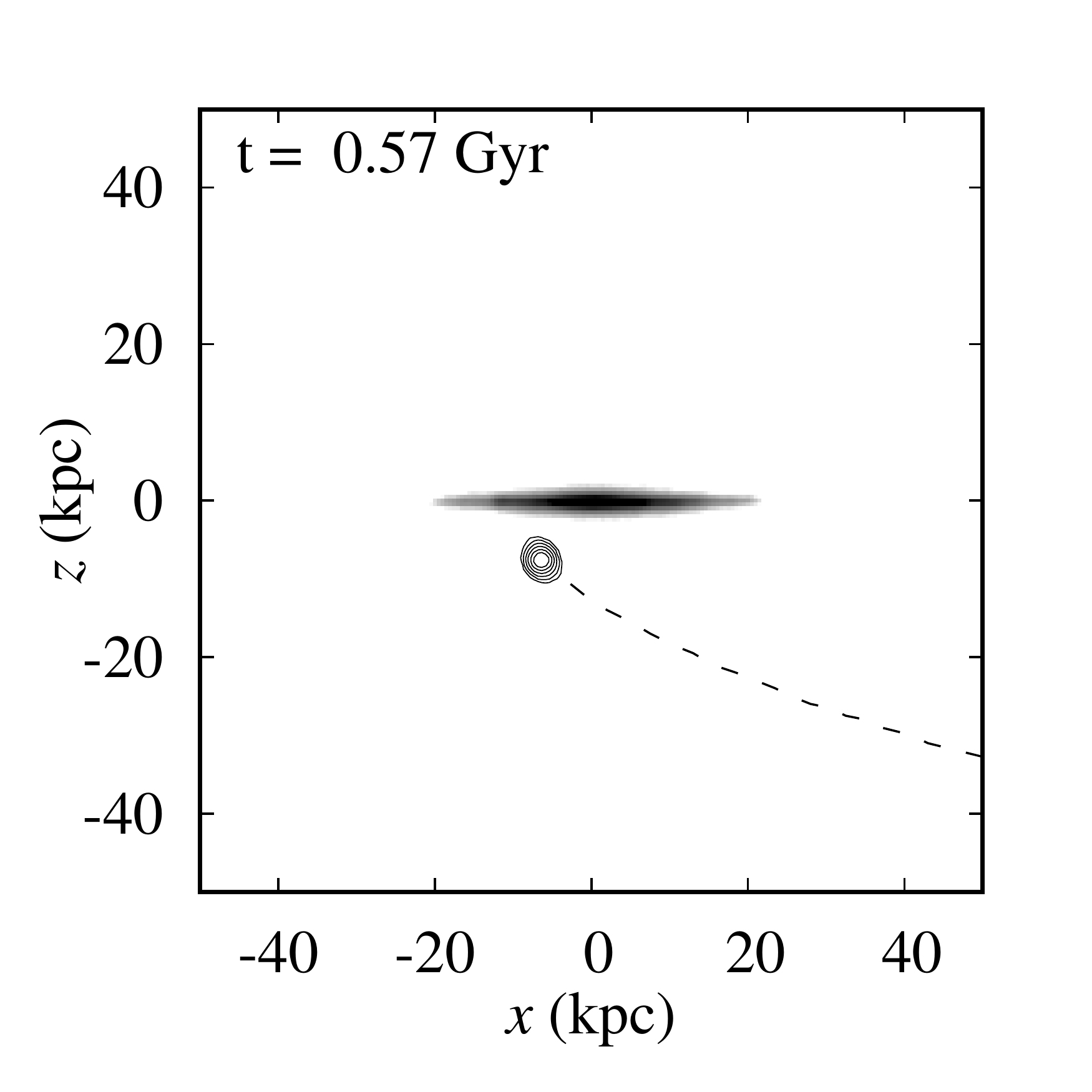}
\includegraphics[width=0.23\textwidth]{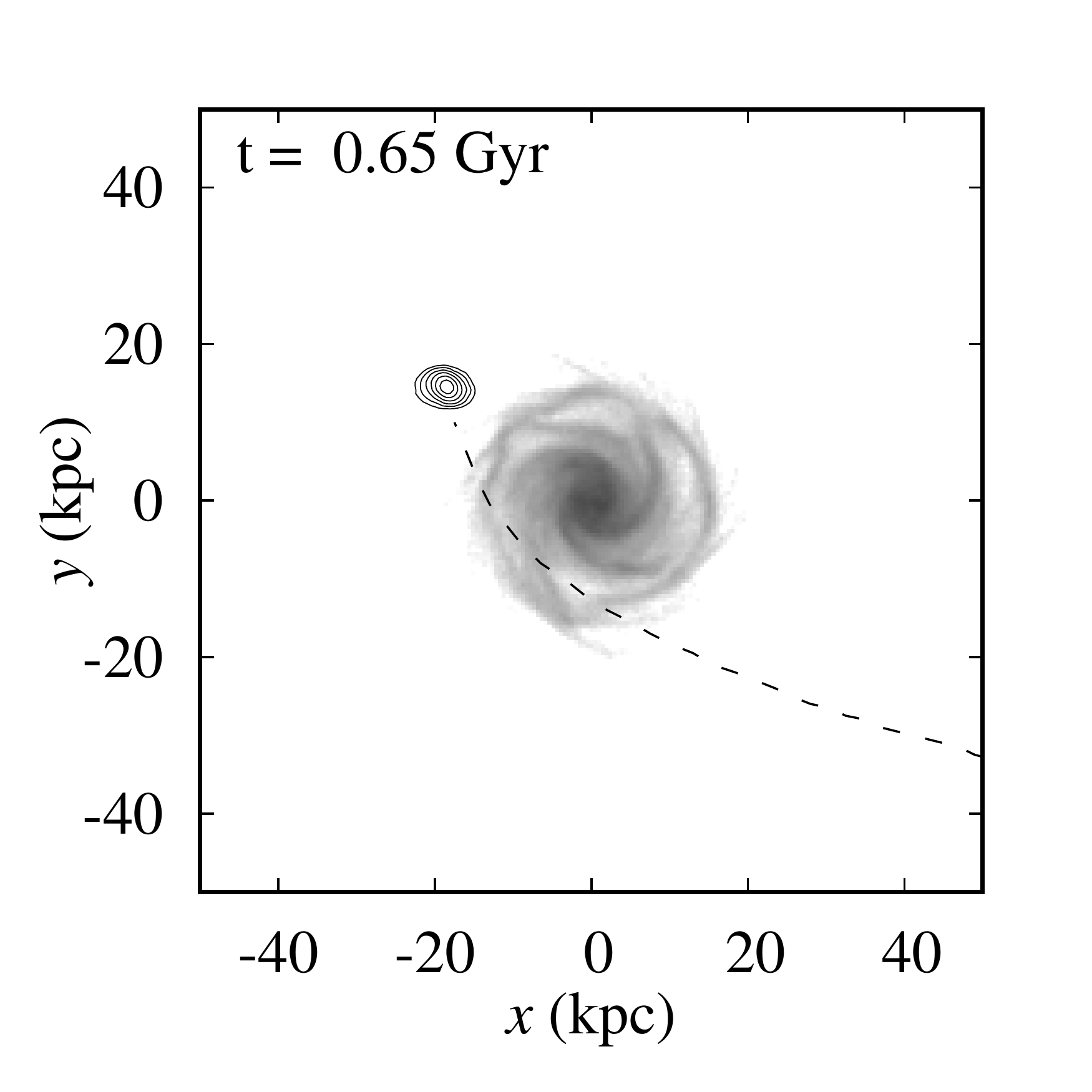}
\includegraphics[width=0.28\textwidth]{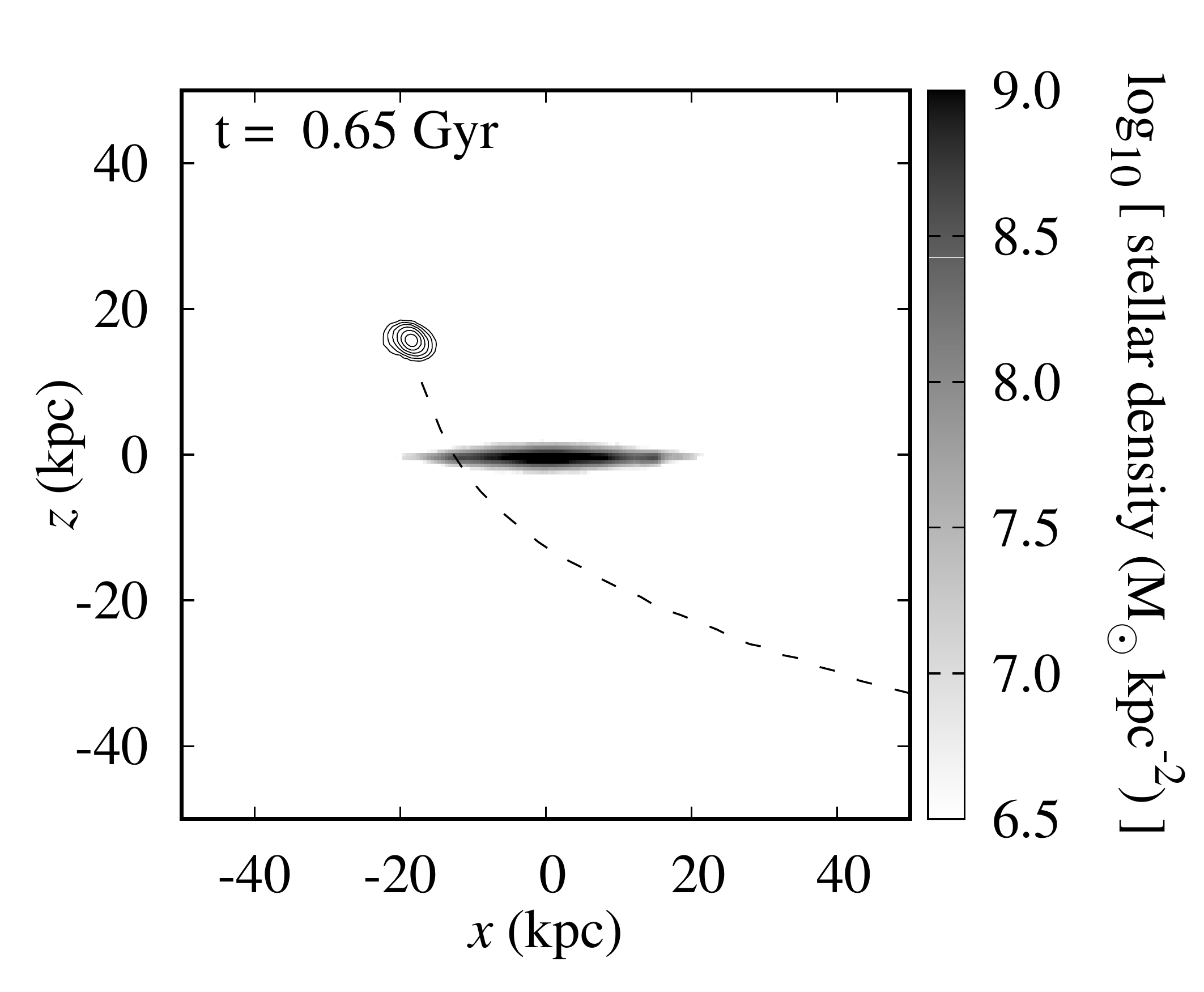}\\
\includegraphics[width=0.23\textwidth]{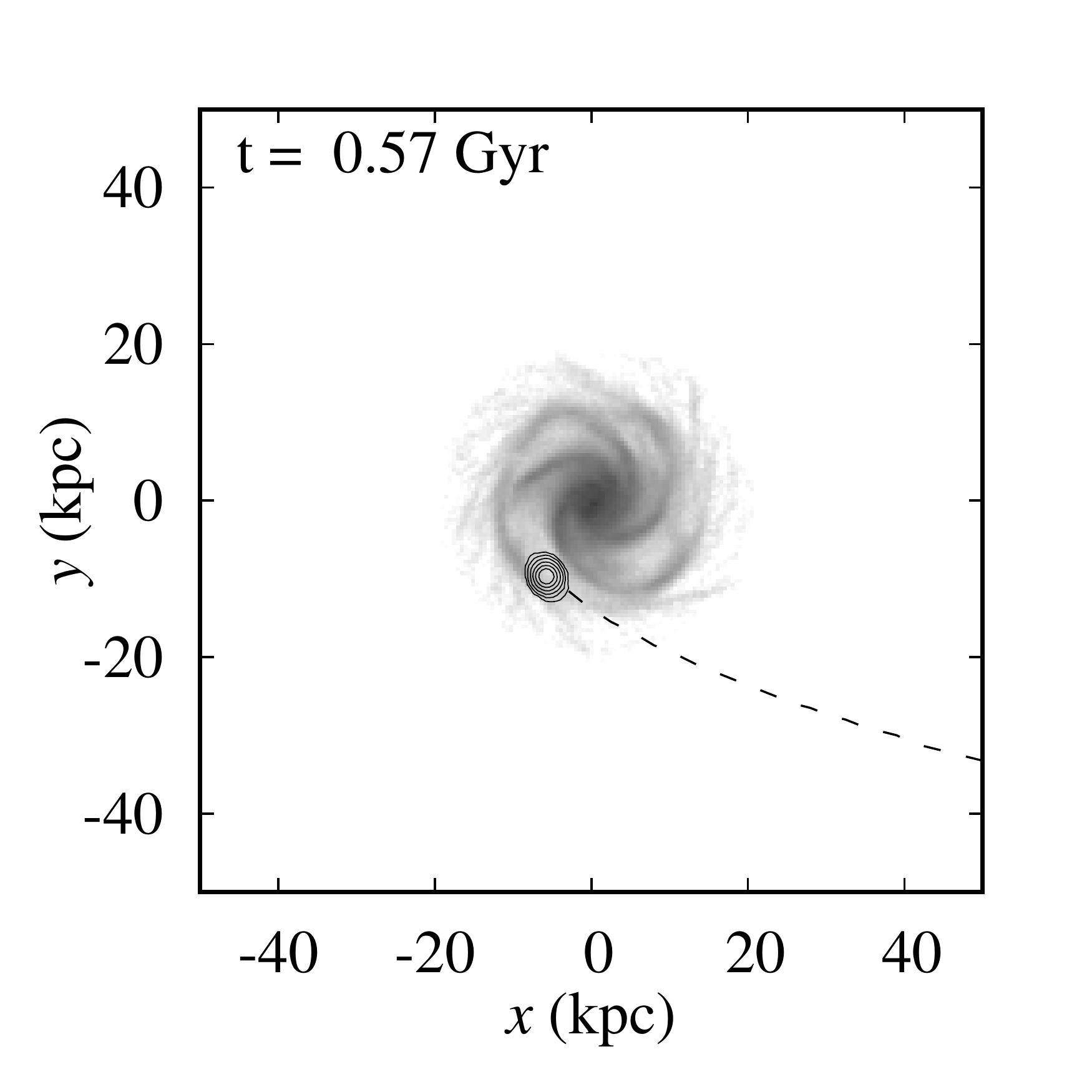}
\includegraphics[width=0.23\textwidth]{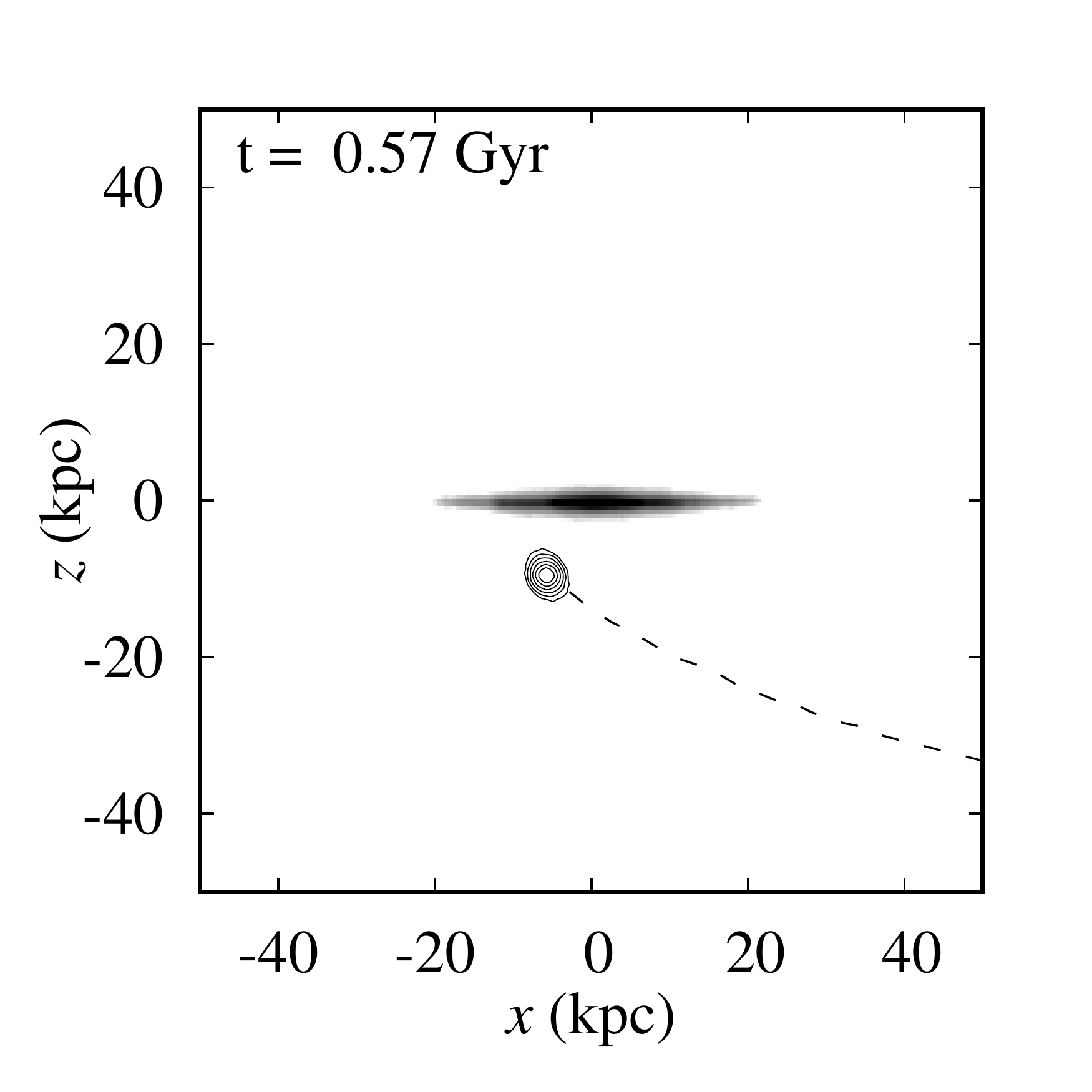}
\includegraphics[width=0.23\textwidth]{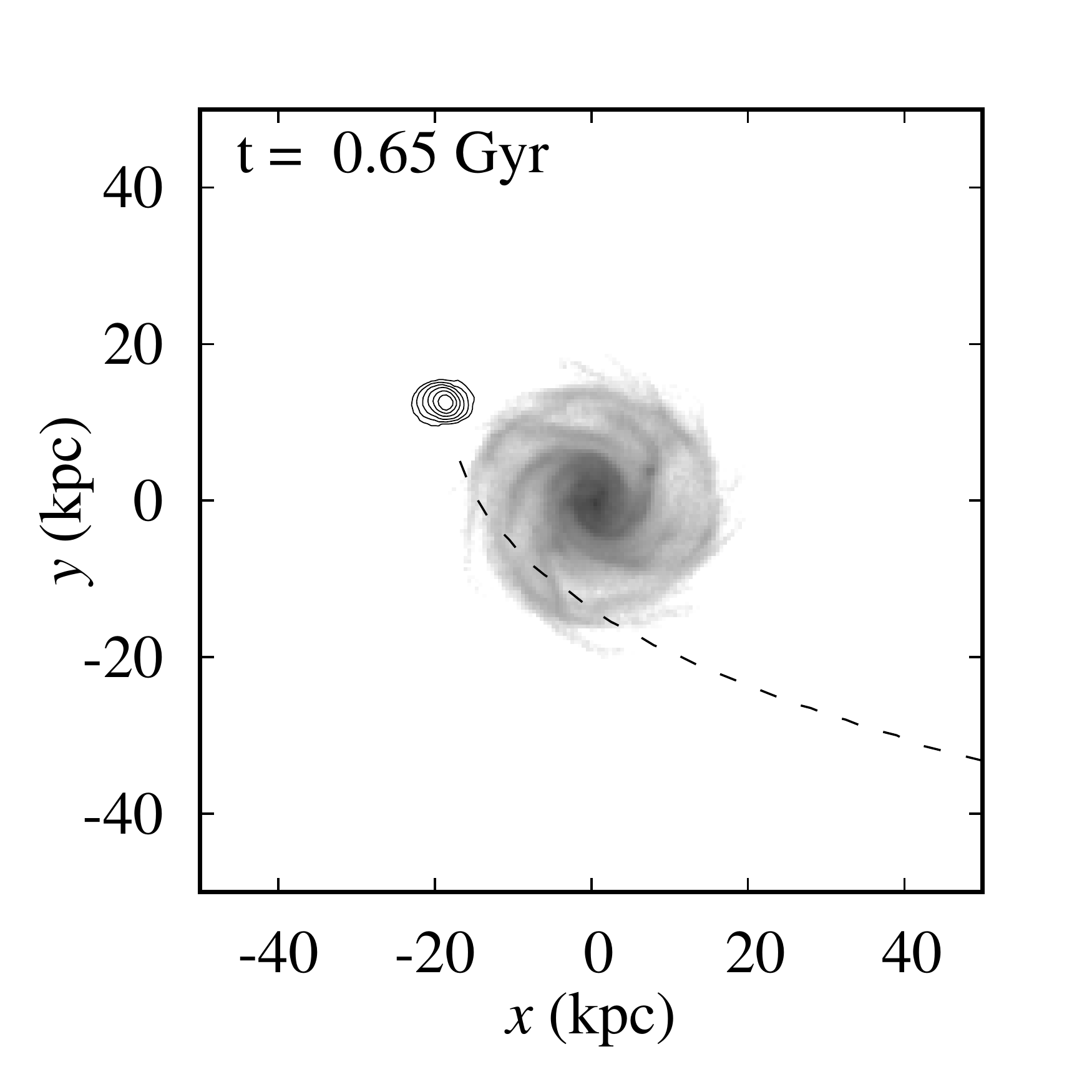}
\includegraphics[width=0.28\textwidth]{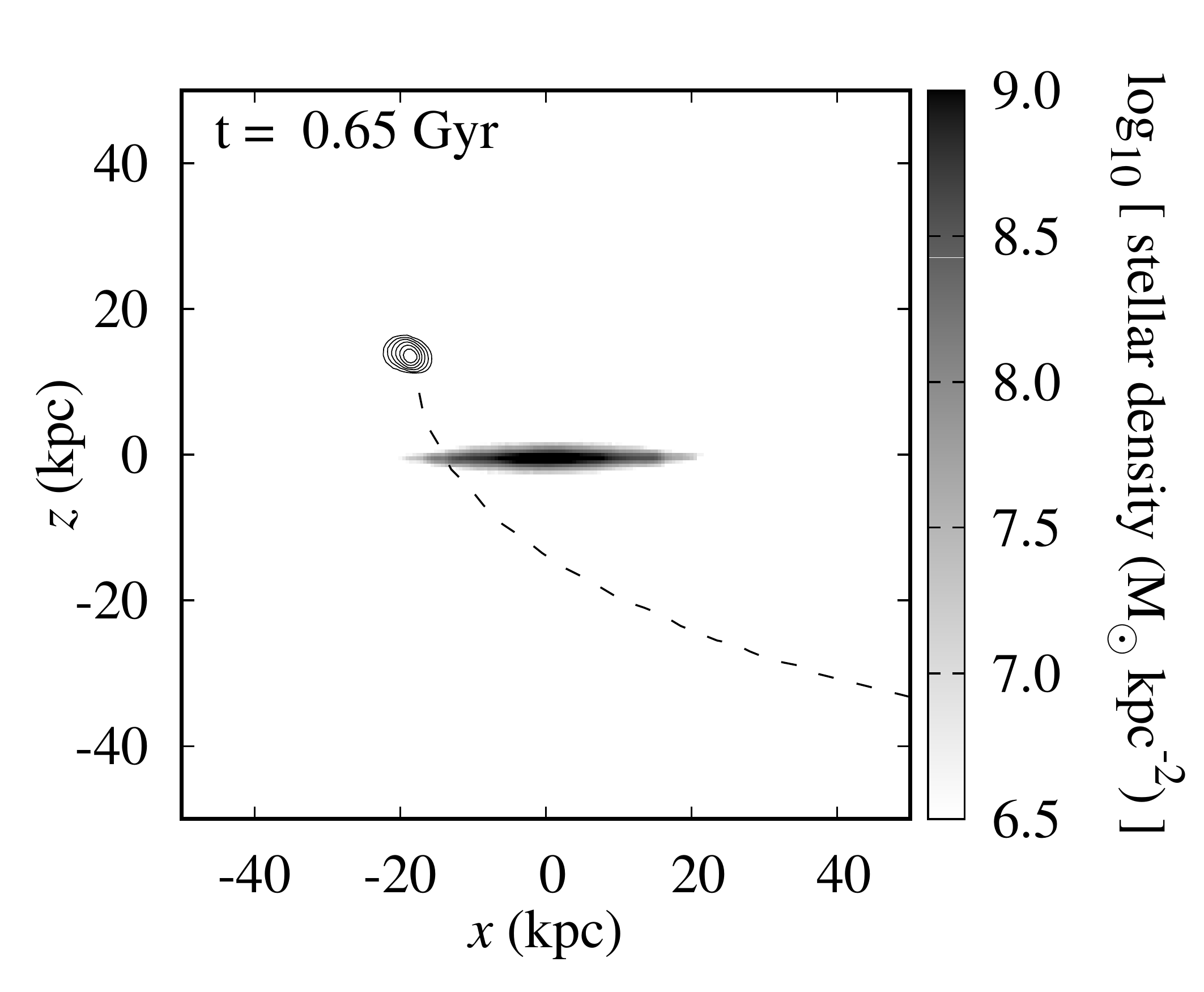}
\caption[ Gas / stellar disc (lo /hi res run) ]{ Same as Figure \ref{fig:trans1_gas}, but displaying the stellar surface density rather than the gas column density. The logarithmic value of the stellar surface density (in units of $\Msun ~\kpc^{-2}$) is indicated by the grey scale to the right of each row, and is the same for all panels. We have imposed a lower limit on the colour scale corresponding to a stellar density (in logarithmic scale) of 6.5 for displaying purposes. Like the gas disc, the face-on projected stellar disc is spinning clockwise. For an animated version of this figures and additional material follow this \href{http://www.physics.usyd.edu.au/~tepper/proj_smith_paper.html\#stars_gc_run2}{link}.}
\label{fig:trans1_star}
\end{figure*}

\begin{figure*}
\centering
\includegraphics[width=0.23\textwidth]{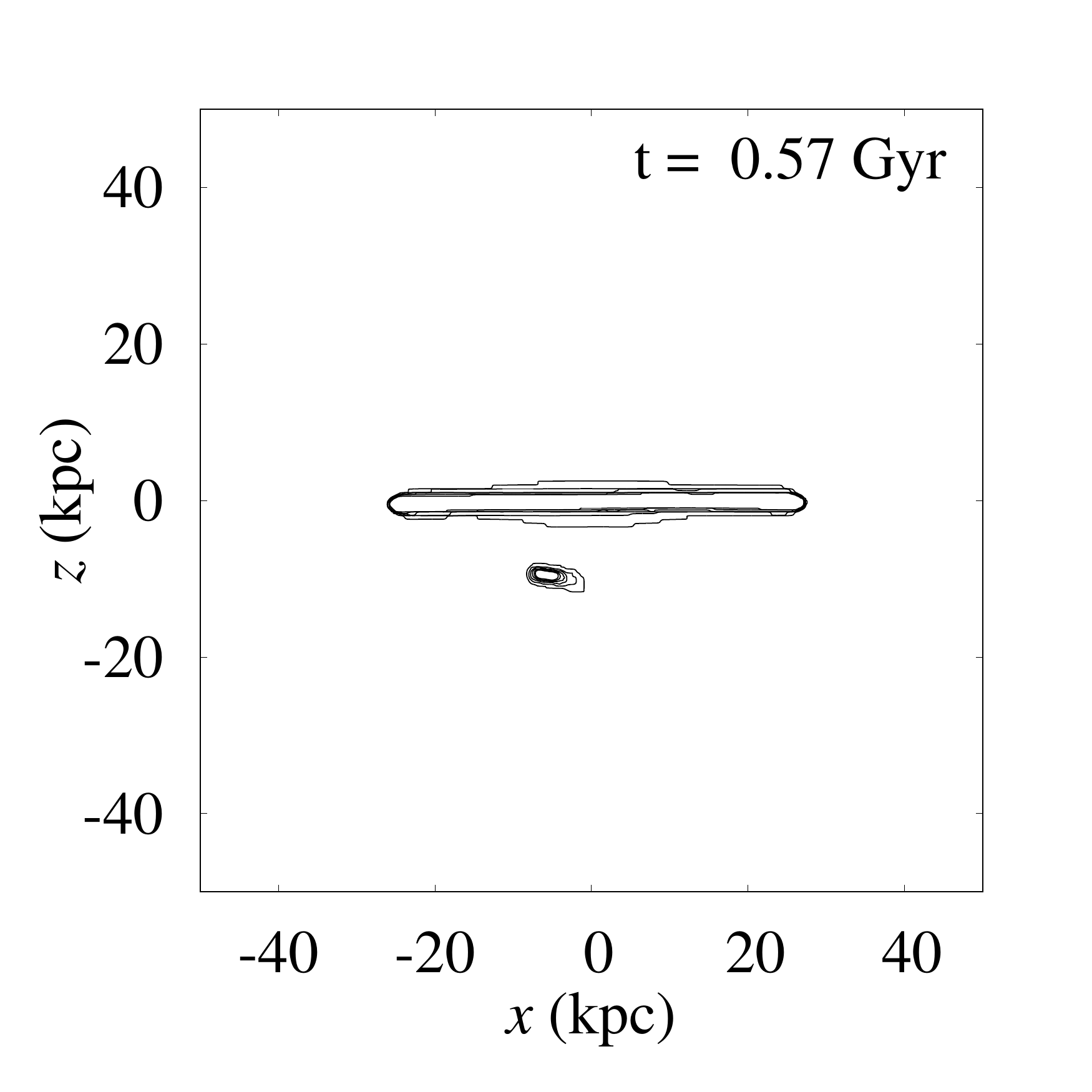}
\includegraphics[width=0.23\textwidth]{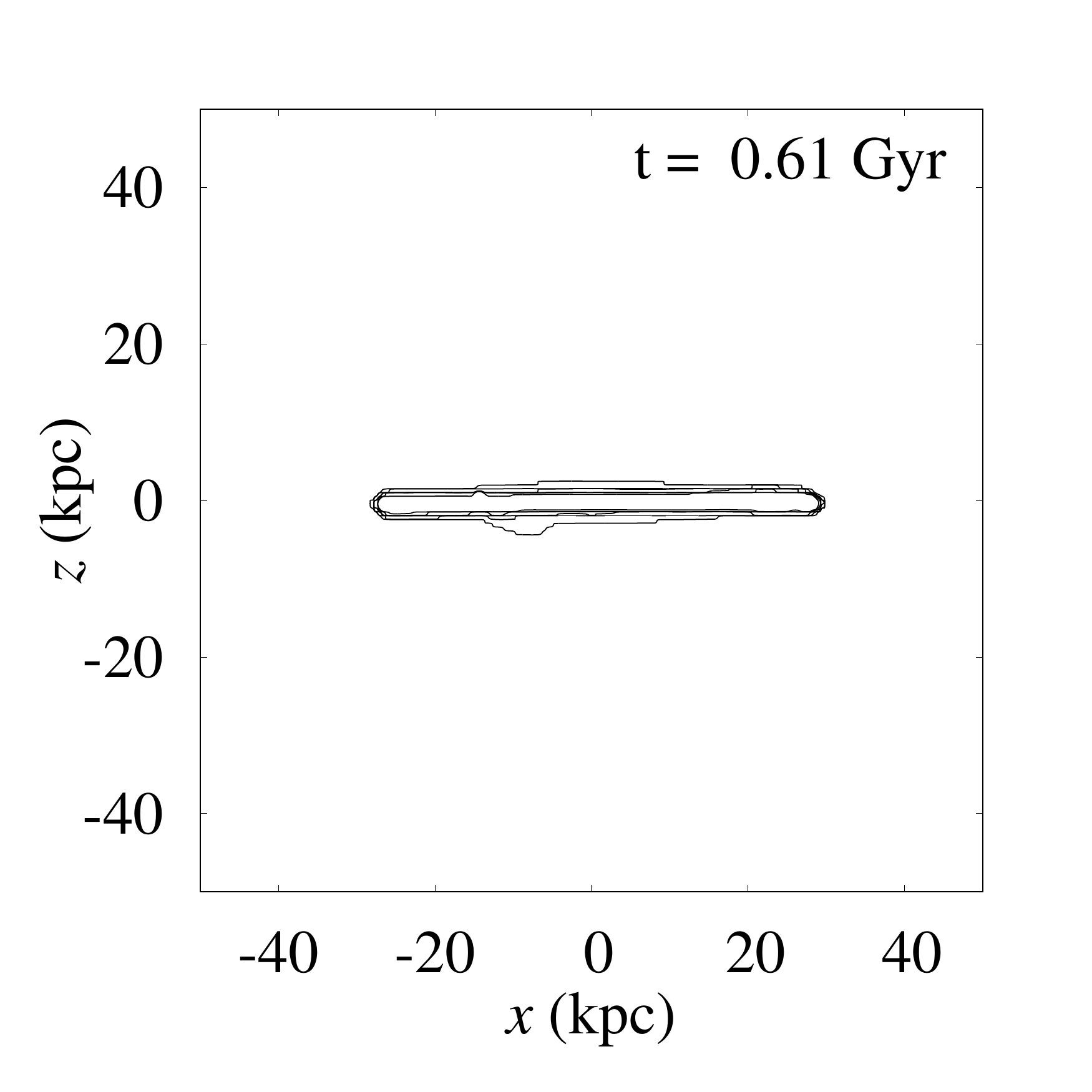}
\includegraphics[width=0.23\textwidth]{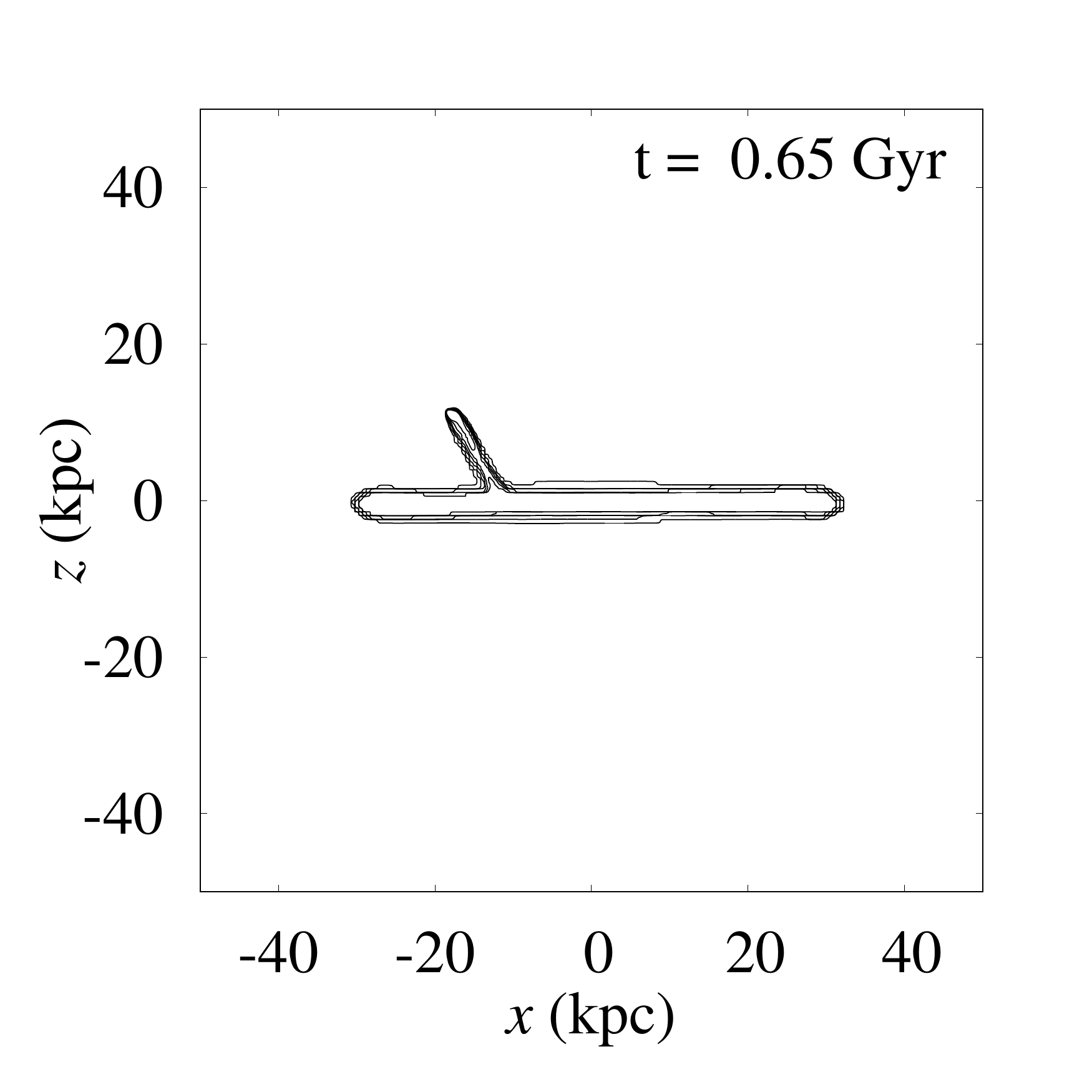}
\includegraphics[width=0.23\textwidth]{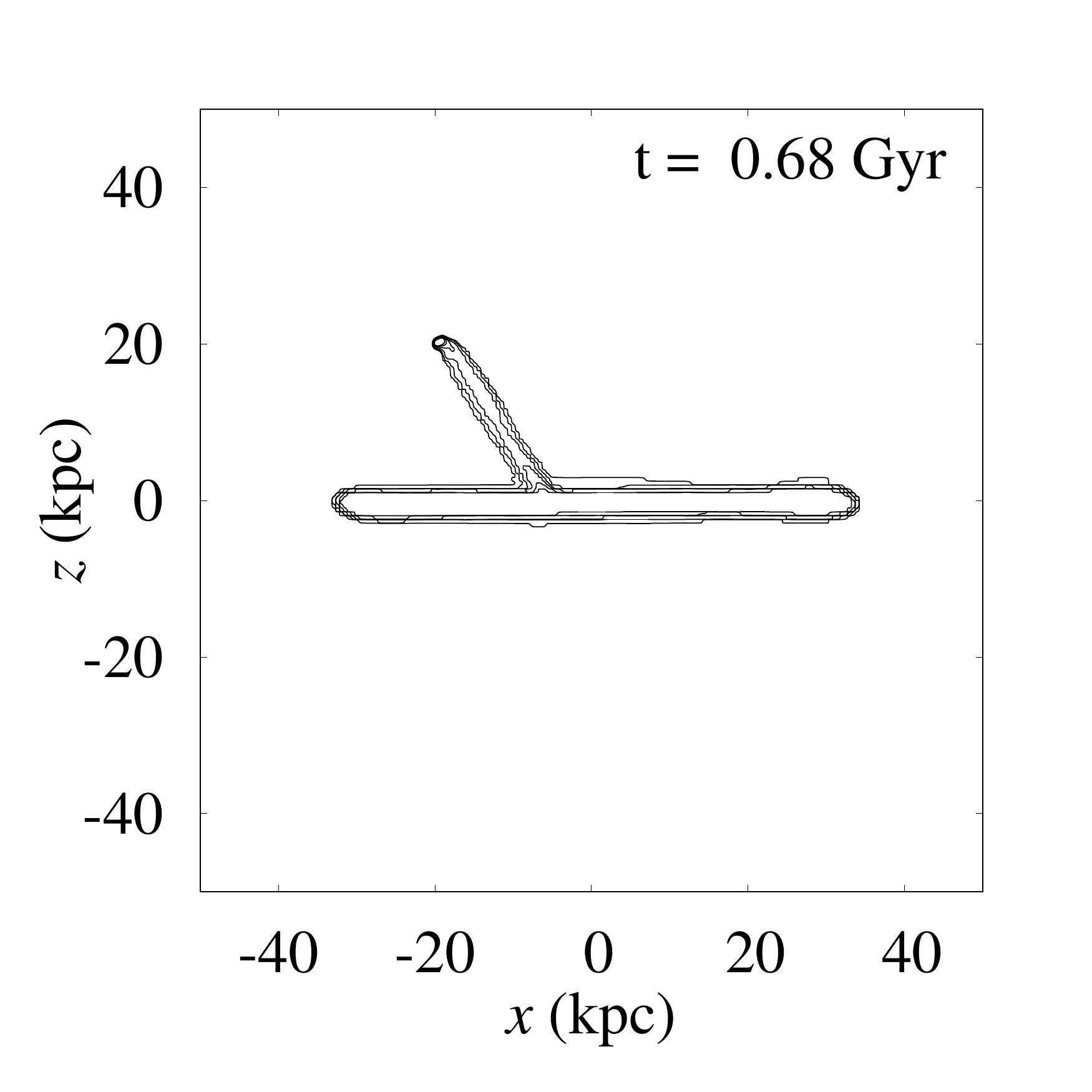}\\
\vspace{-5pt}
\includegraphics[width=0.23\textwidth]{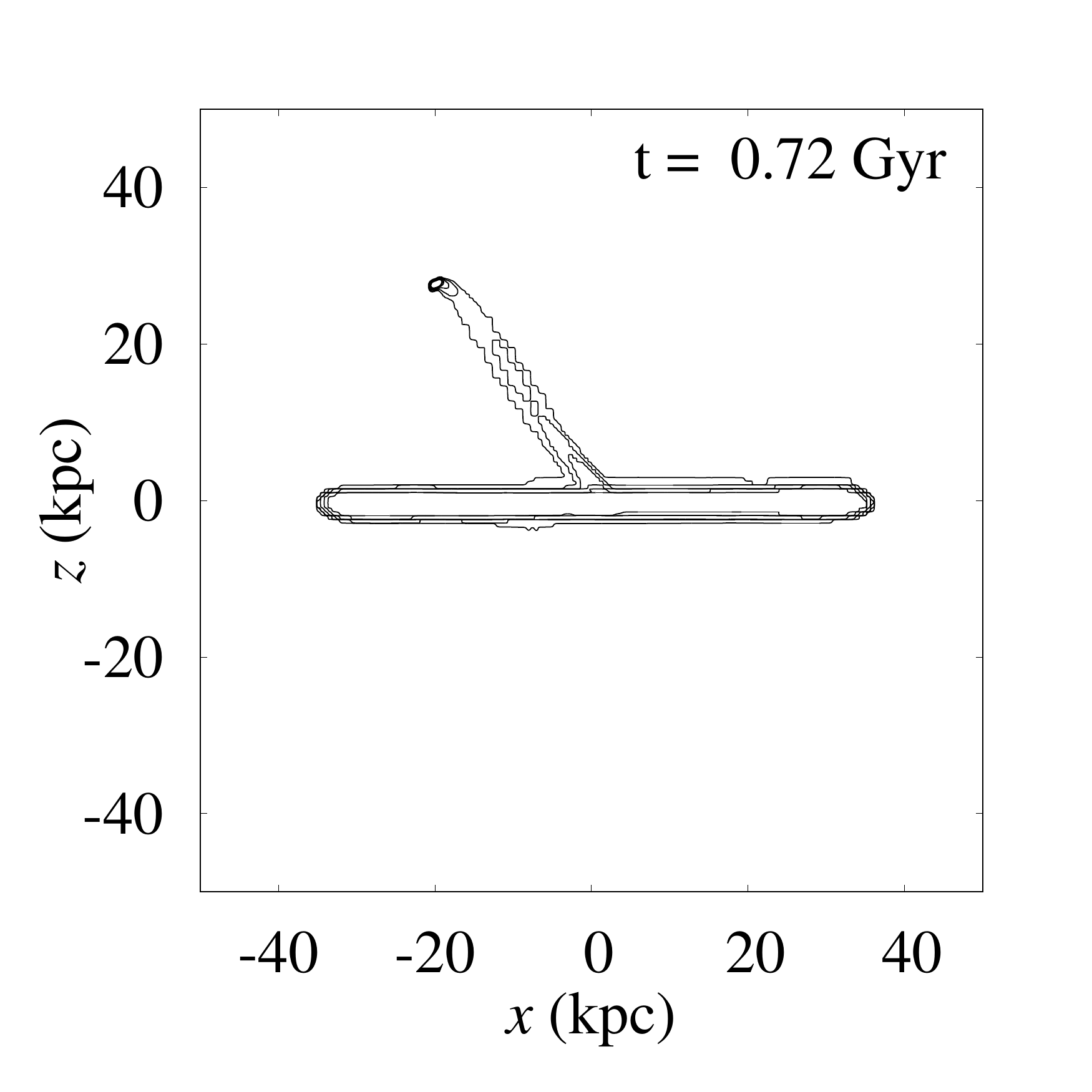}
\includegraphics[width=0.23\textwidth]{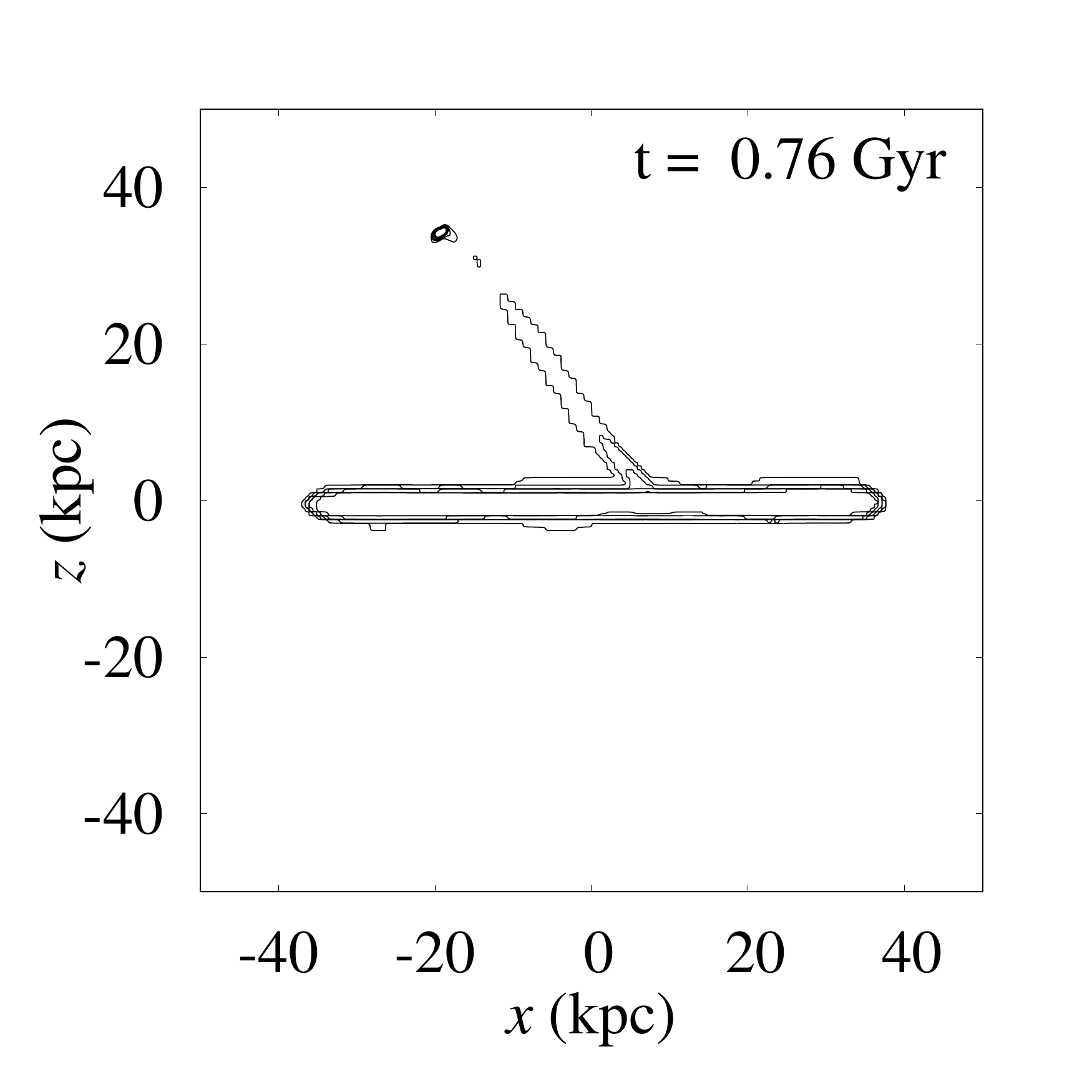}
\includegraphics[width=0.23\textwidth]{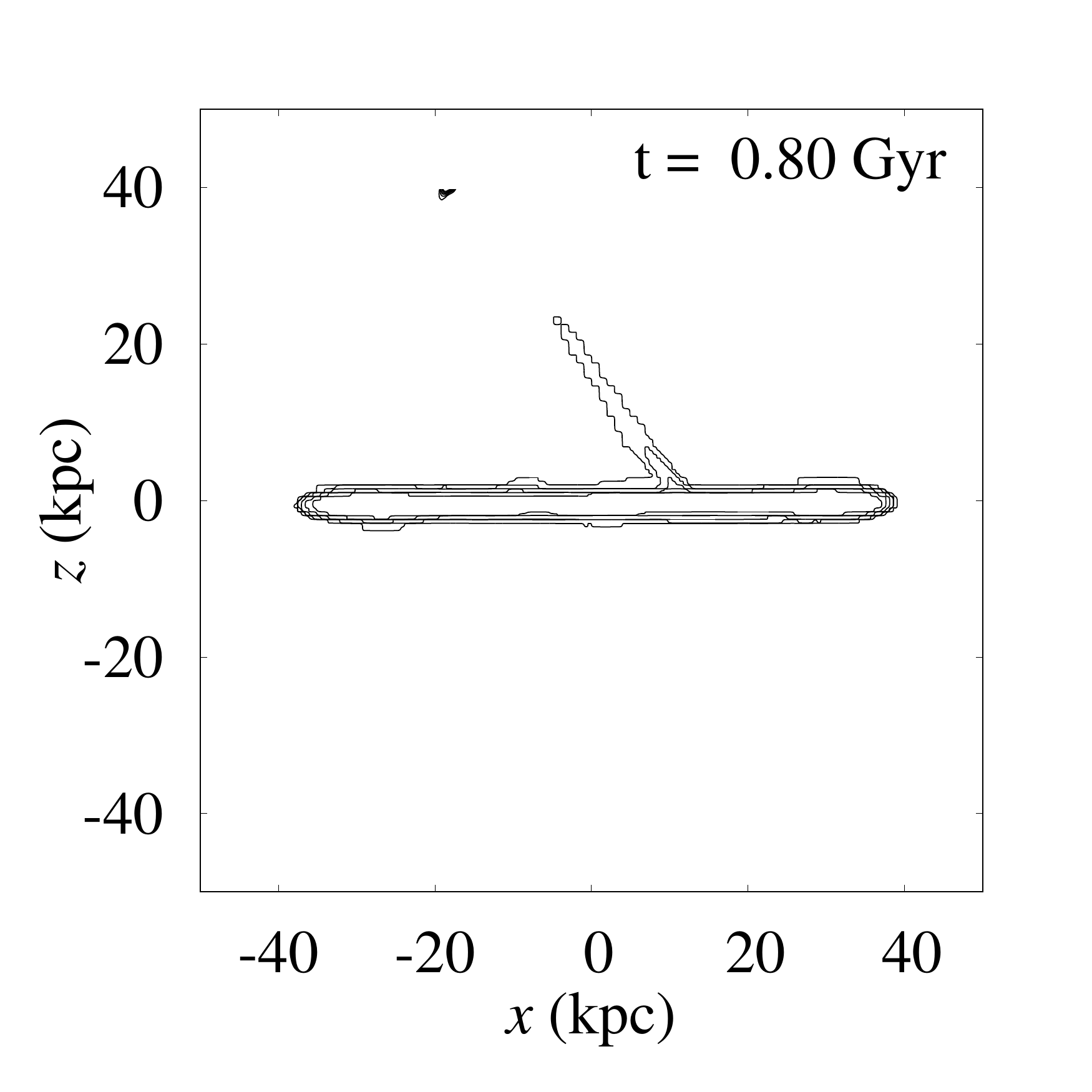}
\includegraphics[width=0.23\textwidth]{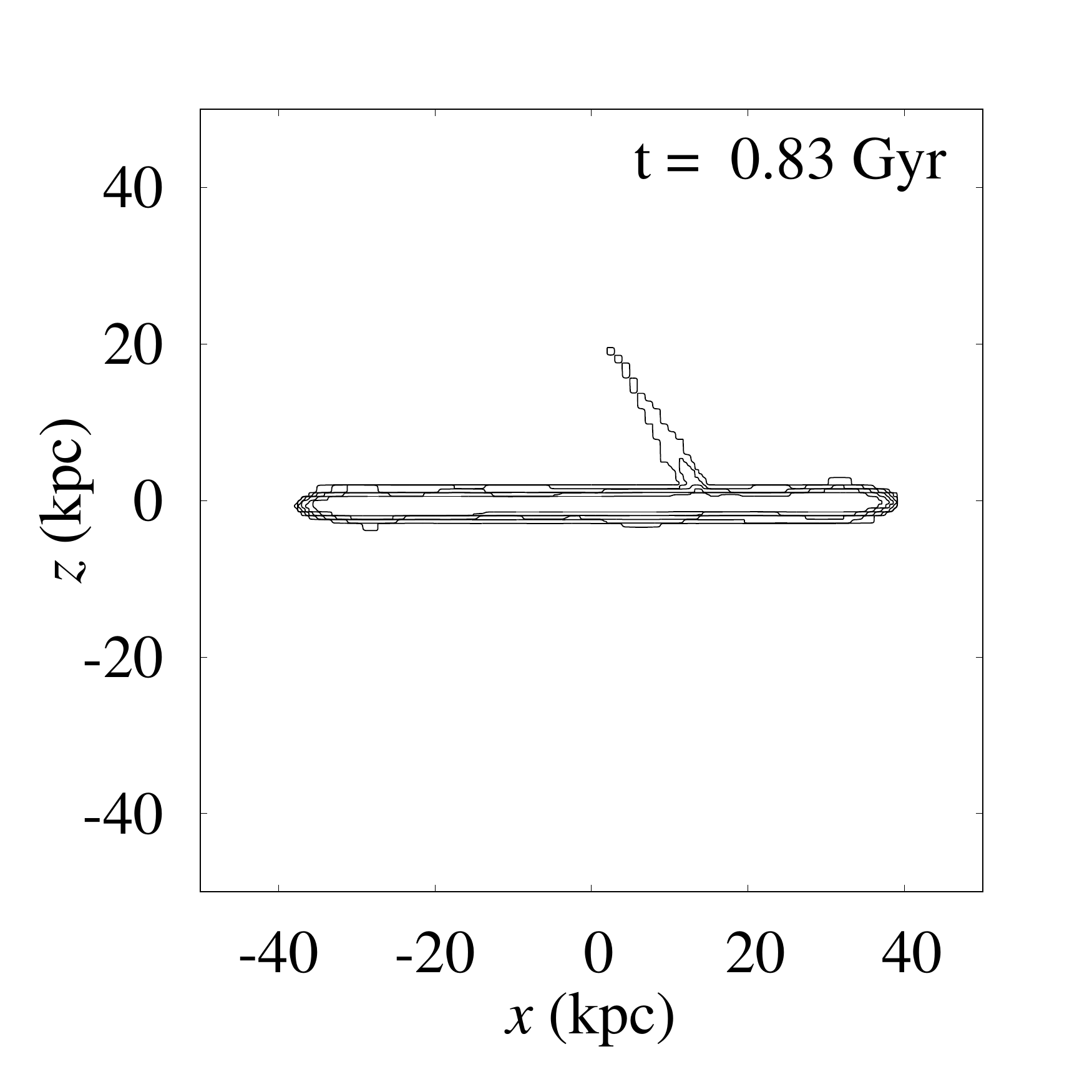}\\
\vspace{-5pt}
\includegraphics[width=0.23\textwidth]{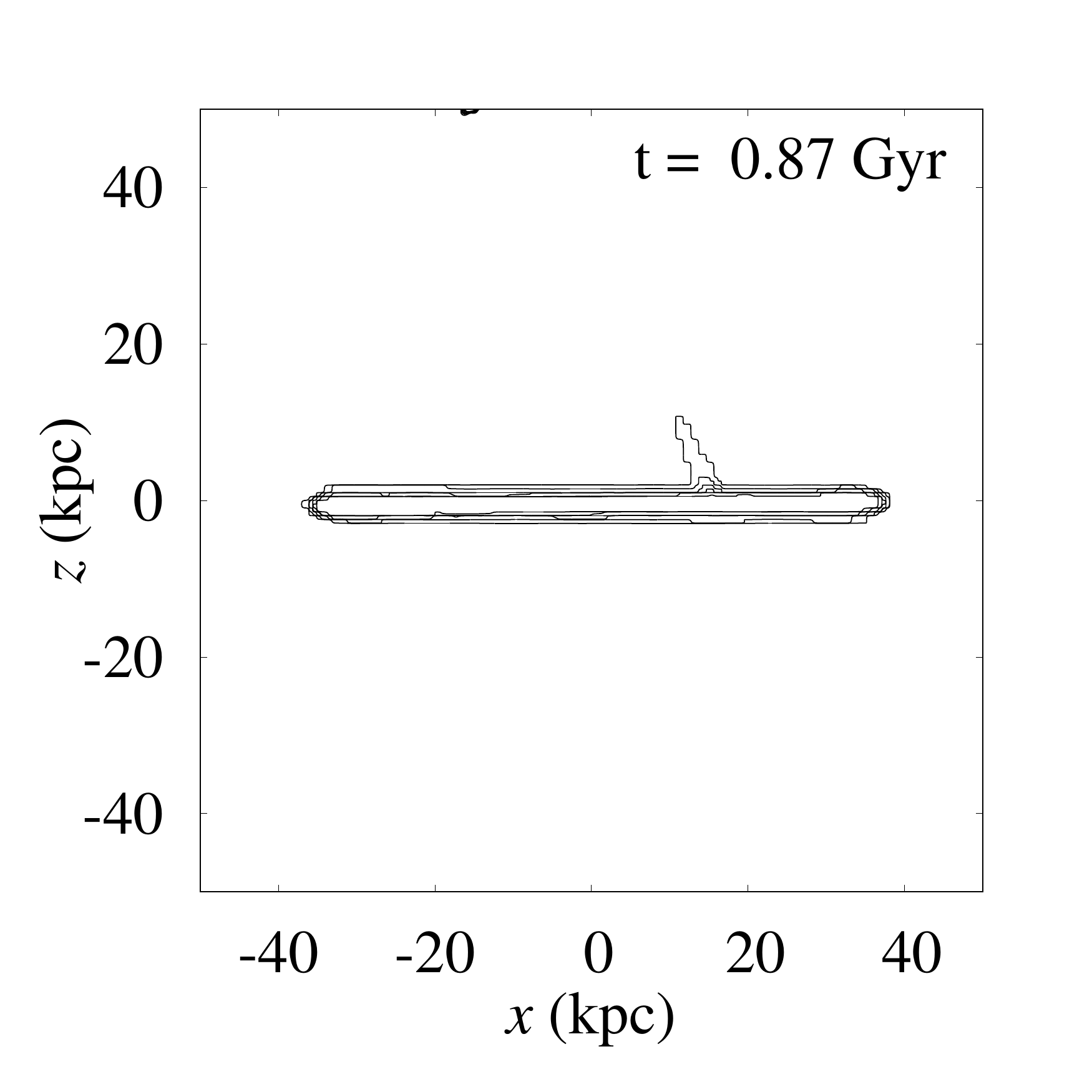}
\includegraphics[width=0.23\textwidth]{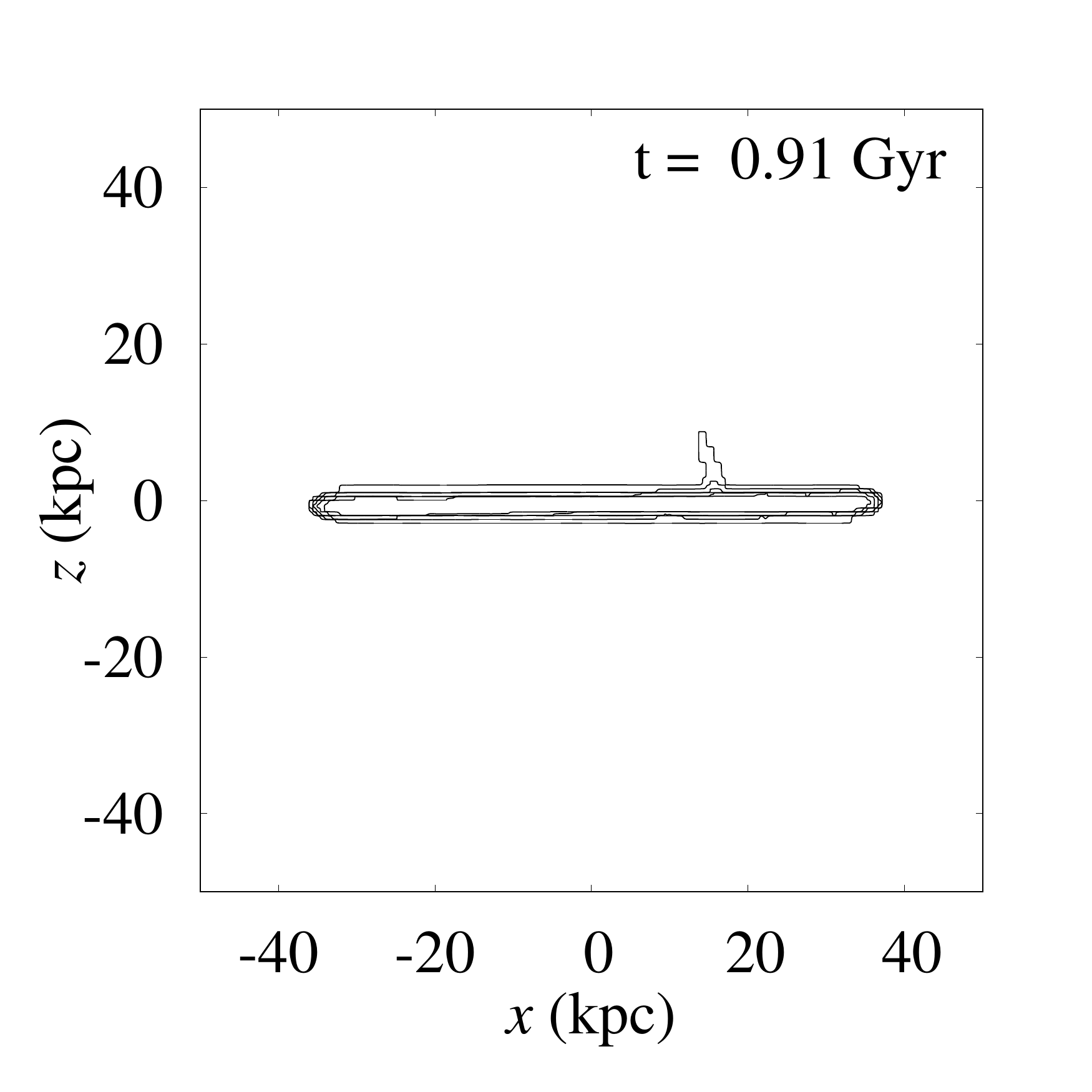}
\includegraphics[width=0.23\textwidth]{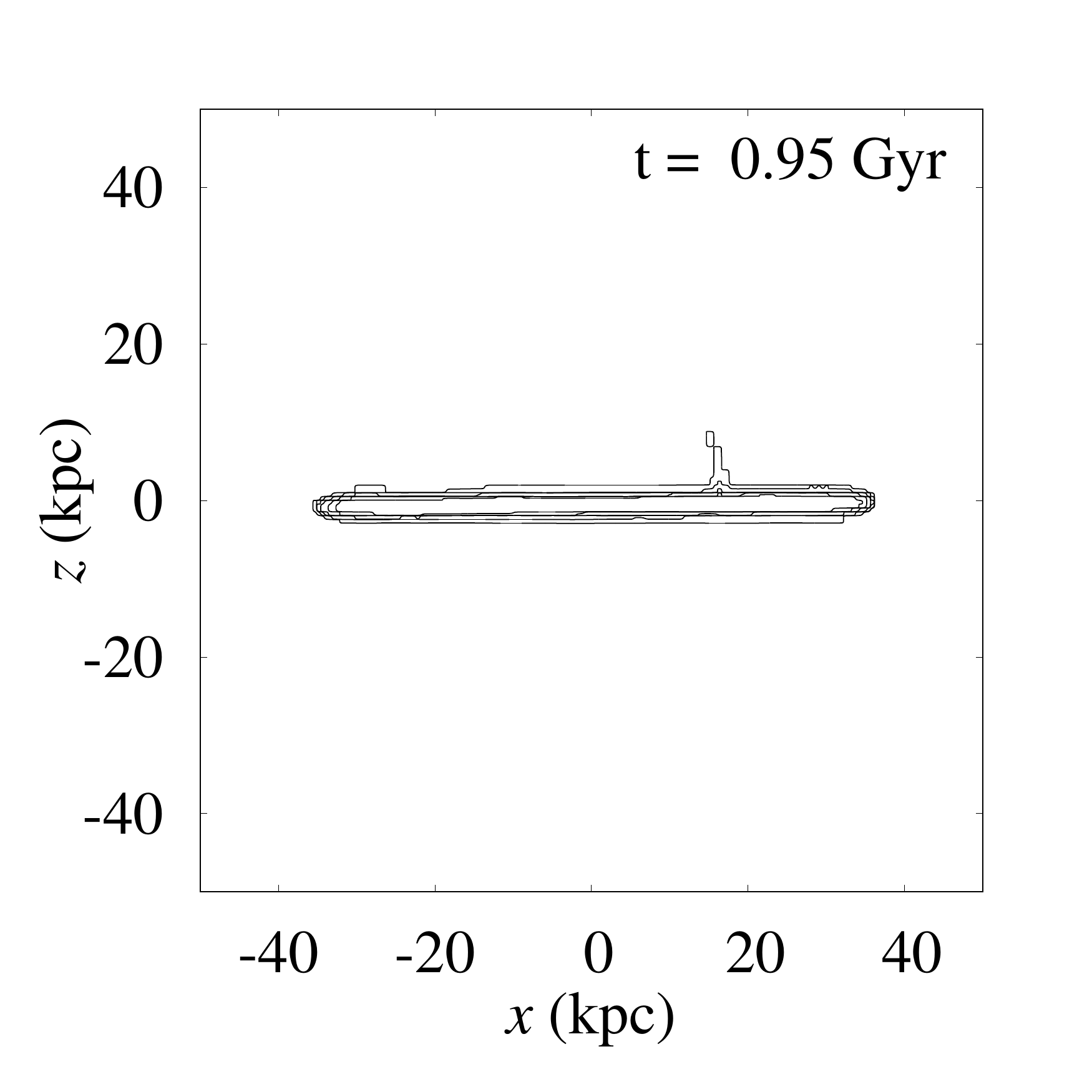}
\includegraphics[width=0.23\textwidth]{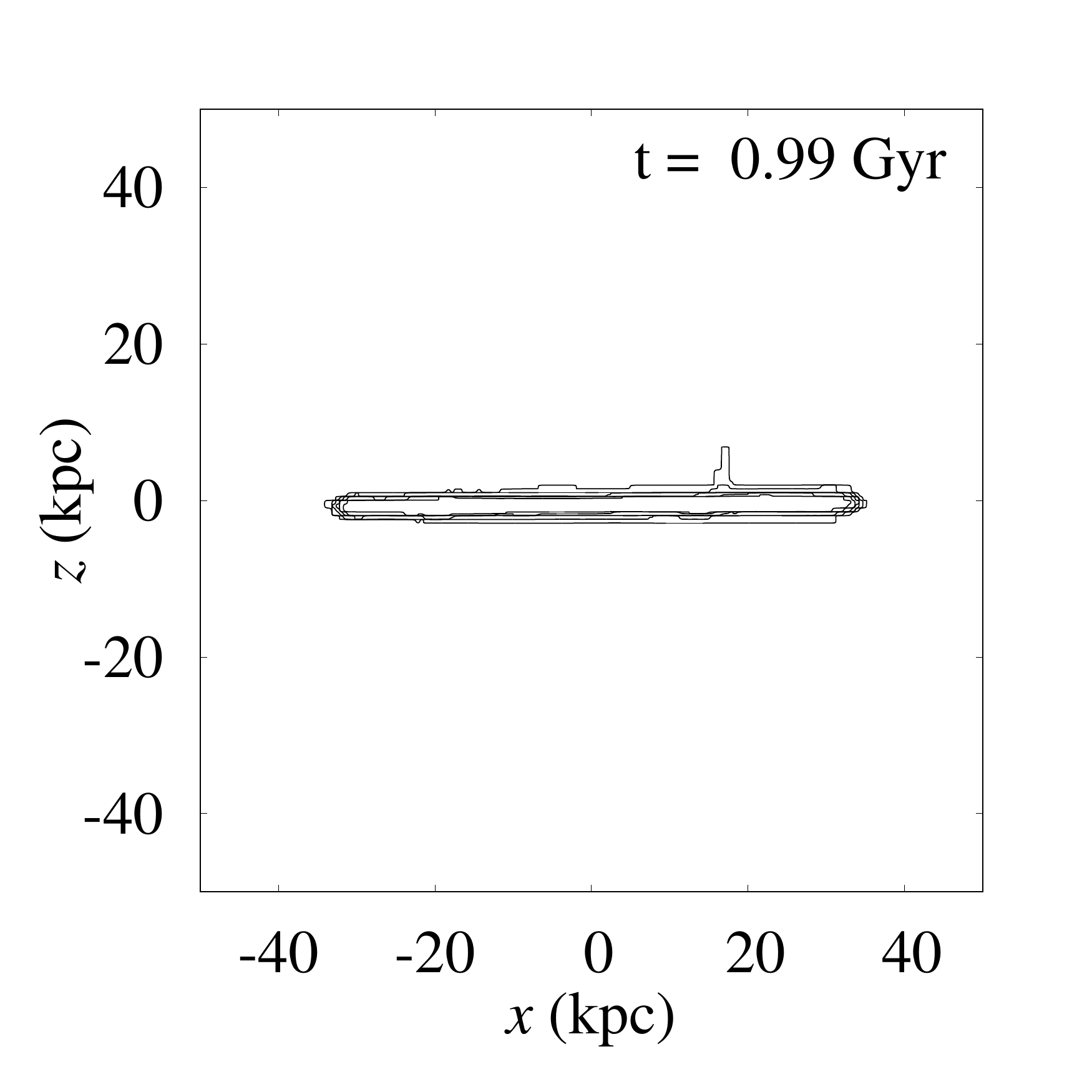}\\
\vspace{-5pt}
\includegraphics[width=0.23\textwidth]{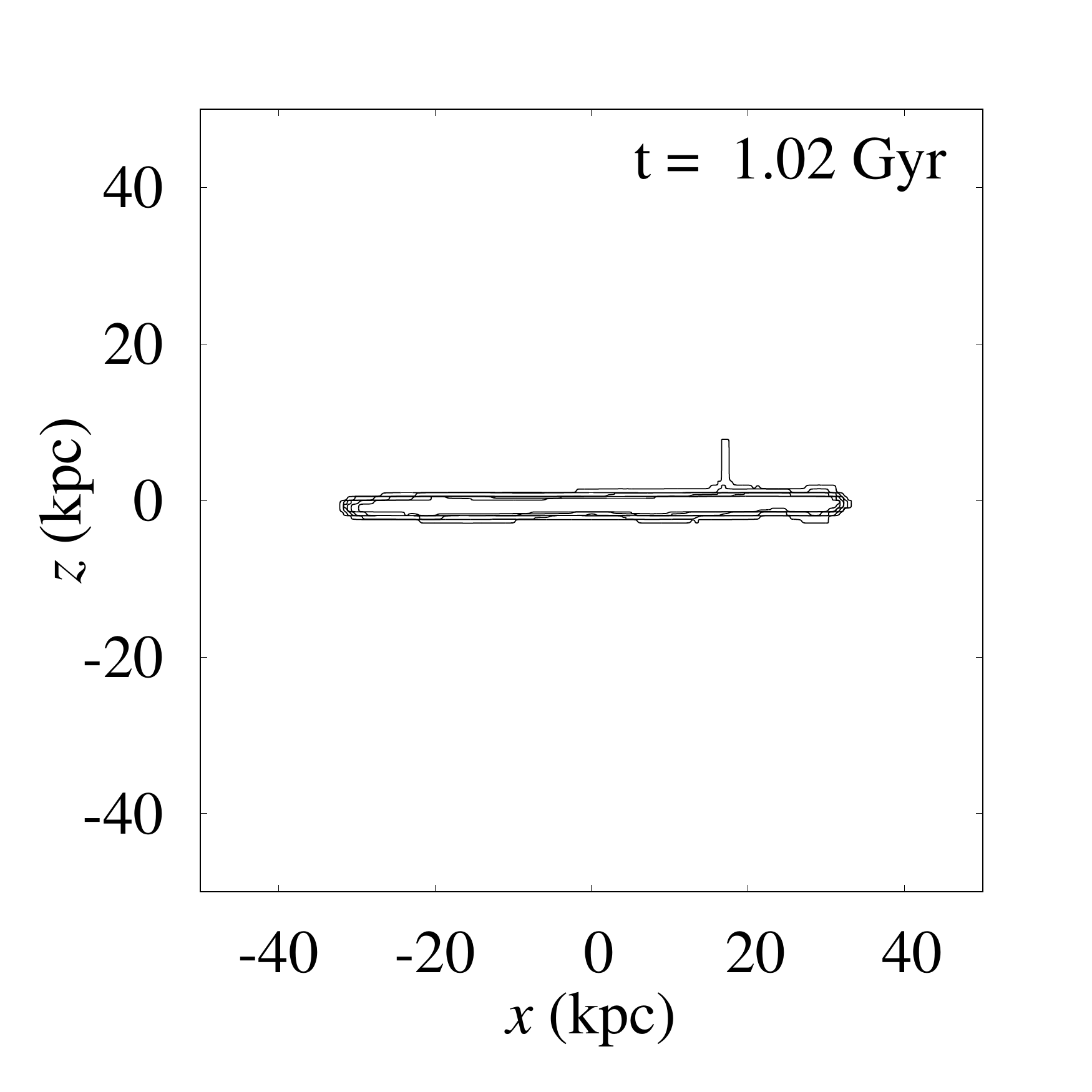}
\includegraphics[width=0.23\textwidth]{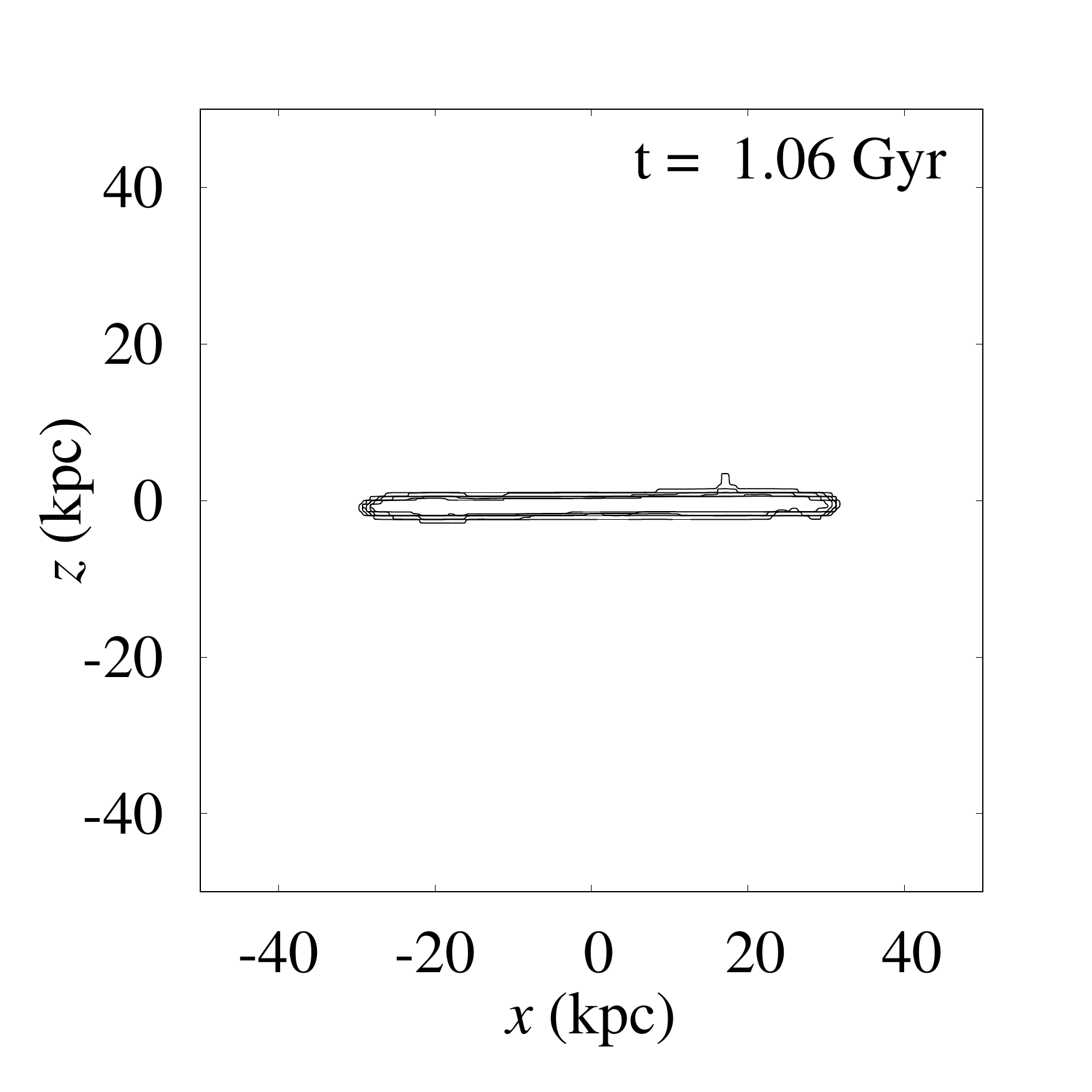}
\includegraphics[width=0.23\textwidth]{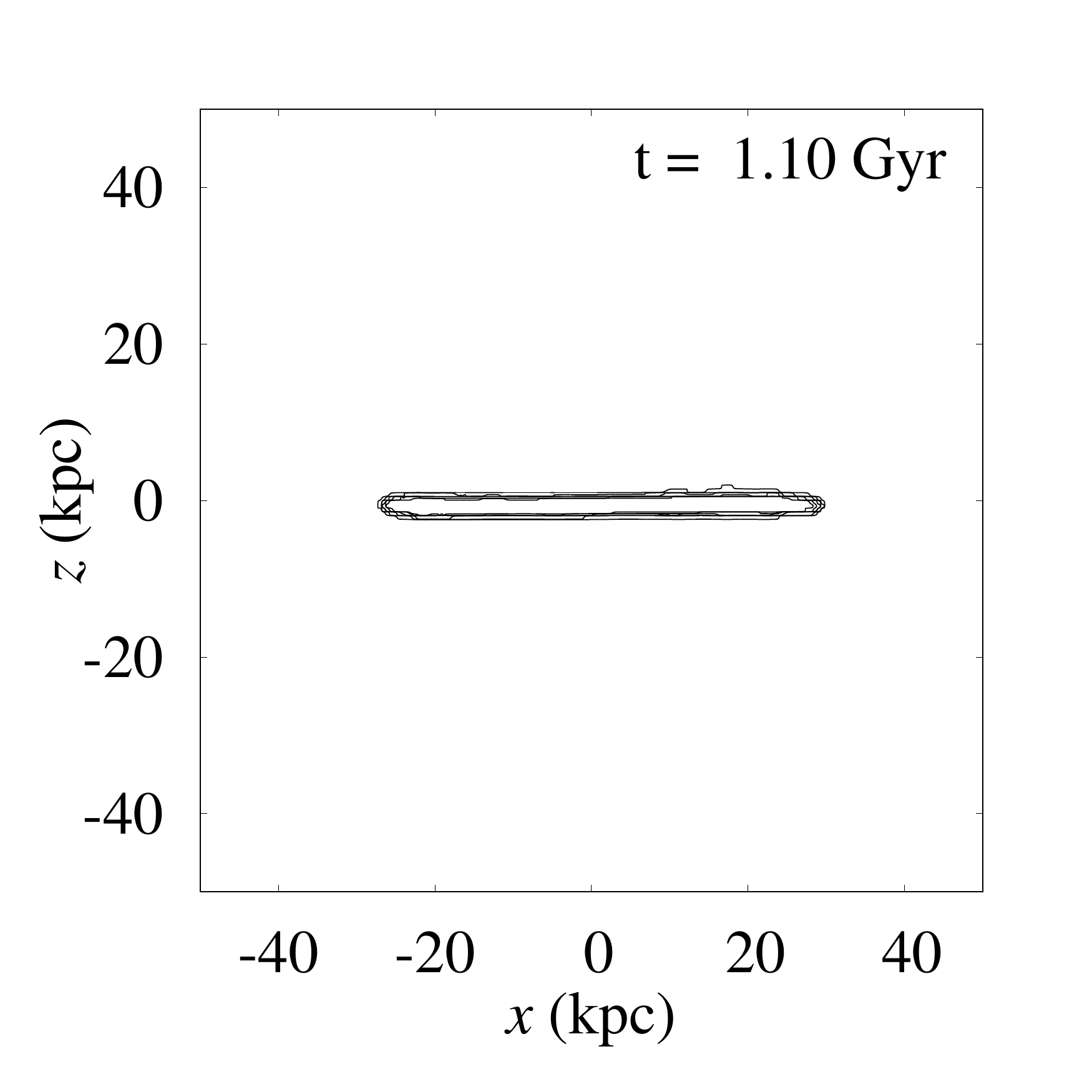}
\includegraphics[width=0.23\textwidth]{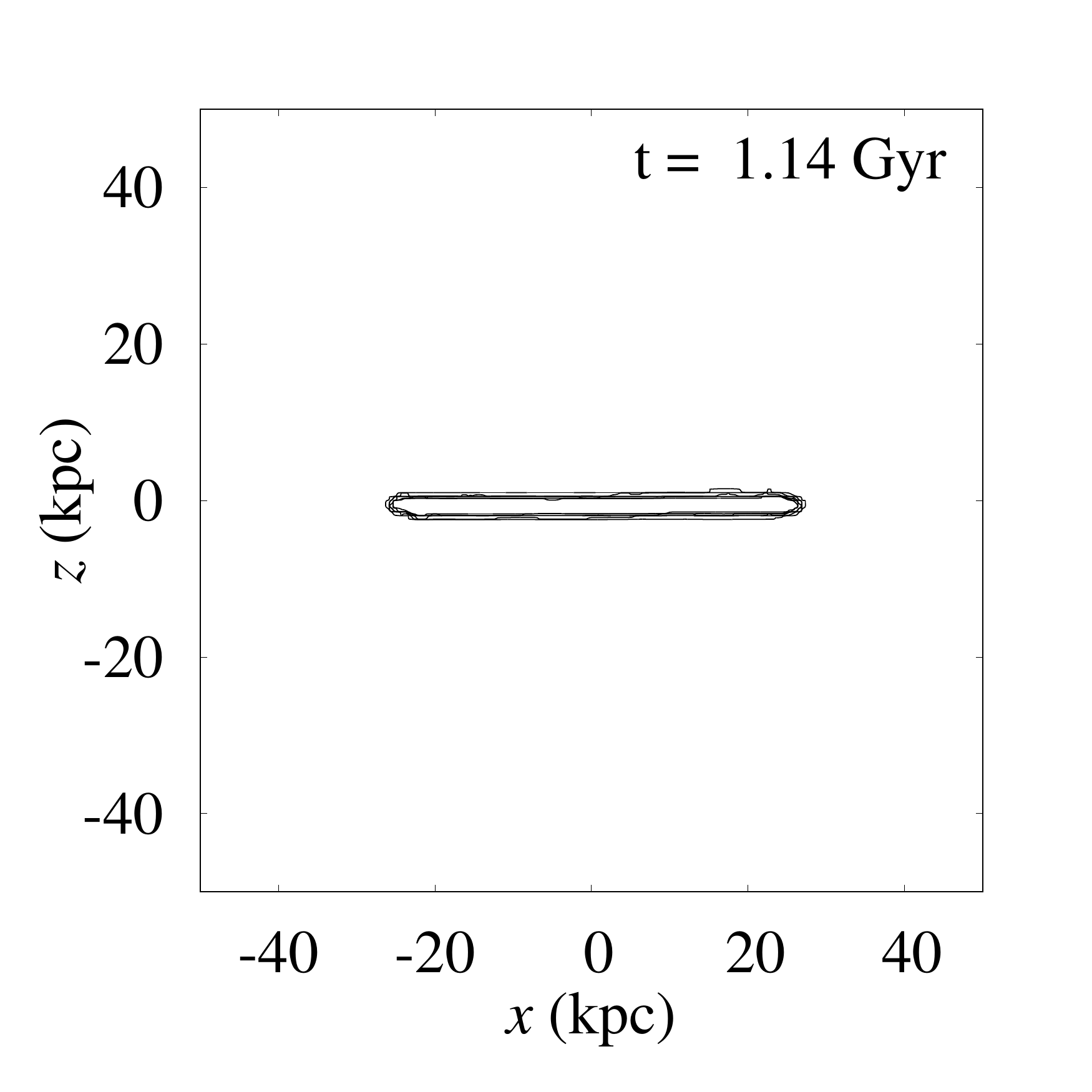}
\caption[ Smith Stream ]{ First approach, transit, rise and fall for the case of a high-mass subhalo (at high resolution, i.e. in RUN 2). Each panel shows an edge-on projection of the system, and contains a series of contours corresponding to total gas column densities in the range $10^{20} ~\psc$ to $10^{21} ~\psc$ in steps of 0.2 dex. The gas disc can be clearly seen as a high-density slab in this projection. The time sequence runs from left to right and top to bottom, and corresponds to a total timespan of roughly 500 Myr. Note that the subhalo has detached from its associated streamer at $t \sim 75 ~\Myr$ and has moved on. The subhalo's second transit features a qualitatively similar, albeit significantly shorter ($\sim 50 ~\Myr$), sequence of events (not shown). For an animated version of this figures and additional material follow this \href{http://www.physics.usyd.edu.au/~tepper/proj_smith_paper.html\#snpart_ctr_gc_run2}{link}.}
\label{fig:str}
\end{figure*}

For each adopted DM subhalo mass we run two simulations: A low resolution with up to 13 refinement levels for the gas component, and a high resolution simulation with a maximum refinement level of 15, both with the coarsest level set at 7. The resulting limiting spatial resolution in each case is \mbox{$500 ~\kpc ~/ ~2^{13} \approx 60 ~\pc$} and \mbox{$500 ~\kpc ~/ ~2^{15} \approx 15 ~\pc$}. The refinement level for collisionless particles is limited\footnote{These values have been chosen to avoid $N$-body relaxations, and are recommended by R.~Teyssier ({\em private communication}). In brief, we run a pure $N$-body simulation with no limit on the refinement level (in practice we set a very high refinement level) and record the maximum level achieved during the run; this gives an idea of what maximum refinement level to impose on the collisionless components when running a full $N$-body GHD simulation.} to 11 and 13 for the low and high-resolution runs, respectively, which translates into a limiting spatial resolution of $\sim 250 ~\pc$ and $\sim60 ~\pc$. The grid in the low resolution simulation is refined (coarsened) at runtime whenever the gas mass in a cell exceeds (falls short of) $\sim 2.5 \times 10^6 ~\Msun$ or the number of collisionless particles is larger (smaller) than 40. The corresponding thresholds for the high-resolution run are $\sim 3 \times 10^5 ~\Msun$ and 25. In addition, refinement is applied in order for the Jeans length to be resolved at all times with at least four cells \citep[][]{tru97a}. Note that in the low-resolution run, the nominal limiting spatial resolution is roughly half the scaleheight of the stellar disc. Nonetheless, this resolution is apparently high enough to keep the stellar disc stable (Figure \ref{fig:trans1_star} and Appendix \ref{sec:sta}). Also, the resolution imposed on the gas will be generally higher than implied by the gas mass threshold alone, as the gas is virtually always spatially coincident with high-density collisionless components. This is particularly true, and important, for the subhalo.

\section{Results}

\begin{table}
\begin{center}
\caption{Summary of simulation runs.}
\label{tab:runs}
\begin{tabular}{lccc}
\hline
\hline
Name 			& DM$^{\,a}$ ($10^{8} ~\Msun$)	& Gas$^{\,a}$ ($10^{7} ~\Msun$)  	& Resolution$^{\,b}$ (pc)	~\\
\hline\\
RUN 1			& 10							& 15							& 60 (13)				~\\
RUN 2			& 10							& 15							& 15 (15)				~\\
RUN 3			& 2.4							& 1.6							& 15 (15)				~\\
RUN 4			& --							& 16							& 15 (15)				~\\
\end{tabular}
\end{center}
\begin{list}{}{}
\item Notes. $^{a}$ Refers to the subhalo's DM or gas mass, respectively. $^{\,b}$ Approximate limiting spatial resolution. The number in parentheses indicates the maximum refinement level $l$. The ratio of the cell size to the box size corresponding to a given limiting resolution is given by $2^{-l}$.
 \end{list}
\end{table}

A disc crossing, or more appropriately, a disc {\em transit}, of a (DM-confined) cloud is accompanied by three distinctive phases which we identify as follows:
\begin{itemize}
	\item[]{\em Approach}: The phase prior to reaching the disc when the cloud is within $\sim 20 ~\kpc$ of the disc plane. During this phase the cloud is moving {\em towards} the disc, and develops an apparent head-tail morphology.
	\item[]{\em Rise}: The phase immediate after the transit, when the cloud is within $\sim 20 ~\kpc$ of the disc plane, for consistency with the above. As discussed below, this phase is characterised by the formation of a `streamer', an elongated gas structure extending between the gas disc and the cloud. During this phase the cloud is moving {\em away} from the disc. 
	\item[]{\em Fall}: The phase during which the streamer formed during rise has detached from the subhalo and falls back to the disc. In this phase, the gas is {\em not} confined by a DM subhalo.
\end{itemize}
If a cloud experiences multiple transits, each will be characterised by the same three phases. Thus, for the sake of clarity, whenever appropriate we will refer to each phase as first approach, first rise, first fall, or second approach, etc. depending on which transit they correspond to.\\

In Appendix \ref{sec:dmf}, we treat the specific case of a pure gas cloud (i.e. without a confining DM subhalo) falling towards the disc. In Figure \ref{fig:dmf} we show using a time
sequence that the massive HVC ($\sim10^8 ~\Msun$) does not survive its first disc transit. Obviously, the HVC does not mix with the ISM on its first approach (Figure \ref{fig:mix2}) and this remains true for the DM-confined case (Figure \ref{fig:trc}). Below we show that the observed metallicity pattern along the Smith Cloud can be understood as a direct consequence of a cloud {\em partially} mixing with the ISM gas on its way through the disc (either at the rise or fall phase). Thus, we do not believe we are observing the Smith Cloud is on its first approach.

Therefore, in what follows we focus only on the discussion of DM-confined clouds. We shall refer to the high-mass subhalo, low-resolution run as `RUN 1'; to the high-mass subhalo, high-resolution run as `RUN 2', and to the low-mass subhalo, high-resolution run as `RUN 3'. The high-resolution run featuring a DM-free cloud is referred to as `RUN 4', for consistency. We will ignore henceforth the low-resolution run featuring a low-mass subhalo, since its corresponding results can be accommodated within the results of the other runs (see Table \ref{tab:runs}).

\subsection{Account of a disc transit} \label{sec:trn}

Quite generally, as the subhalo falls in from a large distance, it speeds up, gradually being deflected from its initial (tangential) direction as a result of the strong gravitational potential at the centre of the Galaxy. Eventually, the cloud and its confining halo, now travelling at $\sim 450 ~\kms$, reach the disc roughly at perihelion at $R \approx 13 ~\kpc$ from the galactic centre (Figure \ref{fig:kin}), encountering gas with column densities in the range ${\rm N}_{tot} \sim 10^{20} - 10^{22} ~\psc$. The subhalo's first approach and first rise in RUN 1 and RUN 2 are illustrated both in Figures \ref{fig:trans1_gas} and \ref{fig:trans1_star}. Notably, the total deflection experienced by the subhalo is so large that it moves away from the disc at a roughly right angle with respect to its initial orbit (see also Figure \ref{fig:stable}).

After the disc transit (i.e. at rise), the DM subhalo detaches from (part of) its gas, leaving behind a {\em streamer} -- a gas tail extending from the disc -- that eventually fades away as it is accreted by the galaxy (i.e. at fall). We emphasise that this behaviour is common to all runs. In RUN 2, and in contrast to the other runs, the DM subhalo {\em retains a significant fraction of its gas} as it moves past the disc towards the galactic pole at high speed (cf. Section \ref{sec:loss}). In fact, the gas cloud within the DM subhalo survives long enough to transit the disc for a second time. Both disc transits along the cloud's orbit are easily identified by the maxima and minima in the velocity and the distance, respectively, displayed in Figure \ref{fig:kin}.

\begin{figure*}
\centering
\includegraphics[width=0.33\textwidth]{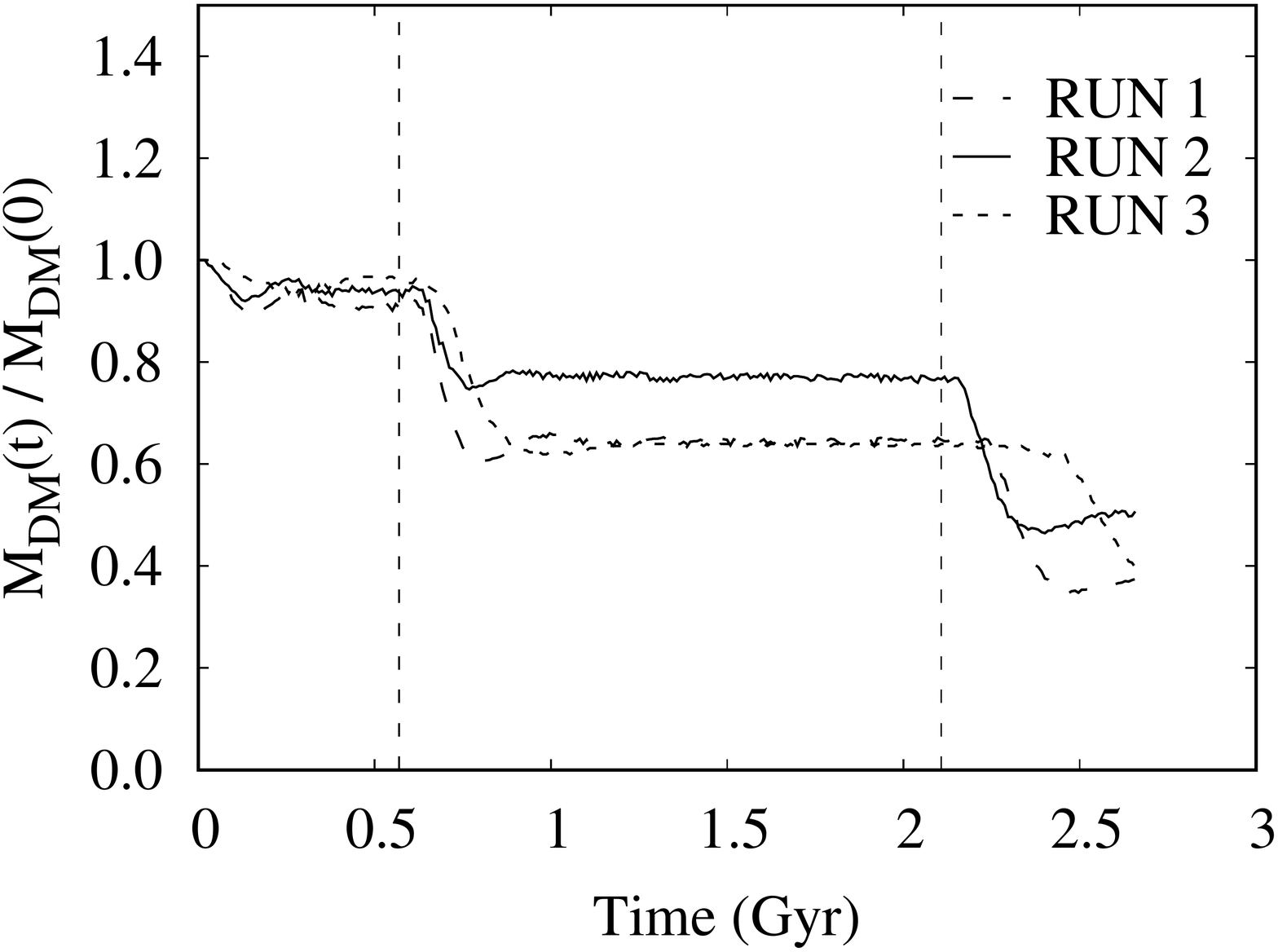}
\hfill
\includegraphics[width=0.33\textwidth]{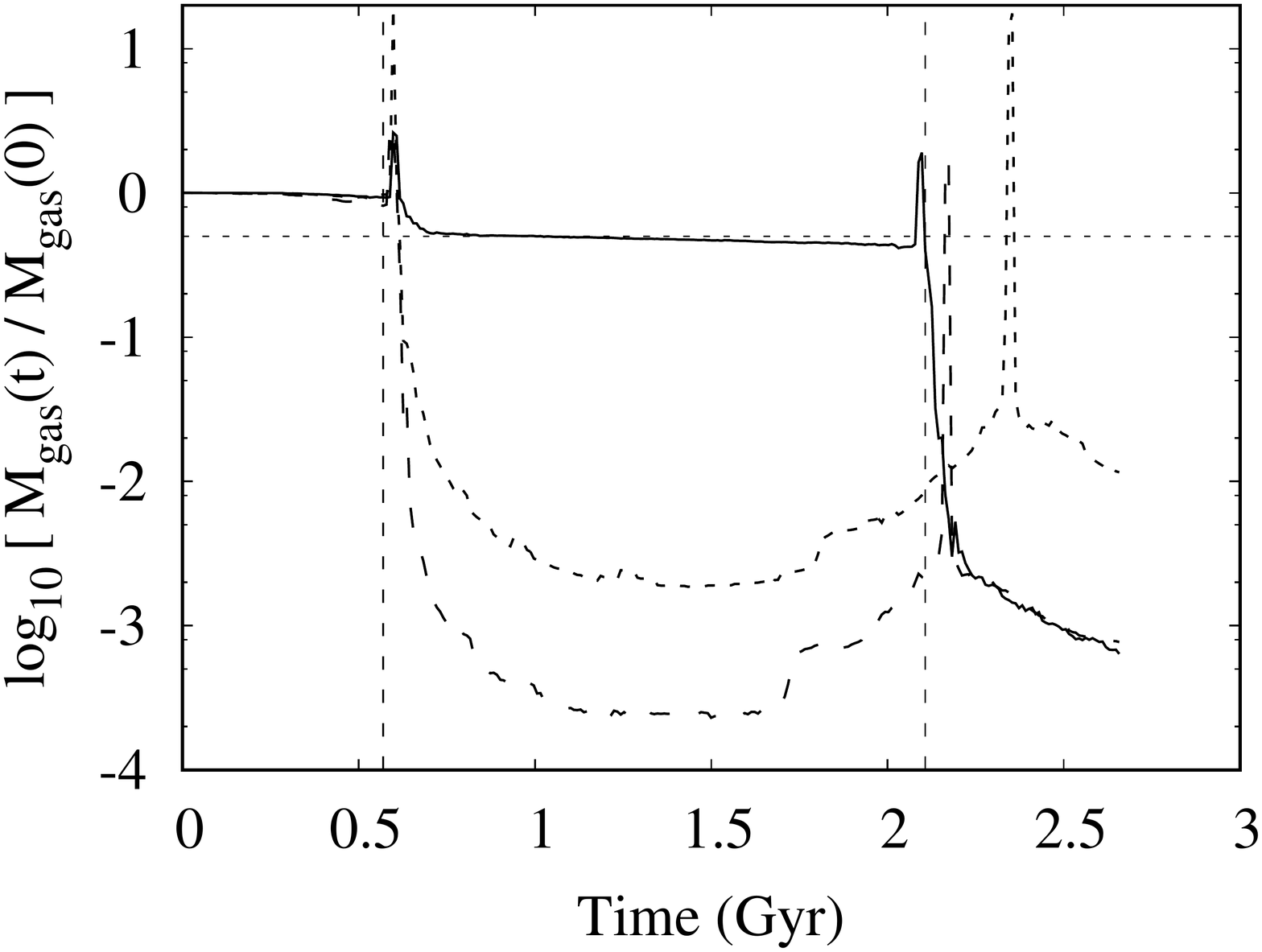}
\hfill
\includegraphics[width=0.33\textwidth]{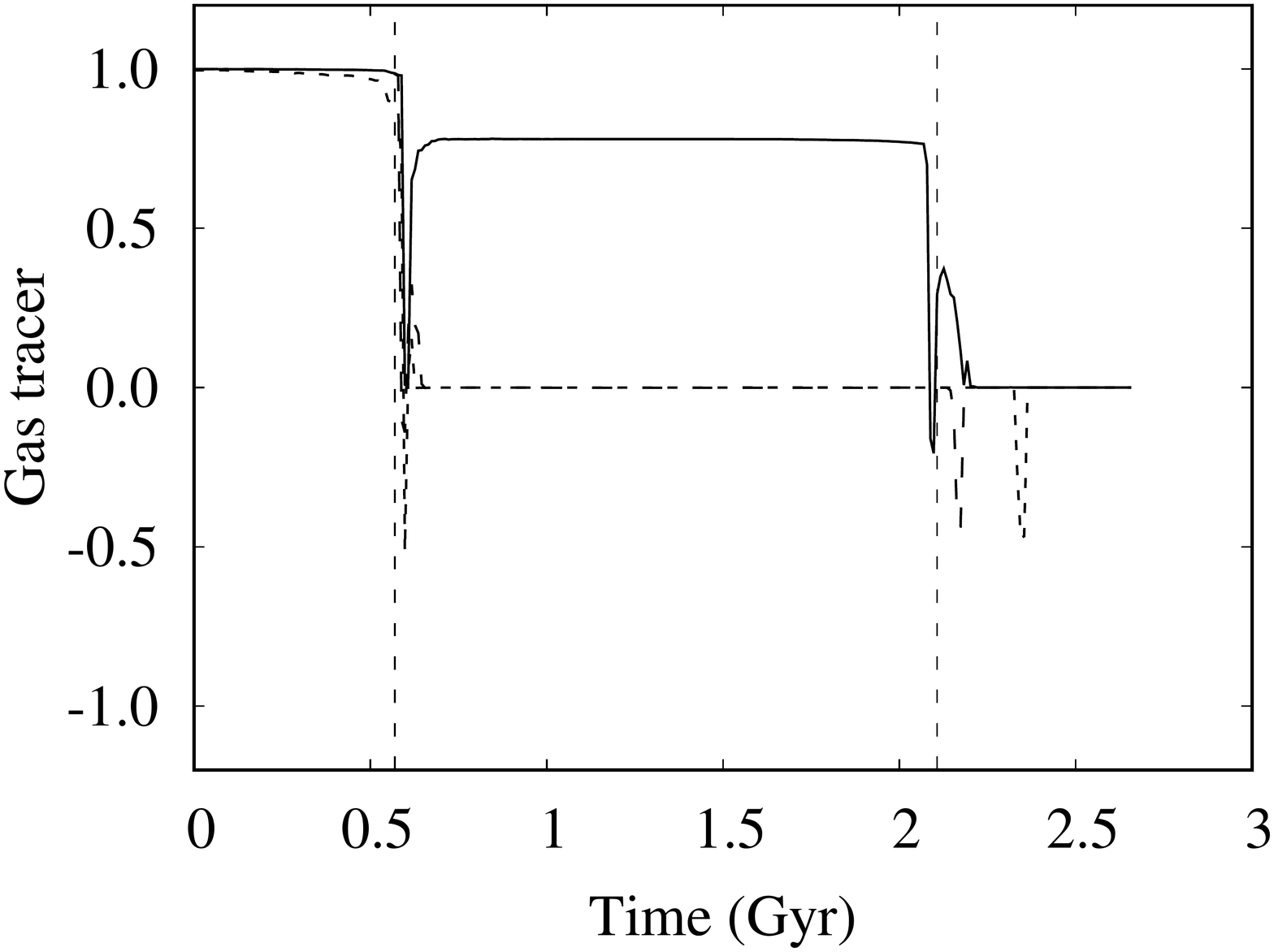}
\caption[ (DM) subhalo masses and tracer ]{ For each run, the lines indicate the DM mass (left), gas mass (middle), and tracer value (right) at a given time within a cubic reference volume of length $5 ~\kpc$ centred on the DM subhalo's core. The masses at a given time are given relative to its corresponding initial value. The horizontal dashed line in the middle panel indicates the 50 percent threshold, for reference (note the logarithmic scale on the vertical axis). The vertical dashed lines indicate the epoch roughly corresponding to each of the disc transits in RUN 2, and are identical in all panels. The spikes (dips) in the relative gas mass (gas tracer) in all runs are caused by the overlap of the subhalo with the gas disc during transits, and can be ignored. Line styles and corresponding meaning are given in the left panel.}
\label{fig:loss}
\end{figure*}

The first disc transit and its associated phases are illustrated in Figure \ref{fig:str} for the case of a high-mass cloud at high resolution (RUN 2). There we show a sequence of snapshots spanning a total time of roughly 500 Myr, showing in particular the evolution of the streamer. Each panel contains a series of contours corresponding to total gas column densities in the range $10^{20} ~\psc$ to $10^{21} ~\psc$ in steps of 0.2 dex. It can be clearly seen that the streamer eventually splits into two main components: one remains attached to the DM subhalo, while the other remains attached to the disc. The latter can be identified as a high-density gas `ridge' at $z \lesssim 5 ~\kpc$ (or $\lesssim$20 deg; see Figure \ref{fig:mix}). The gas mass of the ridge within a cubic volume of 5 kpc at this point in time is $\sim 5 \times 10^7 ~\Msun$, consistent with the result that the subhalo loses roughly half of its mass after the first disc transit, as discussed below. It is worth noting that its total mass can easily accomodate the total current mass of the Smith Cloud.

As the remainder of streamer moves on a prograde orbit around the Galactic Centre, slightly lagging behind the disc, it rises, until it starts to fall back onto the disc after $\sim 100 ~\Myr$ since the transit. We note with interest that the distance to the disc of the high-density fragment never exceeds $\sim 10 ~\kpc$. The streamer moves with a range of velocities corresponding to total speeds $\lesssim 300 ~\kms$ until accretion. To get a sense of the timescale between the formation and accretion of the streamer, we define the `fading time' as the timespan between the transit and the point in time when the density of the subhalo gas drops below a total gas column density of $10^{20} ~\psc$. We thus find that the streamer disappears from view after $\sim 400 ~\Myr$ when it finally falls back onto the disc. A similar behaviour is observed after the subhalo's second transit (not shown). Given the significantly reduced mass of its gas content (see below), the associated streamer fades away after only $\sim 40 ~\Myr$.

\subsection{Mass budget} \label{sec:loss}

In order to quantify the evolution of the subhalo's mass (both the DM matter and its gas content) during a transit, we proceed as follows. We track the subhalo along its orbit, and integrate both the DM matter and the gas density separately over a cubic volume of length $5 ~\kpc$. We estimate the subhalo's position and velocity using its dense core as a proxy for the centre of mass, since the latter deviates significantly from the core's position when the subhalo's tidal distortion becomes important.

The evolution of the subhalo's DM mass along its orbit in each run is shown in the left panel of Figure \ref{fig:loss}. We find that regardless of the initial subhalo's mass and the limiting resolution of the run, the subhalo loses around 20 percent of its initial DM mass after $\sim 100 ~\Myr$ since the transit. The same mass loss is experienced by the DM subhalo after second transit (RUN 2 only). In either case, it is worth mentioning that the mass loss is {\em not} caused by the disc transit, but is rather lost to the tidal tails which result from the gravitational interaction between the subhalo and the strong galactic potential.\footnote{An animation of the subhalo's tidal distortion along its orbit can be found following this \href{http://www.physics.usyd.edu.au/~tepper/proj_smith_paper.html\#dm_sub_gc_run2}{link}.} Despite the significant mass loss and heavy distortion, the DM subhalo recovers its spheroidal shape even before reaching aphelion at $z \sim 100 ~\kpc$ above the disc (Figures \ref{fig:kin} and \ref{fig:stable}).

The gas content of the subhalo experiences a different fate entirely (Figure \ref{fig:loss}, central panel). As briefly discussed above, a low-mass DM subhalo is virtually devoid of gas after its first transit. The gas is simply left behind in the form of a streamer. Deprived of the shielding effect of the DM subhalo, the gas interacts with the hot halo, expanding and becoming more diffuse, its remnant eventually falling back to the disc. The same behaviour is displayed by a high-mass subhalo at {\em low resolution}. In contrast, a high-mass subhalo at high resolution retains roughly half of its initial gas mass after the first transit. After the second transit, however, it is completely lost to the disc, similar to the low-mass halo after its first transit. It is noteworthy that the gas mass in the streamer formed after the first transit in RUN 2 ($\sim 6 \times 10^7 ~\Msun$) is well above the total inferred gas mass of the Smith Cloud ($\gtrsim 4 \times 10^6 ~\Msun$).

{\em We thus find that DM confinement may assist cloud survival during a high-speed impact with the disc, provided the cloud is massive enough and the spatial resolution high enough} (see Table \ref{tab:runs}). This is in agreement with earlier work \citep[][]{nic14b}, but differs fully from more recent results \citep[][]{gal16a}. The clouds we consider are all comparable to the clouds considered in these previous studies both in terms of mass and shape. Our numerical approach is also similar to theirs. Our highest adopted limiting resolution is comparable to the limiting resolution in \citeauthor[][]{nic14b}'s work, but we ignore the resolution in \citeauthor[][]{gal16a}'s simulations. Based on the different results in our high and low resolution models, we speculate that the difference between theirs and our results is most likely due to differences in the adopted limiting resolution.\footnote{We note that \citet[][]{gal16a} have not considered radiative cooling in their models, as have \citet[][]{nic14b} and ourselves, but we cannot tell whether this difference has accentuated the discrepancy.}

Physically, the result that a cloud is destroyed during a transit can be understood in terms of the kinetic argument presented by \citet{bla09a}. It states that, {\em to first order}, a gas cloud will not survive a collision with a gas disc unless its column density is higher than the disc's column density at the point of impact.\footnote{Note that this depends as well on the ratio of the disc's scaleheight to the cloud diameter, which in our case is of order unity.}  Indeed, in RUN 2 the cloud's gas density is slightly higher than the disc's density at the impact point during the first transit, thus it survives; but is slightly below during the second transit and in consequence it is destroyed. The latter is also true for the low-mass cloud (RUN 1) and the low resolution case (RUN 3). The reduced density of the high-mass cloud during its second transit is a direct consequence of the gas mass loss and the reduction in the subhalo's mass. The fact that a {\em slight} difference in the gas density of the cloud with respect to the disc determines whether it survives or not suggests that high-order effects are important.

The fact that our model survives only two transits is due to of our particular choice of initial subhalo parameters. Further transits are not ruled out, but they would require a DM-confined cloud with total mass in excess of $10^9 ~\Msun$, or an orbit that leads to an impact point further out in the disc. In any event, {\em the formation of a streamer at each transit and its eventual accretion onto the Galaxy after its last transit is a general result.}

Regardless of the subhalo's initial mass, or the limiting resolution, the gas disc does not suffer any appreciable damage after the high energy impact; in particular, the cloud {\em does not} `punch' an apparent hole into the disc, {\em nor} does it create a `bubble' as reported by \citet[][]{gal16a}. The reason for the discrepancy between their and our results could be a difference in the gas column density of the cloud relative to the gas column density at the impact point in the disc, although we cannot know for sure, since we ignore their corresponding values. Also, it is unclear whether they set up the ISM in their simulations to be in centrifugal equilibrium or in vertical hydrostatic equilibrium with the total potential, which affects its dynamic stability. Similarly to the gas disc, the stellar disc does not show any apparent memory of the event (Figure \ref{fig:trans1_star}).

Finally, we note with that we {\em do not} see any hint of accretion of gas onto the DM subhalo, either before, during or after its disc transits (Figure \ref{fig:loss}, middle panel). In this respect, we disagree again with \citet[][]{gal16a}. The source of this discrepancy is not clear at this point, but it is puzzling, especially in view of the adiabatic nature of their simulations in contrast to our models including cooling, which may promote gas accretion onto DM-{\em free} HVCs under favourable conditions \citep[][but see Section \ref{sec:mix}]{mar10b,gri17a}.

\subsection{Gas mixing} \label{sec:mix}

\begin{figure}
\centering
\includegraphics[width=0.42\textwidth]{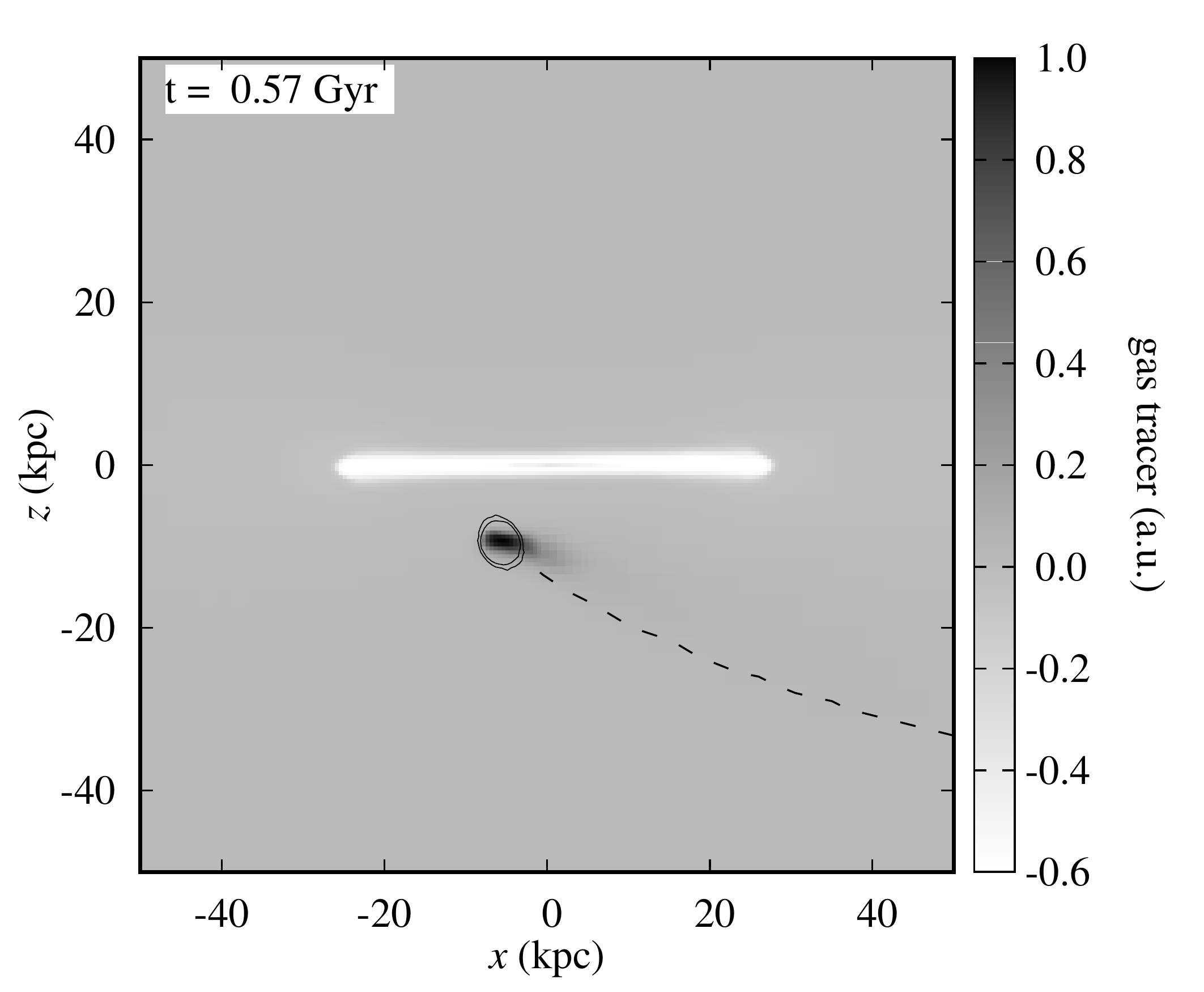}
\includegraphics[width=0.42\textwidth]{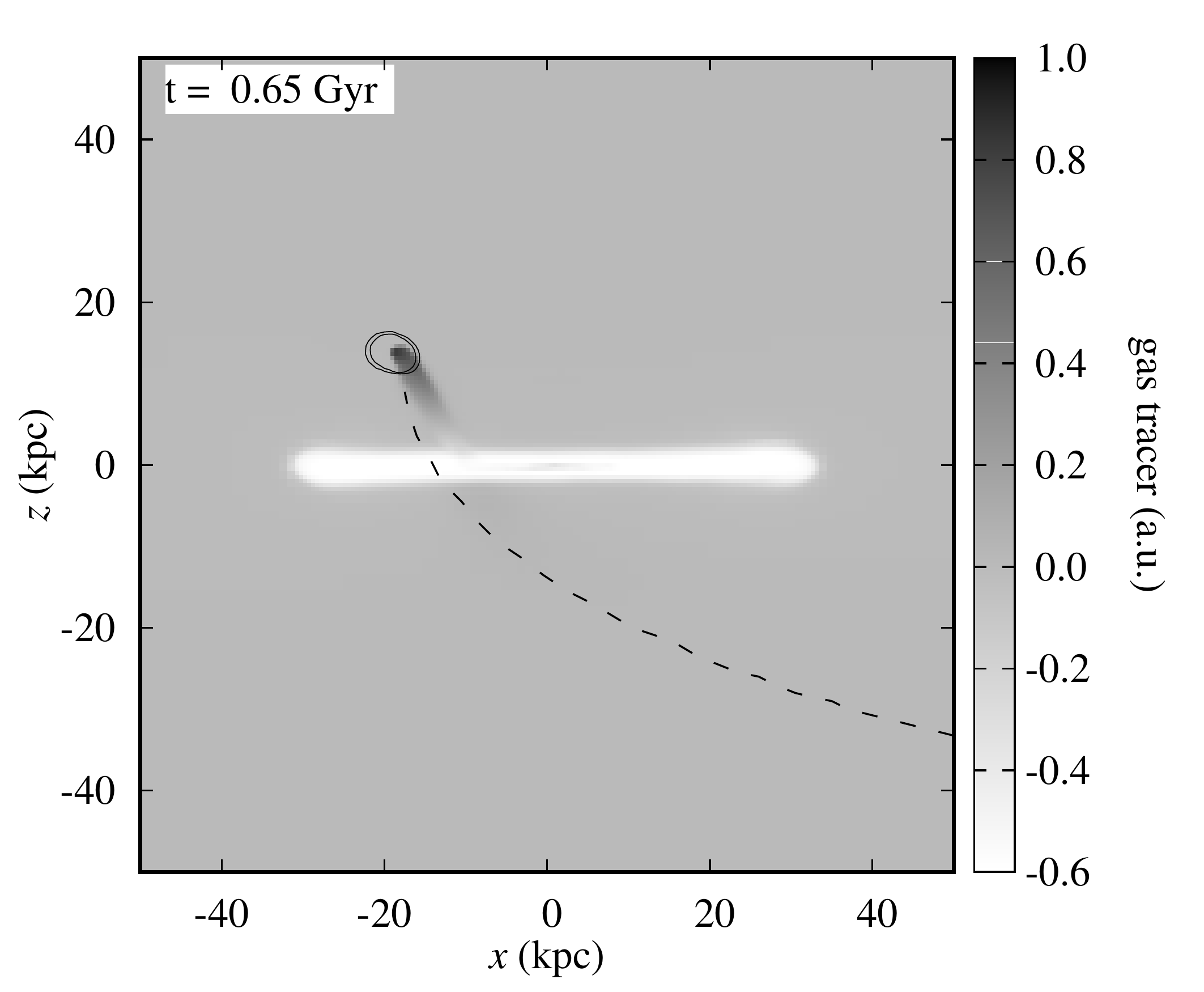}
\caption[ Gas tracer (hi res run) ]{ Edge-on projection of the system prior (top) and after (bottom) the first transit in RUN 2. The colour code indicates the value of a passive tracer characteristic of each gaseous component. The gas disc, the hot halo, and the gas cloud have been initially assigned a value of -1 (`white'), 0 (`grey'), and +1 (`black'), respectively. Dotted lines and time labels as in Figure \ref{fig:trans1_gas}. Note that we have imposed a lower limit at a tracer value of -0.6 for displaying purposes. Also, we have included only the outermost DM contours of the subhalo for clarity. For an animated version of this figures follow this \href{http://www.physics.usyd.edu.au/~tepper/proj_smith_paper.html\#tracer_gc_run2}{link}.}
\label{fig:trc}
\end{figure}

It is generally believed that external gas accreted by the Galaxy, if primordial, cannot contain high amounts of elements heavier than helium ([Fe/H] $\lesssim$ -1) although it can be higher if it is recycled gas due to a Galactic fountain, say. In fact, measurements of the heavy-element content of HVCs around the Galaxy, in particular the Magellanic Stream, suggest that the threshold at the present epoch lies somewhere around 10 percent relative to the solar value \citep[][]{fox13a,ric13a}.

\begin{figure}
\centering
\hfill\includegraphics[width=0.42\textwidth]{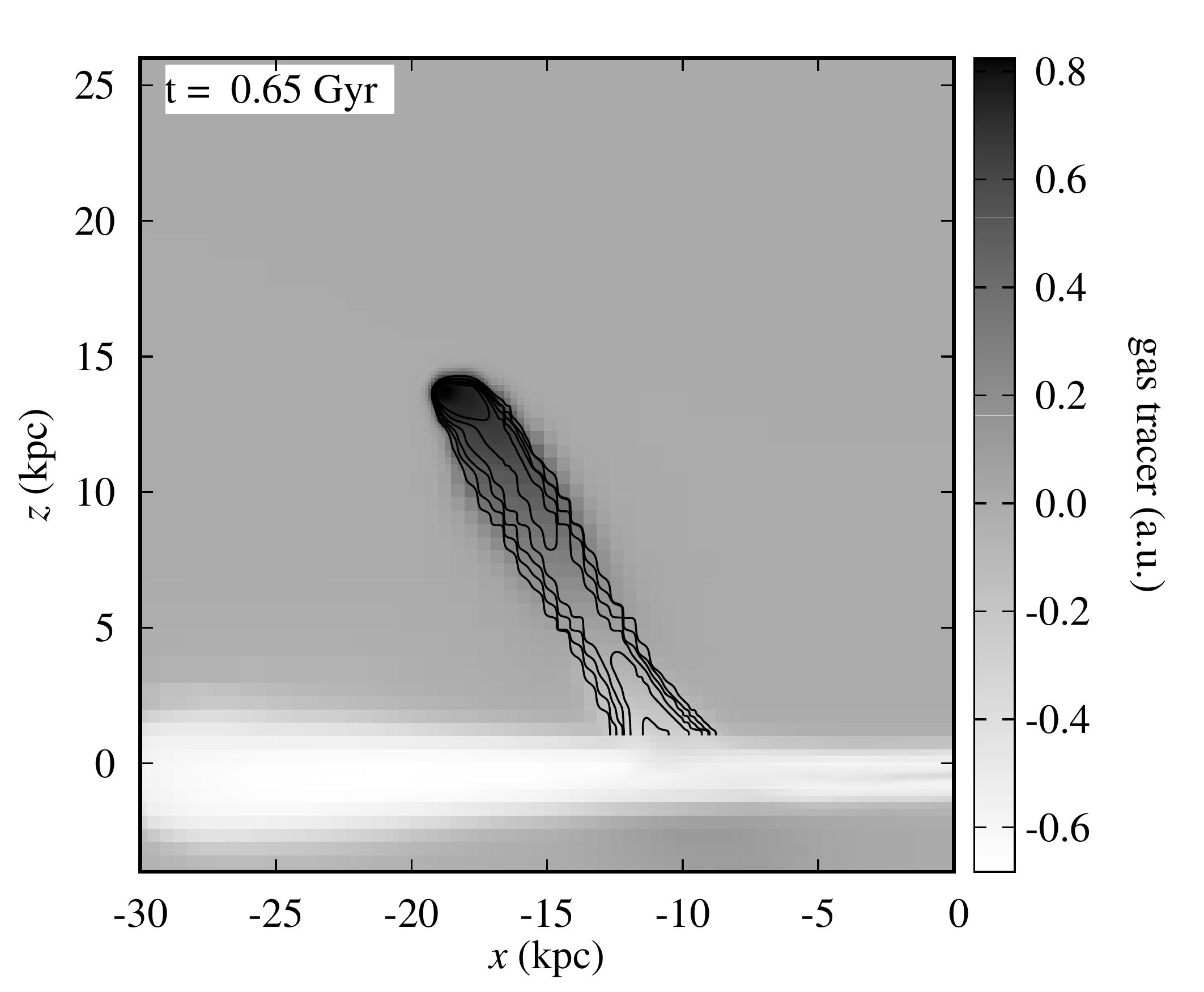}\\
\hfill\includegraphics[width=0.42\textwidth]{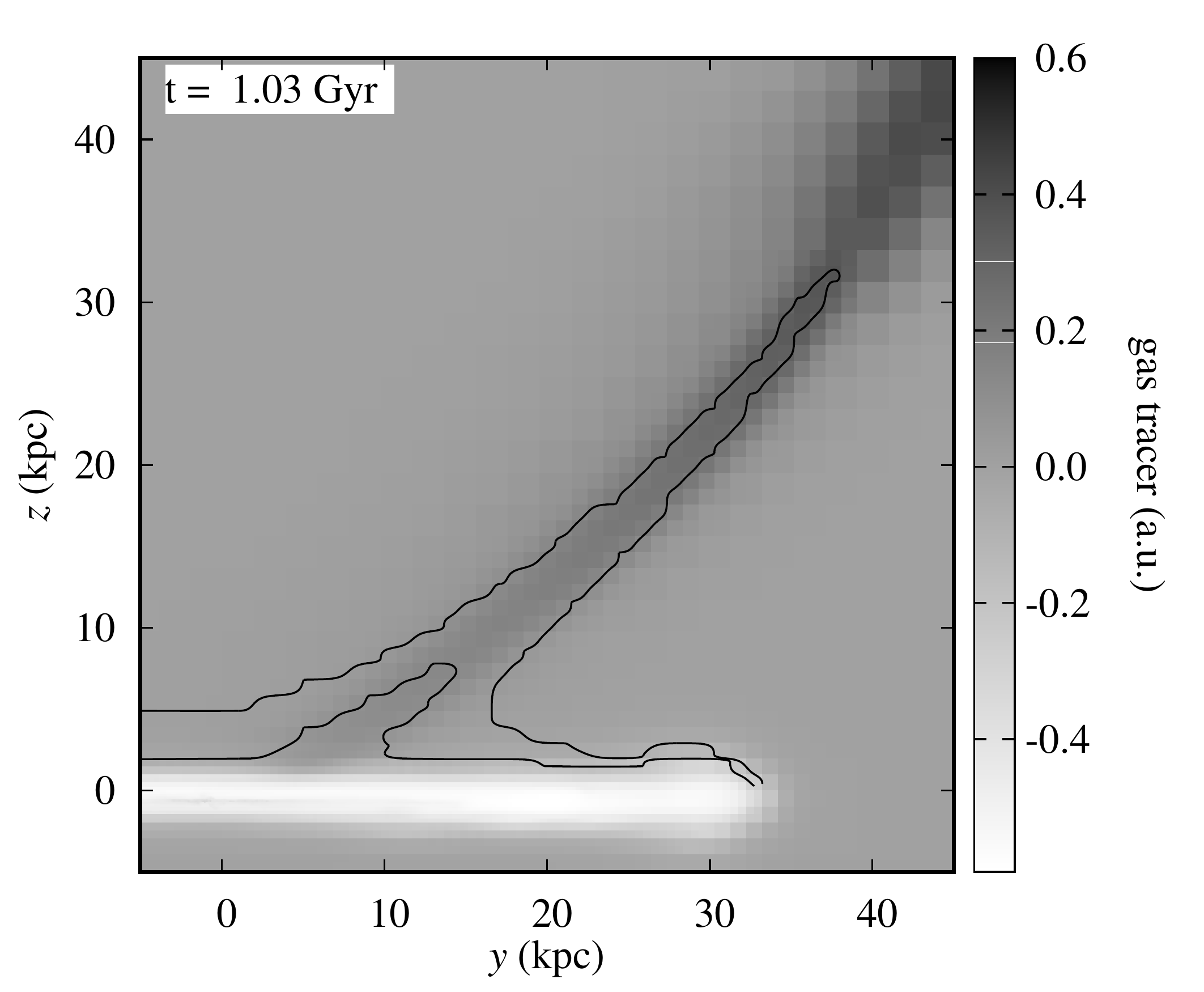}
\includegraphics[width=0.42\textwidth]{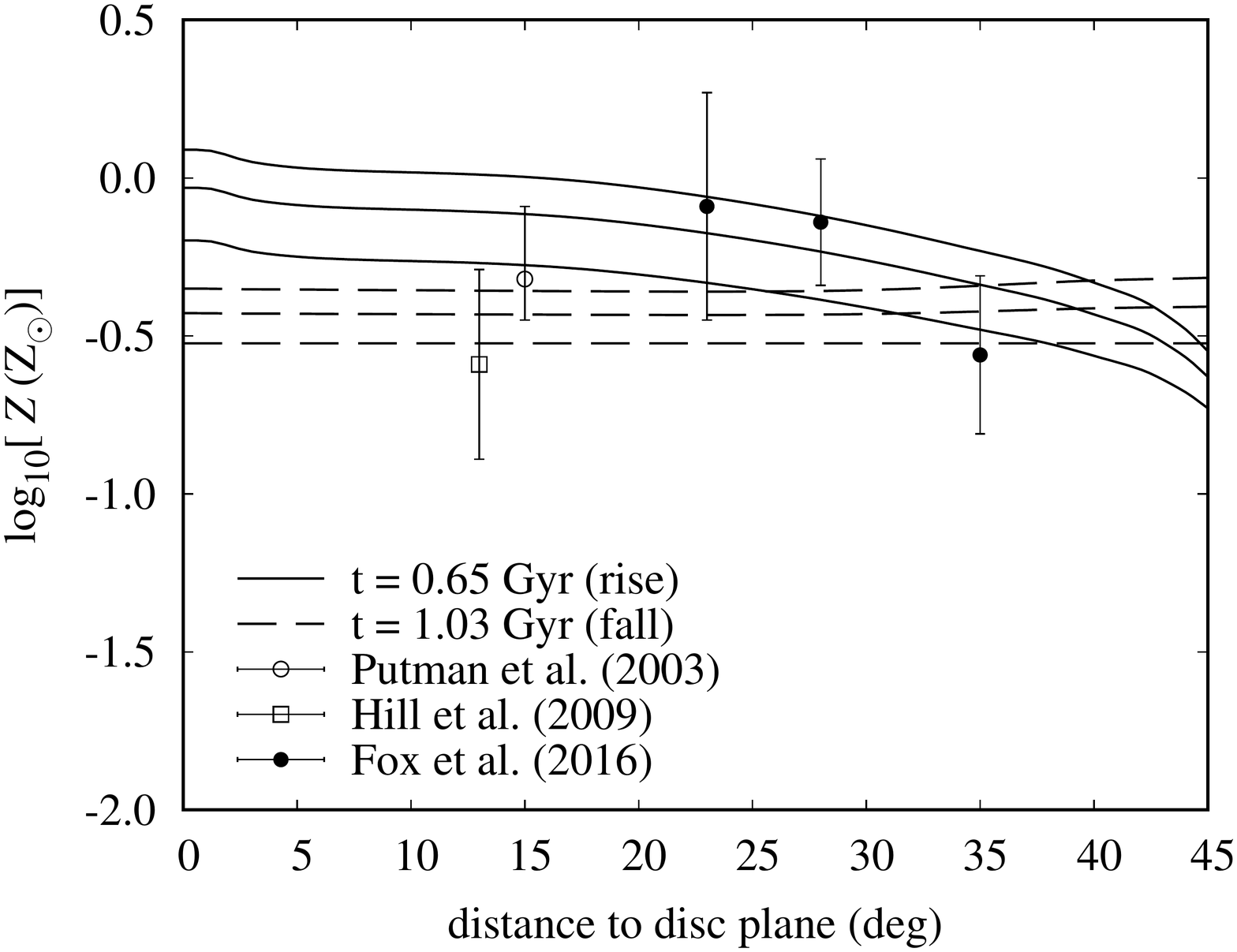}
\caption[ Metallicity estimate ]{ Top: Edge-on projection of the system at $\sim 50 ~\Myr$ after the first transit (rise). Middle: Edge-on projection of the system at $\sim 1 ~\Gyr$ after the first transit (fall). The colour scale indicates in each case the value of the passive tracer (cf. Figure \ref{fig:trc}). The contours correspond to total gas densities in logarithmic scale (in units of \psc) between 20 and 21 in steps of 0.2 dex (top) and at 19.4 and 19.6 (middle).  We have omitted the density contours of the gas disc, for clarity. Note the difference in the spatial scale as well as in the colour range between panels, chosen to enhance the contrast between the various gas components in each case. Bottom: Gas metallicity along the streamer at rise (solid) and fall (dashed). The model curves correspond in each case to an assumed mean ISM metallicity of 2, 1.5, and 1.0  \Zsun\ (from top to bottom). The data points corresponds to different element abundance measurements along the Smith Cloud from the literature. The symbols and error bars indicate the mean value and corresponding uncertainty, respectively. (See also Figure \ref{fig:mix2}.)}
\label{fig:mix}
\end{figure}

\citet[][]{fox16a} have shown that the gas kinematically associated with the Smith Cloud extending over a few kilo-parsec from its tail tentatively displays a gradient in [S/H] -- the higher value being closer to the tip, i.e. to the Galactic plane --, with an average value around half the sulphur abundance of the Sun. However, previously available metallicity estimates based on the nitrogen abundance derived from emission -- rather than absorption -- line measurements \citep[][]{put03b,hil09a} suggest instead a decrease of the metallicity along the Smith Cloud towards the plane (Figure \ref{fig:mix}). Needless to say, any apparent trend is to be taken with caution as the uncertainty in all the available measurements is still large.

In view of these measurements, it has been argued \citep[][]{fox16a,mar17a} that the Smith Cloud is likely of Galactic origin. On the other hand, it has been shown, although under simplified conditions, that a cloud moving through a high-metallicity environment may experience significant mixing, thus washing away its `primordial' signature \citep[][]{hen17a}. The latter is relevant as we believe that the Smith Cloud's kinetic energy ($\gtrsim 4\times 10^{53}$ erg in the frame of the disc's rotation) renders models based on ejection from the disc implausible.

Crucially, \citeauthor{hen17a} did not consider the impact of a cloud with a gas disc, nor they considered DM confinement. In view of the effective shielding provided by DM discussed in the last section, it is important to address whether our simulations support a scenario in which a cloud within a DM subhalo can mix with the ISM during a transit. We accomplish this task in the following way, focusing for now only on RUN 2. We attach to each of the gas components a passive tracer at the beginning of the simulation: The gas disc is initially assigned a value of -1 (`white'); the hot halo, 0 (`grey'); and the gas cloud, +1 (`black'). Thus, we are able to identify at all times the gas initially in a given component and whether it mixes with any other gas component. This is illustrated in Figure \ref{fig:trc}. The panels show the density-weighted value of the tracer across a fraction of the simulation volume before (top) and after (bottom) the first disc transit. The gas within the DM subhalo features a value close to +1 before the transit, but it is clearly below 1 after the transit, signalling mixing with the ISM. A  quantitative assessment of the level of mixing can be obtained by following the same approach we followed to estimate the subhalo's mass loss in Section \ref{sec:loss}.

The result of this exercise is shown in the right panel of Figure \ref{fig:loss}. The relevant case (RUN 2) shows that the cloud significantly mixes with the disc gas during disc crossings, as the the value of its gas tracer drops from 1 to roughly 0.8, after the first transit, and down to 0 after the second (when the DM subhalo completely loses its gas). Remarkably, the run of the tracer is virtually flat between transits, indicating that the gas cloud does not (appreciably) interact with the hot halo gas. Indeed, we find that the gas core within the subhalo retains a temperature of order $10^4 ~\K$ at all times (not shown). This behaviour is opposite to the behaviour of pure gas clouds which interact and may mix significantly when moving through the Galactic corona \citep[e.g.][]{gri14a}. {\em We conclude that DM confinement provides an effective shielding mechanism against hydrodynamic interaction with a diffuse ambient medium.}

A simplistic, back-of-the-envelope calculation shows that, if the transit results in mixing such that the cloud retains a fraction $f_c$ of its gas, the gas cloud's metallicity $Z'_c$ {\em after} the transit is given in terms of the its metallicity $Z_c$ before the transit and the ISM metallicity $Z_d$ by $Z'_c \approx Z_c f_c + Z_d (1 - f_c)$ (all in solar units). Adopting $Z_c = 0.1$ and the mean value found by \citeauthor{fox16a}, $Z'_c = 0.5$, and assuming the mixing at a 20 percent level is representative ($f_c \approx 0.8$), we can estimate the required ISM metallicity at the point of impact during first transit. We get $Z_d \approx 2$ (or 0.3 dex); a rather high, but not implausible value for the Galaxy, in particular within $R \lesssim 10 ~\kpc$ \citep[][]{rud06a}. We note that this value is much higher than the upper limit of $\sim 1 ~\Zsun$ inferred by \citet[][]{hen17a}, although they did not consider mixing with the ISM.

We can actually do better by measuring the metallicity along the cloud and its stream in our model. However, we cannot use the actual metallicity of each component in the simulation, as halo and disc have been initially assigned the same (uniform) metallicity, and it is thus difficult (if not impossible) to distinguish between the mixing of the cloud and halo gas or disc gas. Instead, we estimate the individual contribution of each gas phase to the tracer value in a given cell along the streamer, and compute the total metal mass fraction by weighing the intrinsic metallicity of each component with its tracer contribution. To compare to observations, we transform the position across the cloud into an angular distance. We choose for this experiment two representative snapshots, one at rise, and one at fall after the first transit (Figure \ref{fig:mix}).

The approach outlined above thus requires knowledge of the ISM metallicity at the crossing point, the intrinsic metallicity of the cloud, and its distance. Since both the gas metallicity of the gas disc and the distance to the cloud in our model are to some extent arbitrary, we rescale their values making some educated estimates. Given the uncertainty in the ISM metallicity at the crossing point, we adopt three different values, 1, 1.5 and 2 \Zsun, spanning the range between \citeauthor{hen17a}'s estimate and the value calculated above. The cloud's metallicity is taken to be its initial value in our model, $0.1 ~\Zsun$, which is just slightly below the lowest observed values \citep[taking into account their uncertainty][]{hil09a,fox16a}. A value of zero is assigned to the halo's metallicity, to avoid contamination. The heliocentric distance to the Smith Cloud is fixed by observation \citep[$d = 12.4 \pm 1.3 ~\kpc$;][]{loc08a}.

The above assumptions, together with the mixing it experiences, fix the metallicity of the streamer right after first rise. Indeed, the adopted initial ISM metallicities (1, 1.5 and 2 \Zsun) translate into an initial streamer metallicity of approximately 0.3, 0.4 and 0.5 \Zsun, respectively. These are used to estimate the streamer's metallicity at first fall as per the above approach. In this case, we restore the halo metallicity to its initial value (0.3 \Zsun), since the streamer interacts with it along its orbit.

The run of the gas metallicity roughly along the main axis of the streamer, from its origin in the disc all the way to the cloud's core (the latter only at rise) is shown in the bottom panel of Figure \ref{fig:mix}, together with different element abundance measurements along the Smith Cloud from the literature. It is reassuring that the model results match the {\em mean} observed abundance level. But it seems difficult to match the observations assuming a uniform cloud's metallicity, suggesting that the heavy-element distribution within the Smith Cloud is likely patchy. On the other hand, if the metallicity variations across the cloud are dominated by systematic errors, {\em we can conclude that the enhanced abundance of heavy elements observed in the Smith Cloud may well be the result of the cloud \textnormal{partially} mixing with the ISM during transit, provided the ISM metallicity at the impact point is sufficiently high}, in qualitative agreement with \citet[][]{hen17a} 

The above result does not depend heavily on the particular snapshot chosen at rise or fall. Indeed, we have checked that performing the same calculation at different epochs (not too far in the cloud's evolution though) turns out to be somewhat redundant, as the multi-epoch metallicity measurements more or less fall on top of each other, reaching only different (higher or lower) angular distances at different times. Note that a cloud on its second approach would feature a chemical signature very similar to its signature during first rise (i.e. dashed lines in the bottom panel of Figure \ref{fig:mix}). This is a direct consequence of the lack of mixing of the DM-confined cloud with the halo gas along its orbit between transits (see Figure \ref{fig:loss}, right panel).

\begin{figure}
\centering
\includegraphics[width=0.42\textwidth]{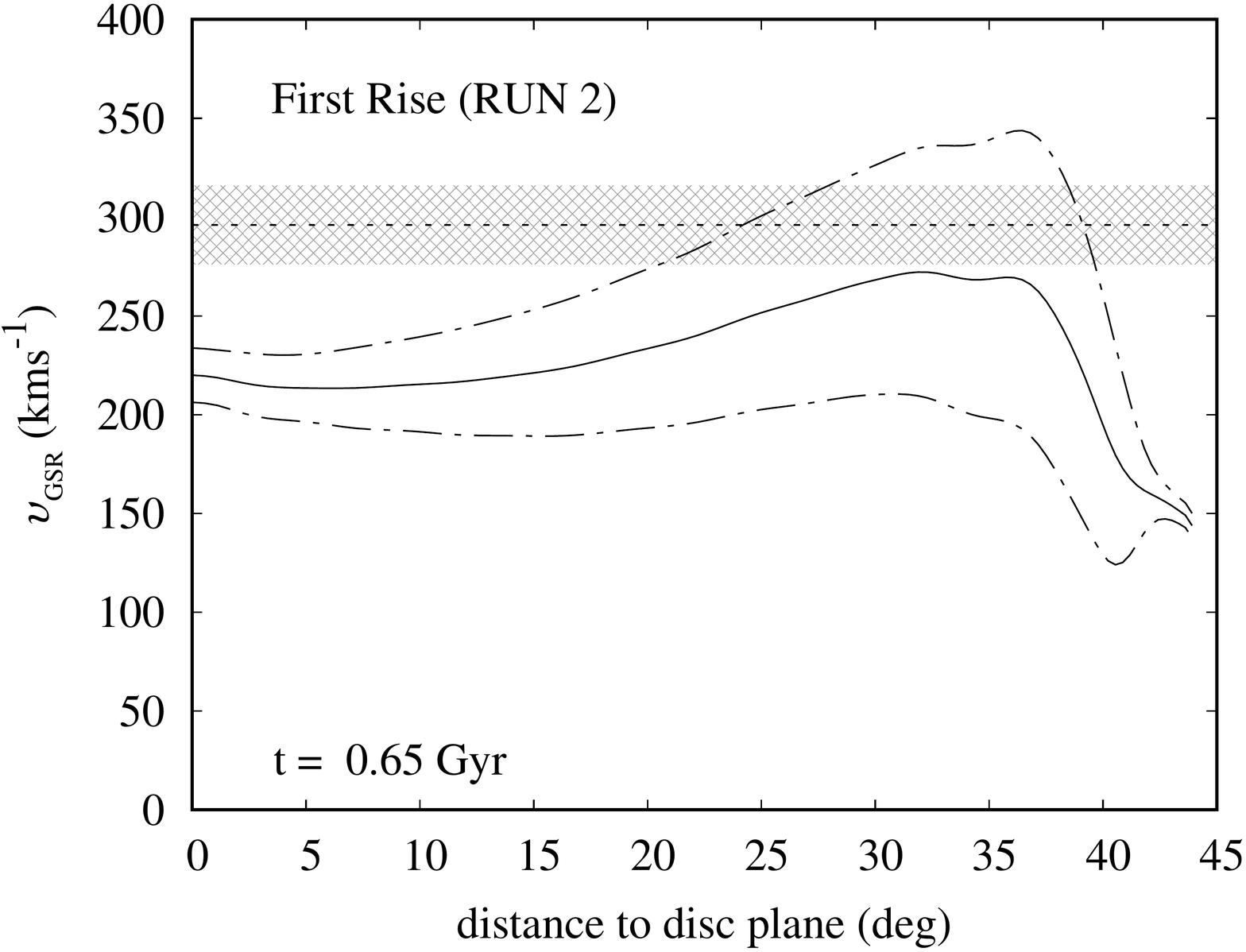}\\
\includegraphics[width=0.42\textwidth]{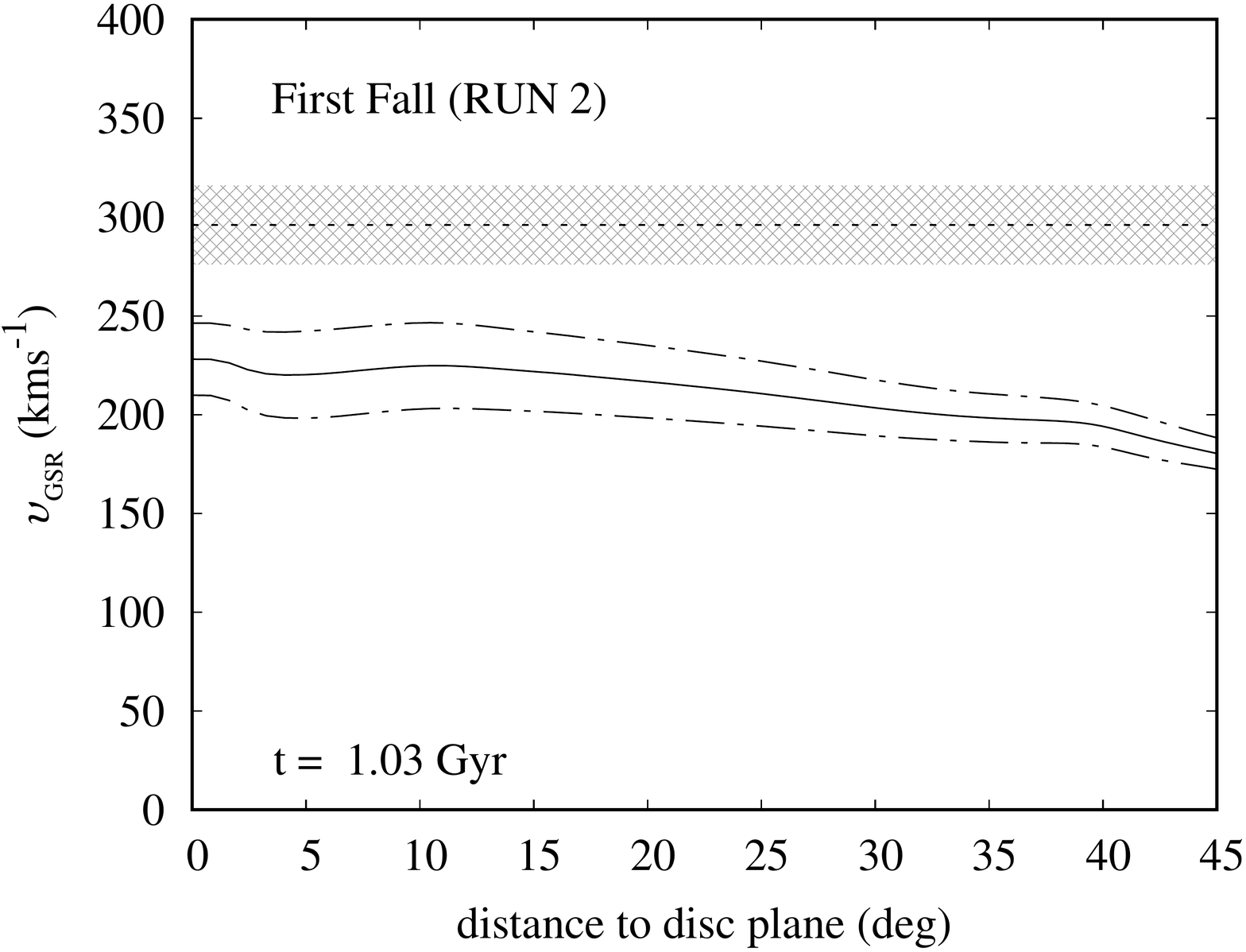}
\caption[ Gas kinematics ]{ Velocity gradient along the streamer in the Galactic Standard of Rest (GSR) at first rise (top) and first fall (bottom; see Figure \ref{fig:mix}). The solid line indicate the mean at each position and the dashed lines, the standard deviation. The dotted horizontal line and the hatched area correspond to the mean value and the uncertainty in the total space velocity of the Smith Cloud \citep[$\sim 296 \pm 20 ~\kms$;][]{loc08a}. See text for details. An animated version of the line-of-sight velocity of the gas and additional material can be found following this \href{http://www.physics.usyd.edu.au/~tepper/proj_smith_paper.html\#stars_gas_kin_gc_run2}{link}.}
\label{fig:gkin}
\end{figure}

\subsection{Gas structure and kinematics} \label{sec:struct}

The Smith Cloud displays an apparent cometary morphology, with a bullet like, dense core (closest to the Galactic \HI\ disc) surrounded by filamentary shreds of gas \citep[][]{loc16a}. In our models, the DM confined gas shows a clear head-tail structure both when approaching the disc and moving away from it after its transit (see e.g. Figure \ref{fig:trc}). Although with a less apparent 'head' and a more prominent tail, the streamer displays a similar geometry. In all cases, however, the gas shows a rather smooth appearance, in disagreement with \citet[][]{nic14b}'s corresponding result. The reason for this discrepancy is not fully clear; it could be due to differences in the numerical scheme adopted by each group. More specifically, the Riemann Solver we have used, based on the {\em Local Lax-Friedrics} method \citep[LLF; ][see also \citealt{tor09a}]{lax54a}, is perhaps diffusive enough to wash out any small-scale structure, despite our relatively high spatial resolution. It is also possible that our adopted refinement strategy has lead to a `smearing' of structure close to the resolution limit, especially when the gas detaches from its confining DM subhalo. In any case, we do not believe that this circumstance has negatively affected our general results, as they likely do not rely on small-scale effects.

To conclude, we briefly consider the kinematics of the gas streamer. The total space velocity of Smith Cloud ($\sim 300 ~\kms$; measured at its tip) comprises a vertical component of $73 \pm 26 ~\kms$, and an azimuthal component of $270 \pm 21 ~\kms$ \citep[][]{loc08a}, thus moving significantly faster than the {\em local} (i.e. at $z \approx 3 ~\kpc$) Galactic ISM, which lags with respect to the  disc rotation \citep[][]{lev08b,fra08b}. The kinematic measurements of the streamer in our model are presented in Figure \ref{fig:gkin}. The top panel shows the total speed (in the rest-frame of the Galaxy) of the gas along the main axis of the streamer at first rise as shown in the top Figure \ref{fig:mix}, from the tail all the way to the subhalo's core and beyond. The speed increases from $\sim 230 ~\kms$ at the base of the streamer to $\sim 280 ~\kms$ at the location of the subhalo. Beyond, the gas velocity drops as the we start to probe the cloud-halo interface. The bottom panel shows the corresponding result for the gas streamer at first fall shown in the bottom panel of Figure \ref{fig:mix}. In either case, the speed is only marginally consistent with the observed speed of the Smith Cloud, indicated by the hatched area in both panels. We note that, by definition, the vertical velocity of the cloud at rise is positive, while it is negative during fall, and is on the order of 100 \kms\ in either case. The cloud's velocity during second approach is identical to the subhalo's velocity (see Figure \ref{fig:kin}), and thus much higher than the total space velocity of the Smith Cloud. A better match between model and observation may be obtained by imposing a different kinematic initial condition on the cloud's progenitor, although it is not obvious what an appropriate orbit and initial velocity one should choose.

\section{Discussion}

Based on our results, we conclude that the Smith Cloud may be explained as the result of a highly energetic disc transit of an infalling DM confined gas cloud in the recent past. This explanation offers in turn a number of possible interpretations. The first of these, which corresponds to a DM-confined cloud {\em on first approach} is deemed unlikely, as it requires  the {\em intrinsic} metallicity of the Smith Cloud to be roughly equal to or higher than $0.5 ~\Zsun$ in order to explain its observed enrichment level. We note that this metallicity is significantly higher than the mean metallicity {\em today} of the ISM in the Small Magellanic Cloud (0.2 \Zsun) and comparable to the present-day mean ISM metallicity of the Large Magellanic Cloud \citep[][]{rus92a}. Note that these systems have halo masses well in excess of $10^9 ~\Msun$ \citep[][]{sta04a} and $10^{10} ~\Msun$ \citep[][]{van14a}, respectively, i.e. are currently more massive than our Smith Cloud progenitor.

The second possible interpretation of the Smith Cloud as a DM-confined cloud {\em on a second approach} appears problematic given its velocity prior to impact which is too high ($\sim 450 ~\kms$) compared to the observed orbital velocity of the Smith Cloud ($\sim 300 ~\kms$). The kinematic properties of our model cloud may be affected by the choice of orbit and initial velocity. But given that the Galaxy's DM halo mass is well constrained from observation \citep[e.g.][]{zar17a,die17a}, and thus the overall Galactic potential, it remains difficult to conceive an orbit that would allow a {\em distant}, infalling cloud to have an impact velocity comparable to the observed speed of the Smith Cloud.

Thus we are left with the most plausible scenario where the Smith Cloud corresponds to (part of) a gas streamer, a long-lived, elongated gas structure extending to the Galactic disc which formed right after the subhalo's transit. But it remains ambiguous whether such a scenario better describes the Smith Cloud as the streamer is moving away from the disc, or falling back onto the Galaxy.

The observed cometary morphology of the Smith Cloud, with its `head' pointing in the direction of the Galactic plane, is often invoked as the main argument in favour of motion towards the Galactic disc. But to date its true proper motion remains unknown. The definite answer may come from a detail study of the magnetic fields around the cloud. As we have shown in our related study of magnetised HVCs \citep[][]{gro17a}, the motion of a gas cloud within a magnetised medium results in a distinctive field configuration around the cloud, consisting essentially of a compressed magnetic skin at its {\em leading} edge and a current sheet {\em behind} the cloud, whose relative strength varies along the cloud's orbit. In view of this result, determining whether and where the Smith Cloud features a magnetic skin and tail will be helpful in constraining its proper motion. The recently discovered enhanced Faraday rotation measures associated with the Smith Cloud \citep[][]{hil13a} are not as yet enough, and more sensitive measurements at and around the Smith Cloud are now needed to elucidate the full structure of its associated field.

To sum up, in our interpretation the Smith Cloud is likely a fragment of a massive, gaseous stream, the `Smith Stream', composed {\em in part} from gas dislodged from the Galactic ISM in the past by the passage through the Galactic disc of a massive subhalo filled with gas. Tentative evidence for the association of the Smith Cloud with a larger gas structure has been briefly discussed by \citet[][]{loc12a}. There, it appears to be part of a filament of gas at an angle with respect to the plane of the Galaxy, extending for several kilo-parsec from the location of the Smith Cloud to the disc and beyond. We conjecture that the gas structure opposed in direction to the Smith Cloud with respect to the disc plane may have a tidal origin, created by ram pressure stripping of gas off the subhalo as it moved through the Galactic hot halo, combined with the extreme tidal forces acting upon it along its orbit. We note that the former requires some form of internal heating, as otherwise the gas would remain bound to the subhalo.

\begin{figure}
\centering
\includegraphics[width=0.42\textwidth]{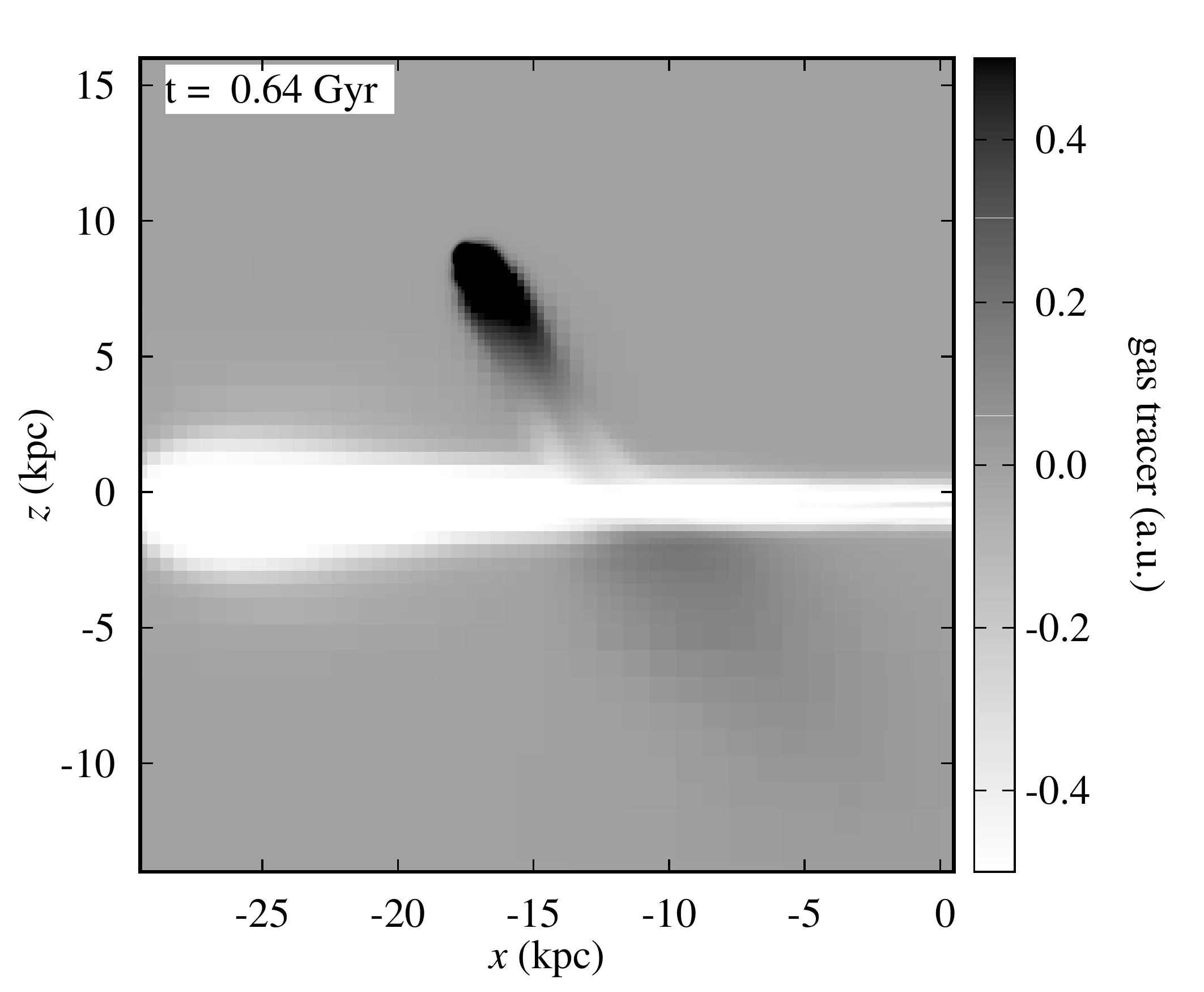}
\includegraphics[width=0.42\textwidth]{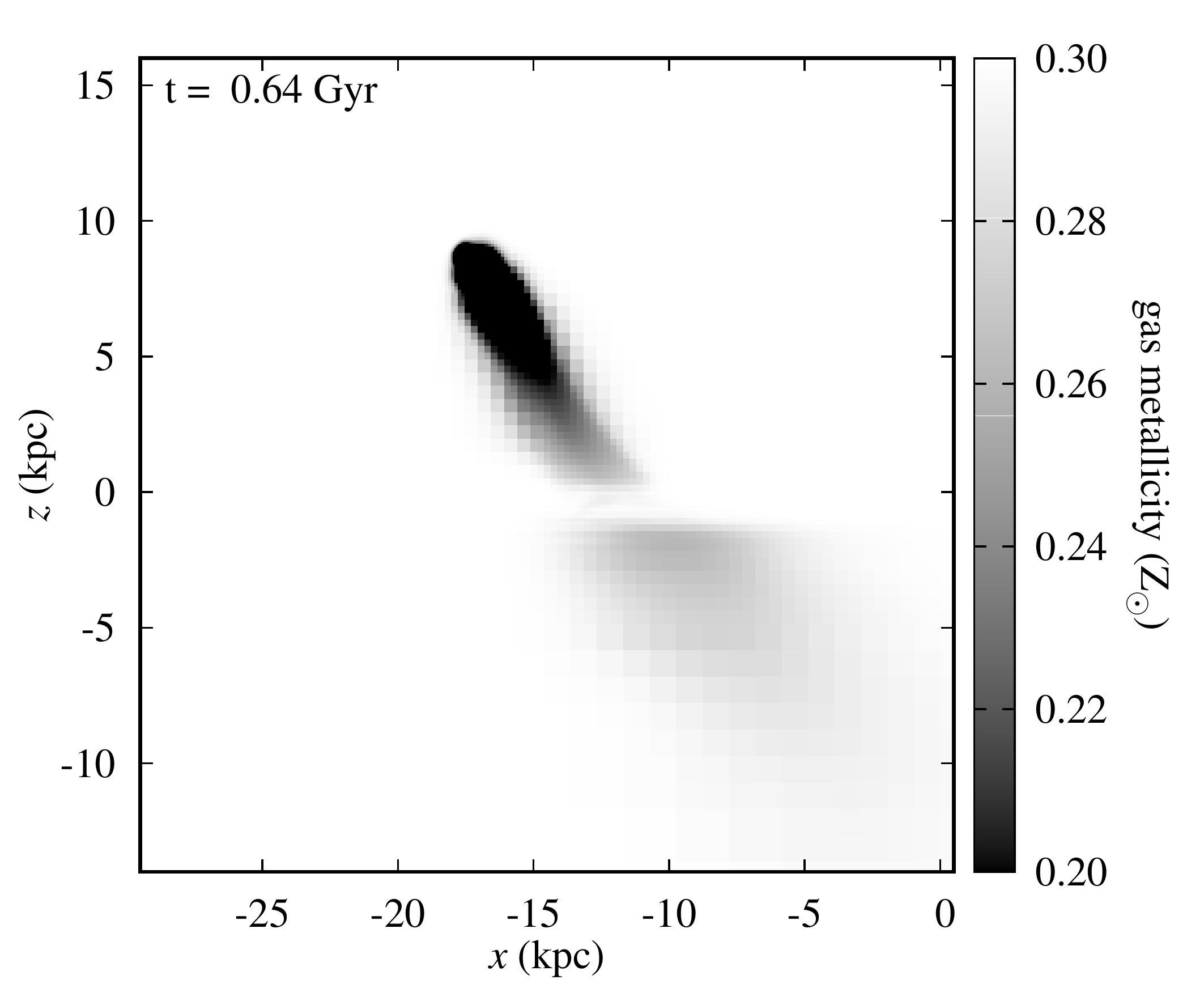}
\caption[ Gas tracer and metallicity (hi res run) ]{ Edge-on projection of the system at $\sim 40 ~\Myr$ after the first transit (RUN 2). The top panel shows the gas tracer (see also Figure \ref{fig:mix}); the bottom panel, the gas metallicity. The gas cloud is easily identifiable, moving upwards away from the disc. Note that we have restricted the colour scale to an absolute tracer value of 0.5 in the top panel, and to a minimum gas metallicity of $0.2 ~\Zsun$ in the bottom panel to enhance the contrast and render visible the tail below the disc formed prior to the transit. Recall that the both hot halo and the gas disc have been initially assigned a metallicity of $0.3 ~\Zsun$, and are thus indistinguishable from one another in the bottom panel.}
\label{fig:str2}
\end{figure}

We have shown that the transit of a massive enough subhalo through the Galactic disc results in a stream on one side of the disc plane. Our model also shows the formation of another tail along the subhalo's orbit on its approach to the disc (Figure \ref{fig:str2}). This tail is probably too diffuse to be detected at a limiting gas column density of $10^{19} ~\psc$. Our model does not produce a prominent tidal tail, likely due to the absence of internal heating that prevents the diffuse hot halo from removing a significant fraction of the gas confined by the subhalo \citep[][]{nic11a}.

The significance of our results is best appreciated within a broader cosmological context. High-resolution, $N$-body simulations of Galaxy-sized halos show that the number of subhalos with masses between $\sim 10^8 \Msun$ and $\sim 10^9 \Msun$ is on the order of 1000 and 100, respectively \citep[][]{spr08a}. However, the overwhelming majority of these subhalos (or of any mass, for that matter) are found in the outer parts of the halo, at $r \gtrsim 100 ~\kpc$, in spite of the fact that their number {\em density} is highest in the central region. In fact, only about 10 percent of halos with masses above $10^8 \Msun$ are found within $r \sim 20 ~\kpc$ \citep[][]{gao04a}. Thus, there are between 10 and 100 possible `Smith-sized' halos over cosmic time inside the optical disc radius of the Galaxy, i.e. roughly one every 0.1 -- 1 Gyr. Assuming they all have gas, our models indicate that, regardless of their mass, these subhalos will survive long enough to be accreted by the Galaxy, bringing in a total gas mass of order $10^9 \Msun$ at a rate of roughly $0.1 ~\Msun ~\yr^{-1}$, an order of magnitude lower than the present-day star formation rate of the Galaxy \citep[e.g.][]{rob10a}.

We recognise that our results may have been affected by the lack of two important features in our galaxy model: 1) a truly {\em fast}-rotating hot halo; and 2) feedback processes within the Galaxy. Both these are expected to have a significant impact on the evolution of gas on its way to be accreted by the Galaxy. A fast-rotating corona may affect the hydrodynamic evolution of gas depending whether this gas is coming in on a prograde or a retrograde orbit. And it may also affect the evolution of gas streamers as these rise from the disc. Similarly, feedback (e.g. winds) may assist gas removal by increasing the gas internal energy, thus affecting the survivability of (DM-confined) clouds \citep[][]{coo09a}. Any improvement of our model needs to address these shortfalls. In addition, it may be necessary to move away from a single subhalo-Galaxy interaction, and to include a number of such subhalos featuring e.g. a range of masses, gas fractions, and orbits to cover a wider parameter space and thus better assess their survivability as well as their collective effect on the Galactic disc. It may be of further interest to consider prescriptions that allow the gas to cool well below our temperature floor \citep[$T = 10^4 ~\K$, e.g.][]{tan16b}, not considered here due to the very high resolution required.

In any case, the model introduced here -- in particular of the Galaxy -- represents a significant step forward in the study of gas processes within isolated systems at galactic scales, which is an important and necessary complement to large-scale, full cosmological simulations of galaxy formation.

\section*{Acknowledgements}

We thank the referee for carefully reading our manuscript and providing insightful comments that helped improving the presentation of our results. TTG acknowledges financial support from the Australian Research Council (ARC) through an Australian Laureate Fellowship awarded to JBH. We acknowledge the Sydney Informatics Hub and the University of Sydney's high performance computing (HPC) cluster Artemis for providing the HPC resources that have contributed to the research results reported within this paper. We thank Romain Teyssier and Valentin Perret, respectively, for making {\sc ramses} and \dice\ freely available. All figures and movie frames created with {\sc gnuplot}, originally written by Thomas Williams and Colin Kelley.\footnote{Version 5.0, currently available at: \url{http://www.gnuplot.info}.} All animations assembled with {\sc splicer}, written by R.~S. Sutherland.\footnote{Version 1.2.2, currently available at :\\ \url{https://miocene.anu.edu.au/splicer}.}\\

\clearpage

\bibliographystyle{mnras} 
\input{manuscript_mnras.bbl} 

\appendix

\section{DM-free clouds} \label{sec:dmf}

\begin{figure*}
\centering
\includegraphics[width=0.23\textwidth]{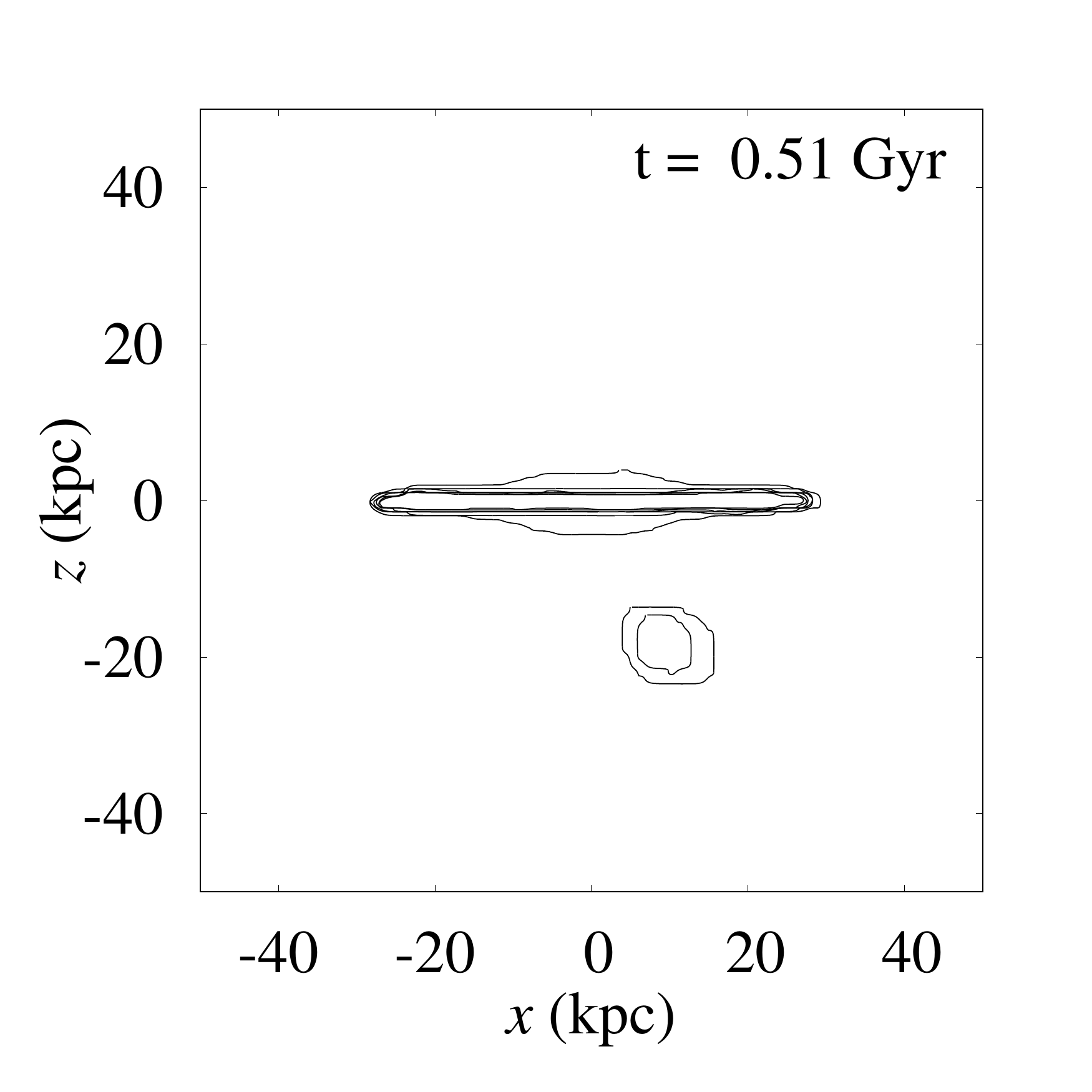}
\includegraphics[width=0.23\textwidth]{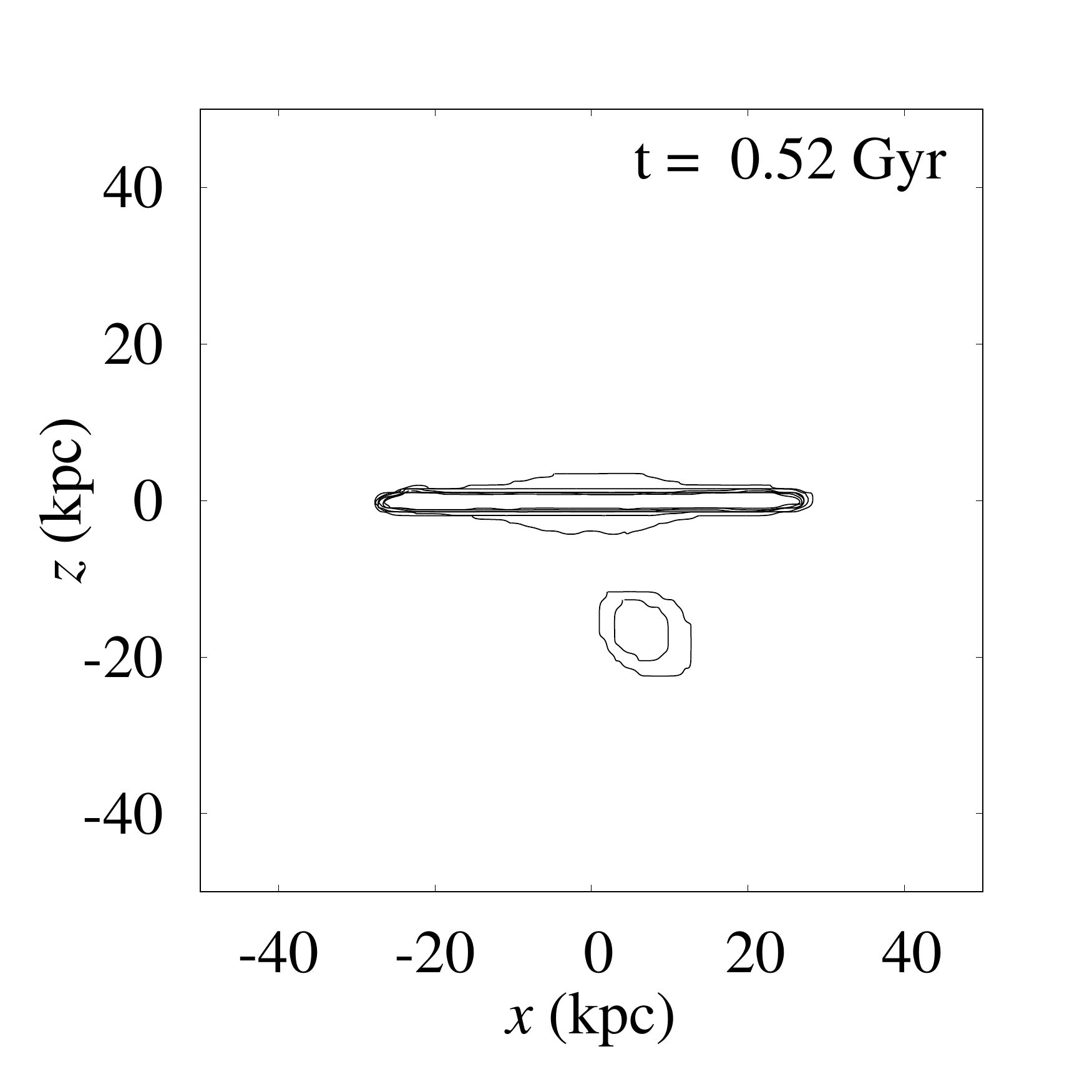}
\includegraphics[width=0.23\textwidth]{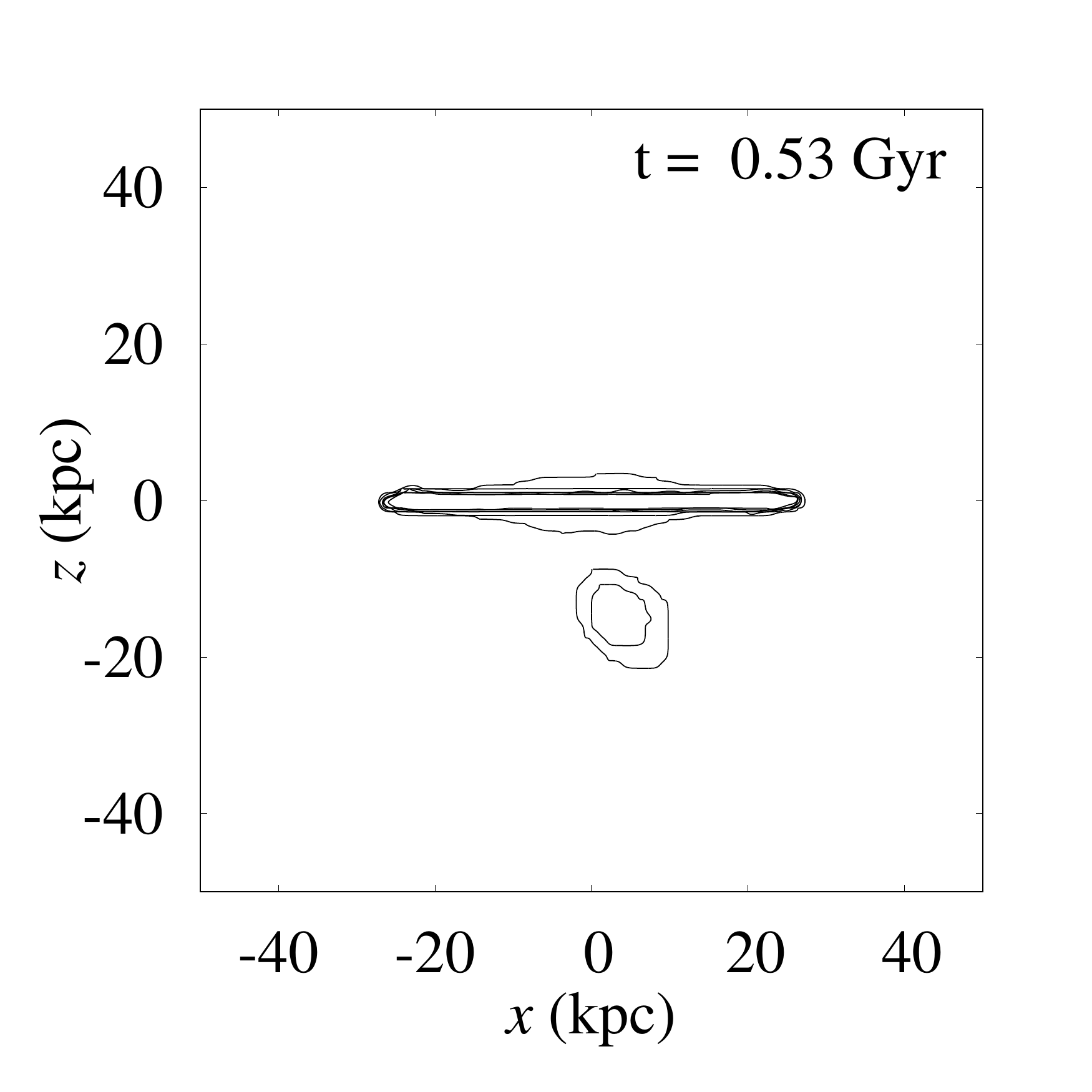}
\includegraphics[width=0.23\textwidth]{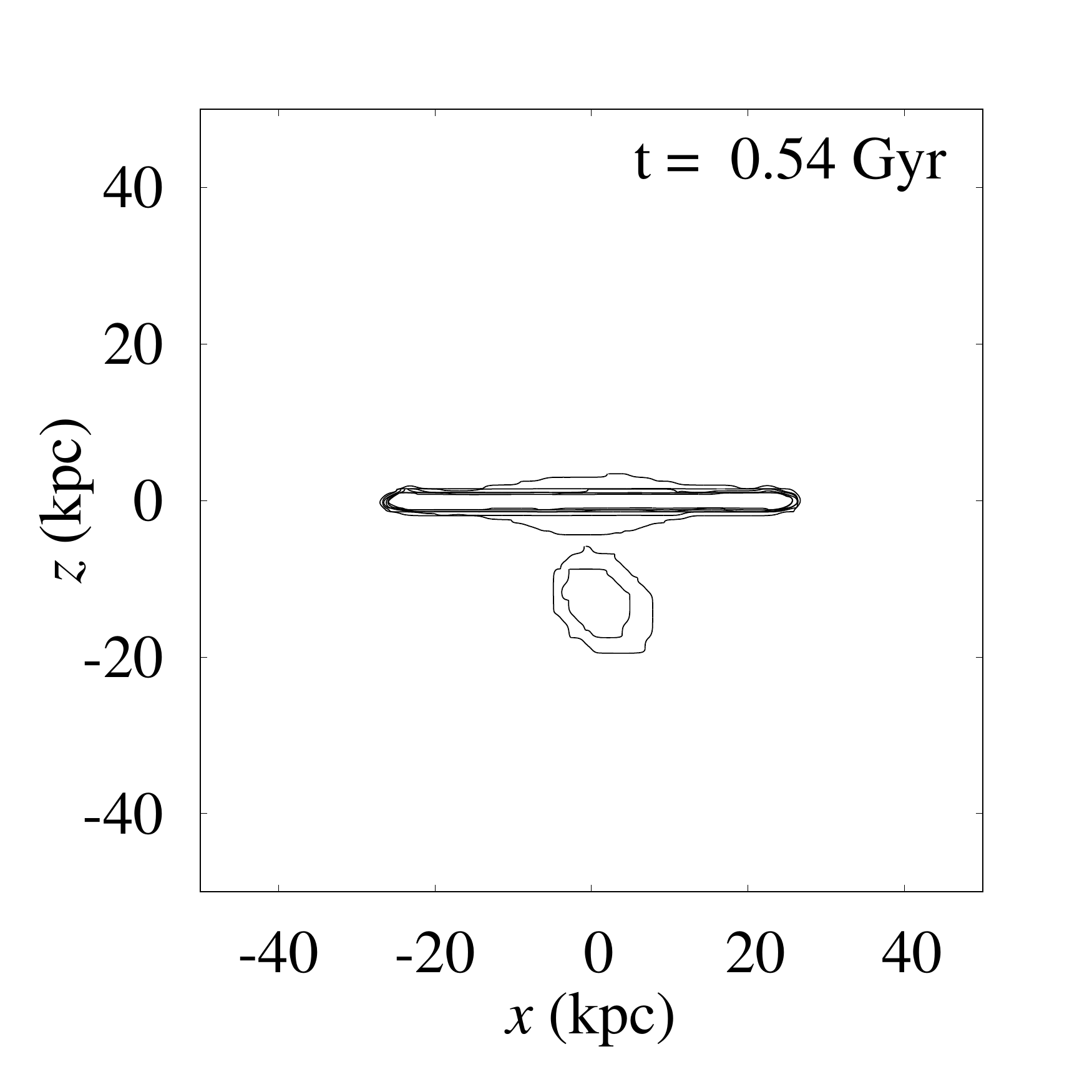}\\
\vspace{-5pt}
\includegraphics[width=0.23\textwidth]{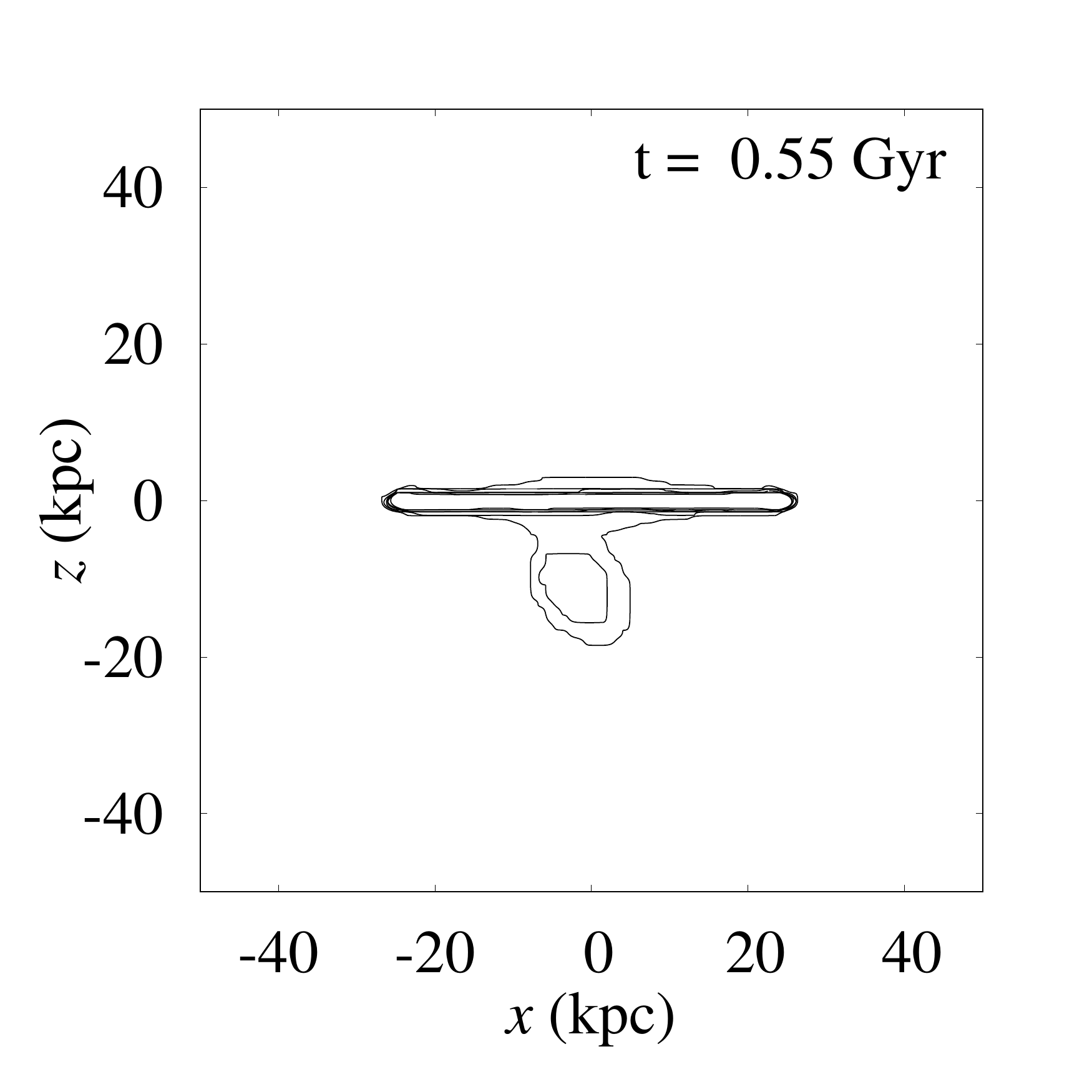}
\includegraphics[width=0.23\textwidth]{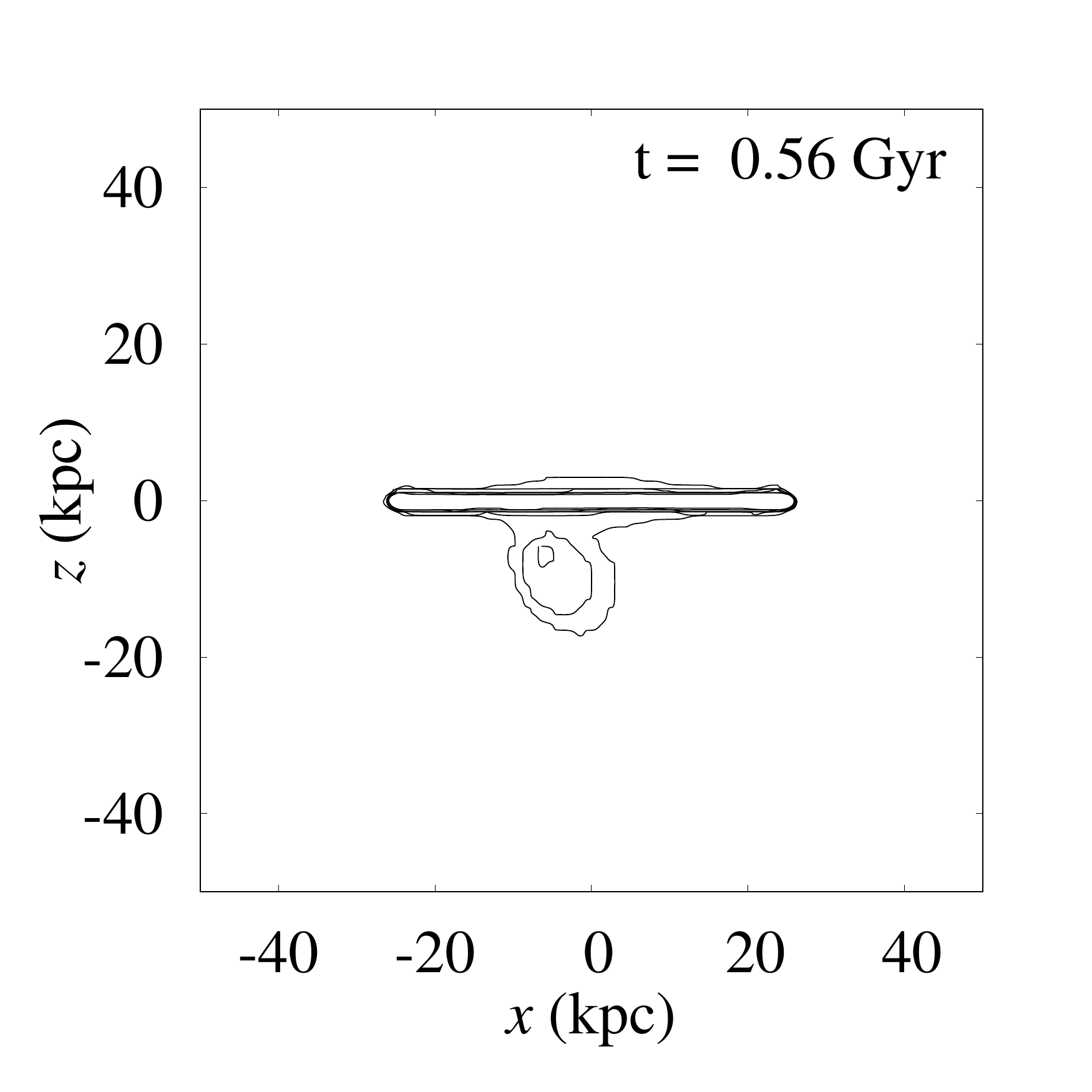}
\includegraphics[width=0.23\textwidth]{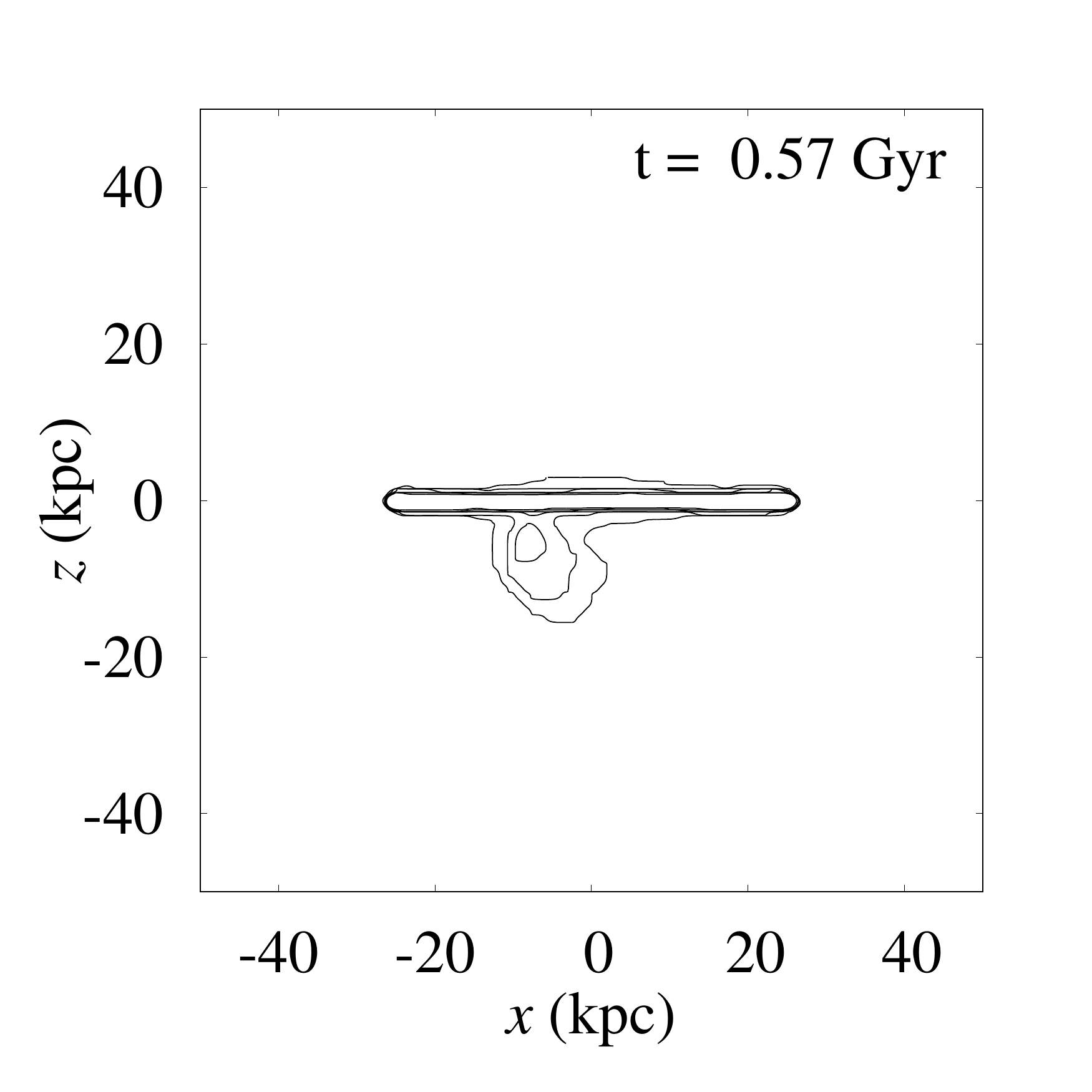}
\includegraphics[width=0.23\textwidth]{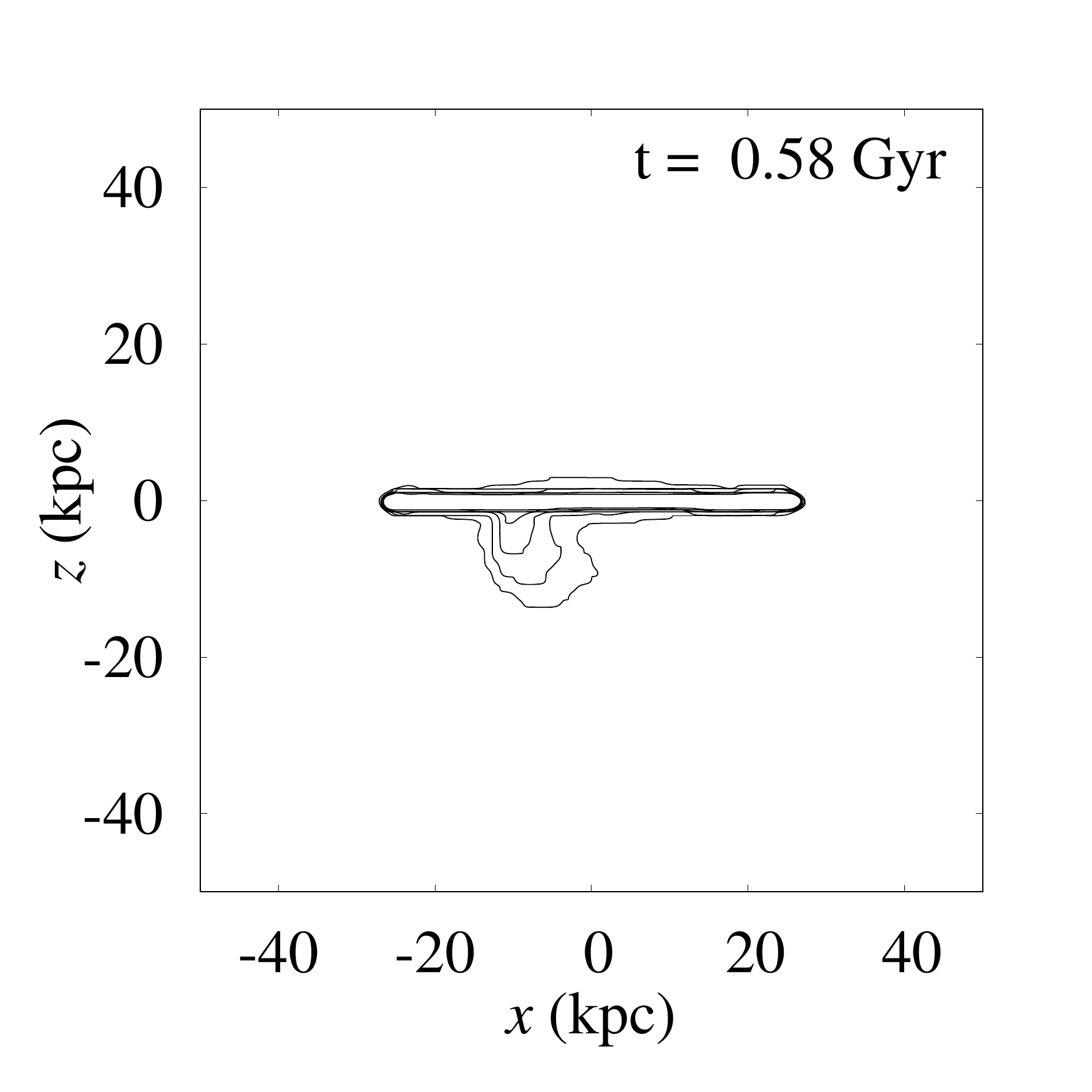}\\
\vspace{-5pt}
\includegraphics[width=0.23\textwidth]{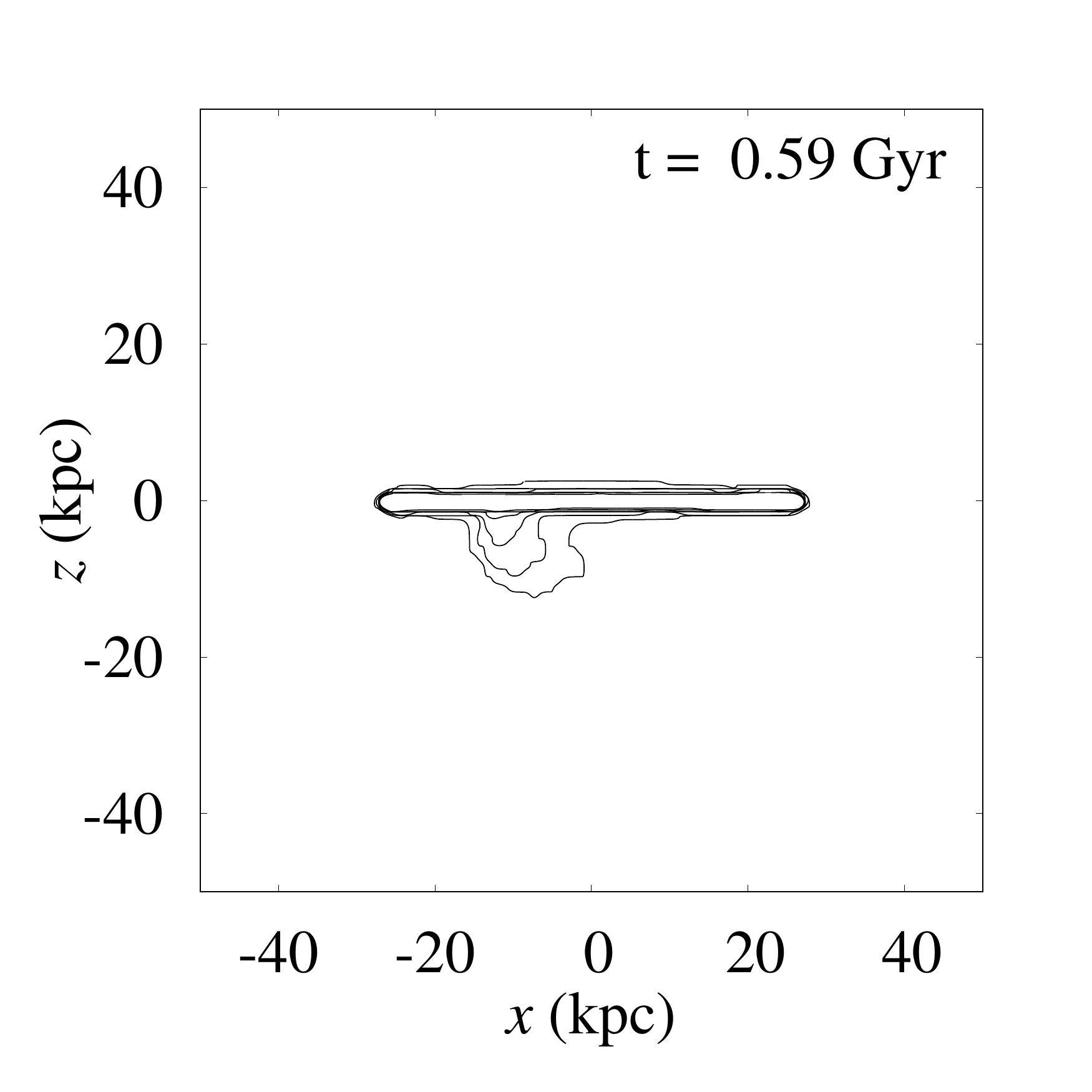}
\includegraphics[width=0.23\textwidth]{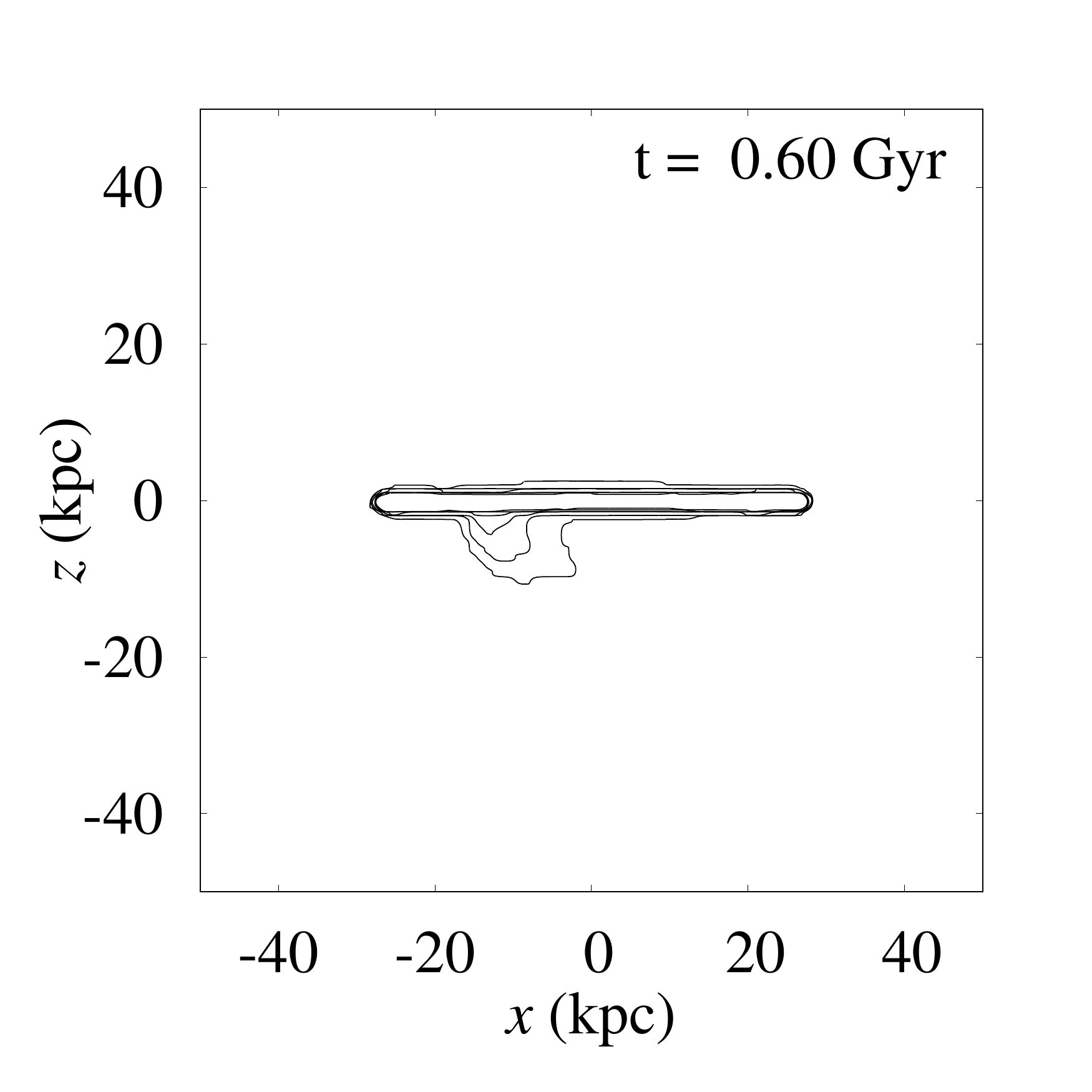}
\includegraphics[width=0.23\textwidth]{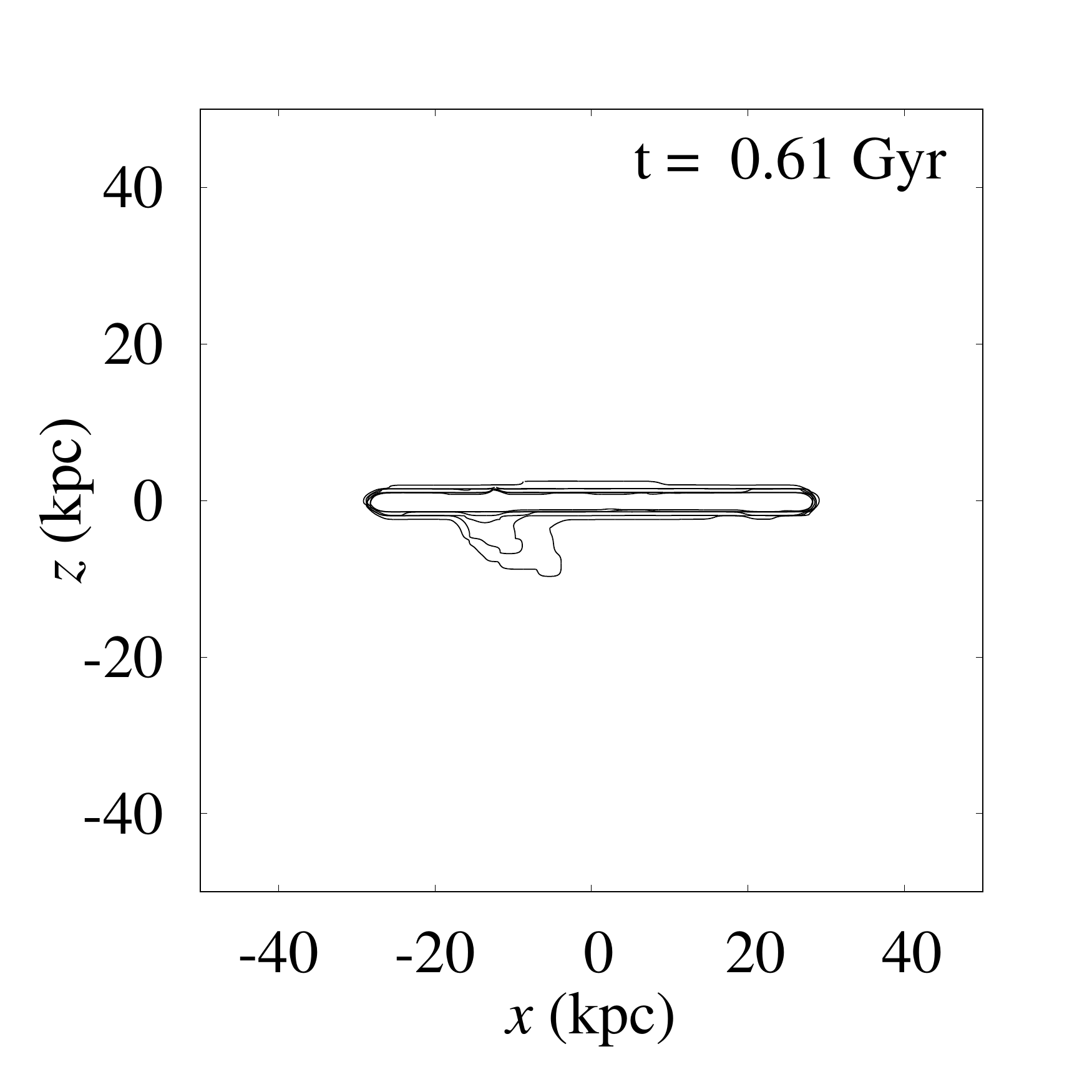}
\includegraphics[width=0.23\textwidth]{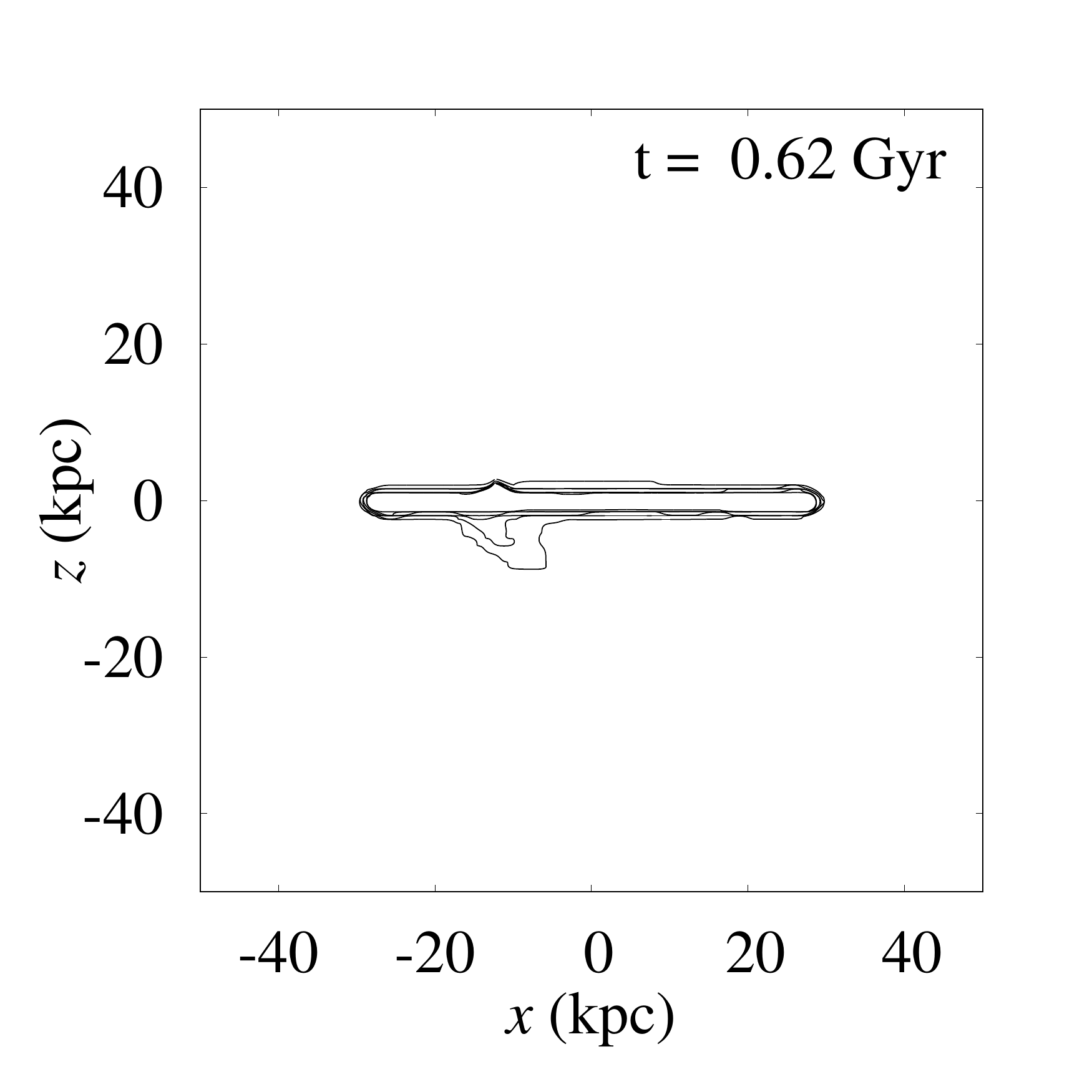}\\
\vspace{-5pt}
\includegraphics[width=0.23\textwidth]{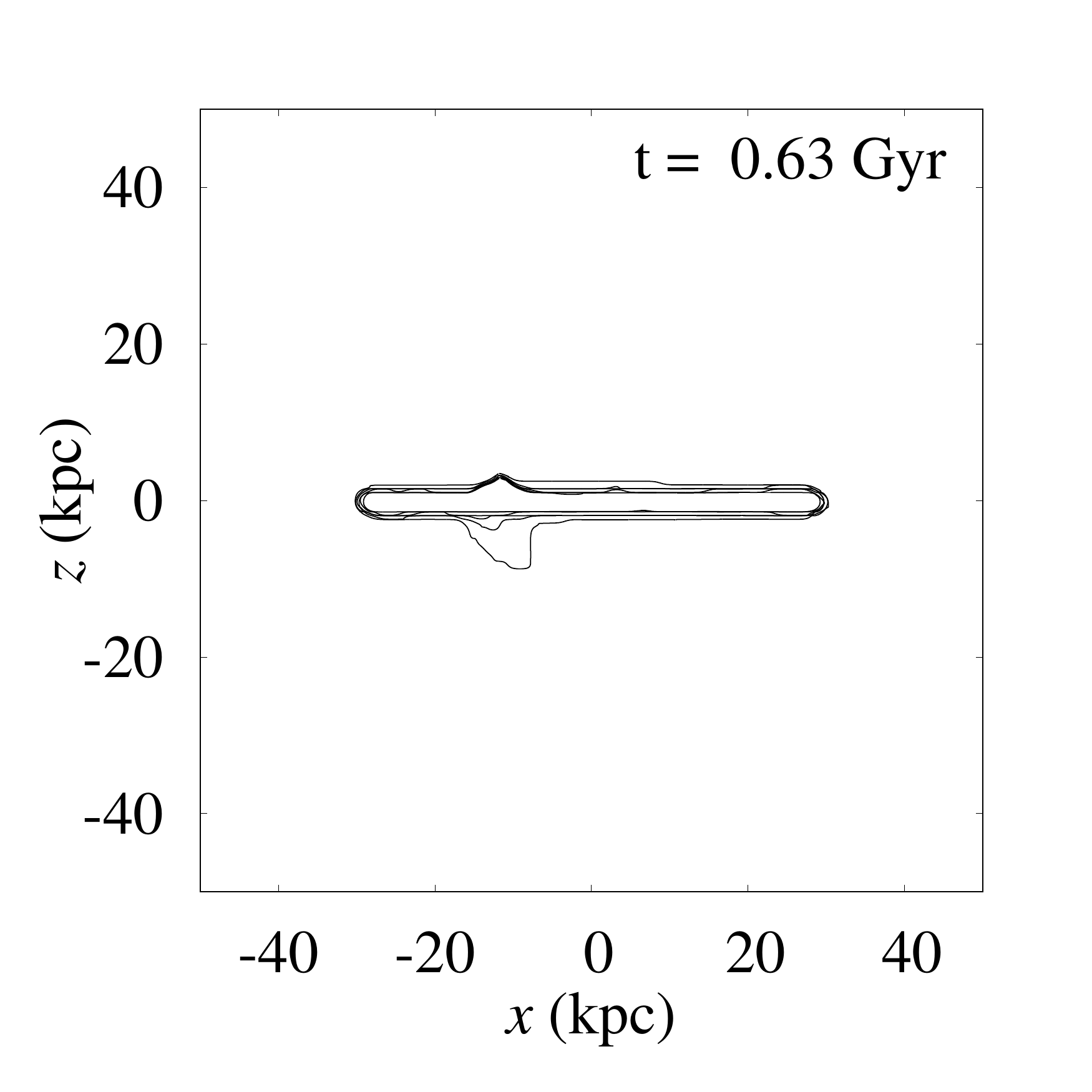}
\includegraphics[width=0.23\textwidth]{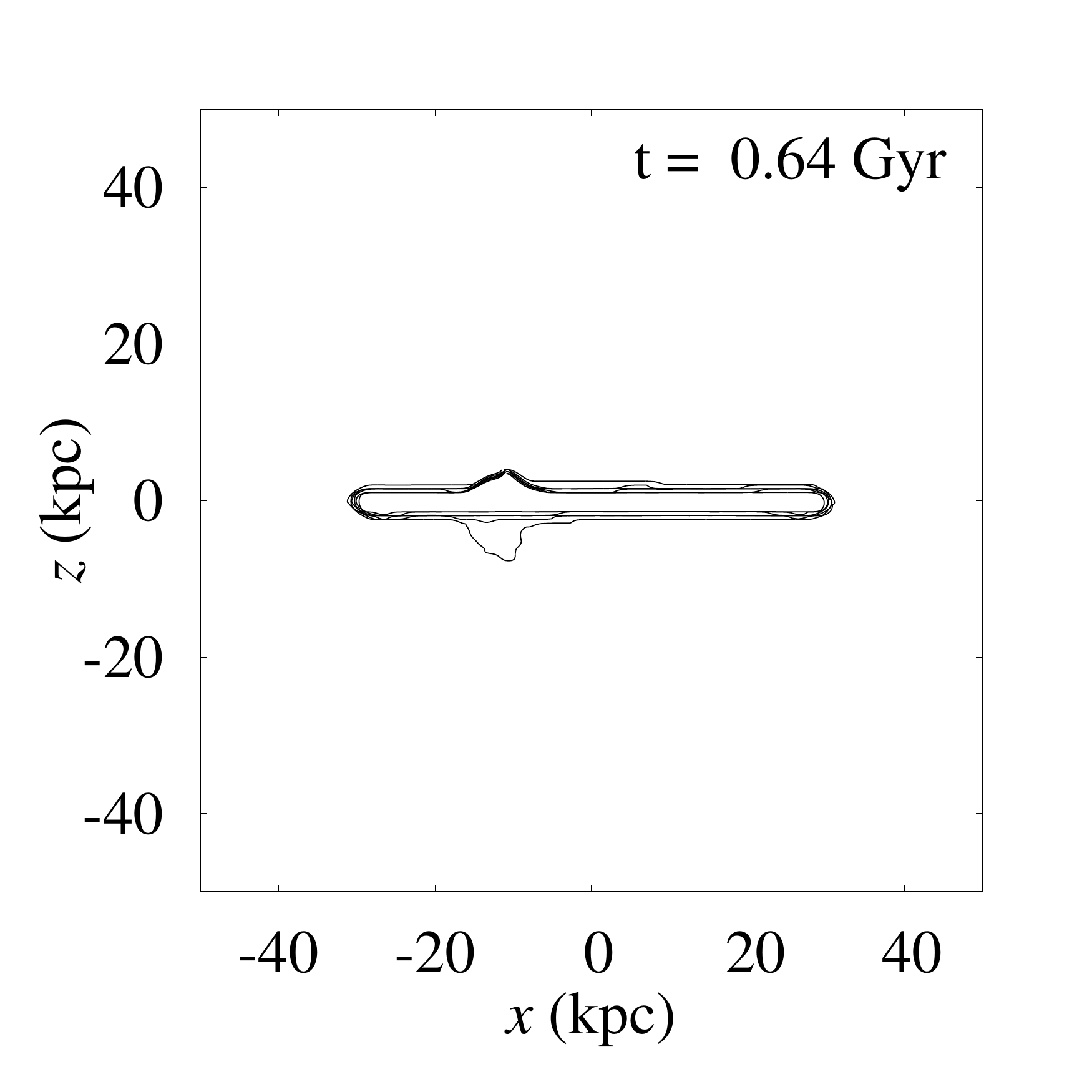}
\includegraphics[width=0.23\textwidth]{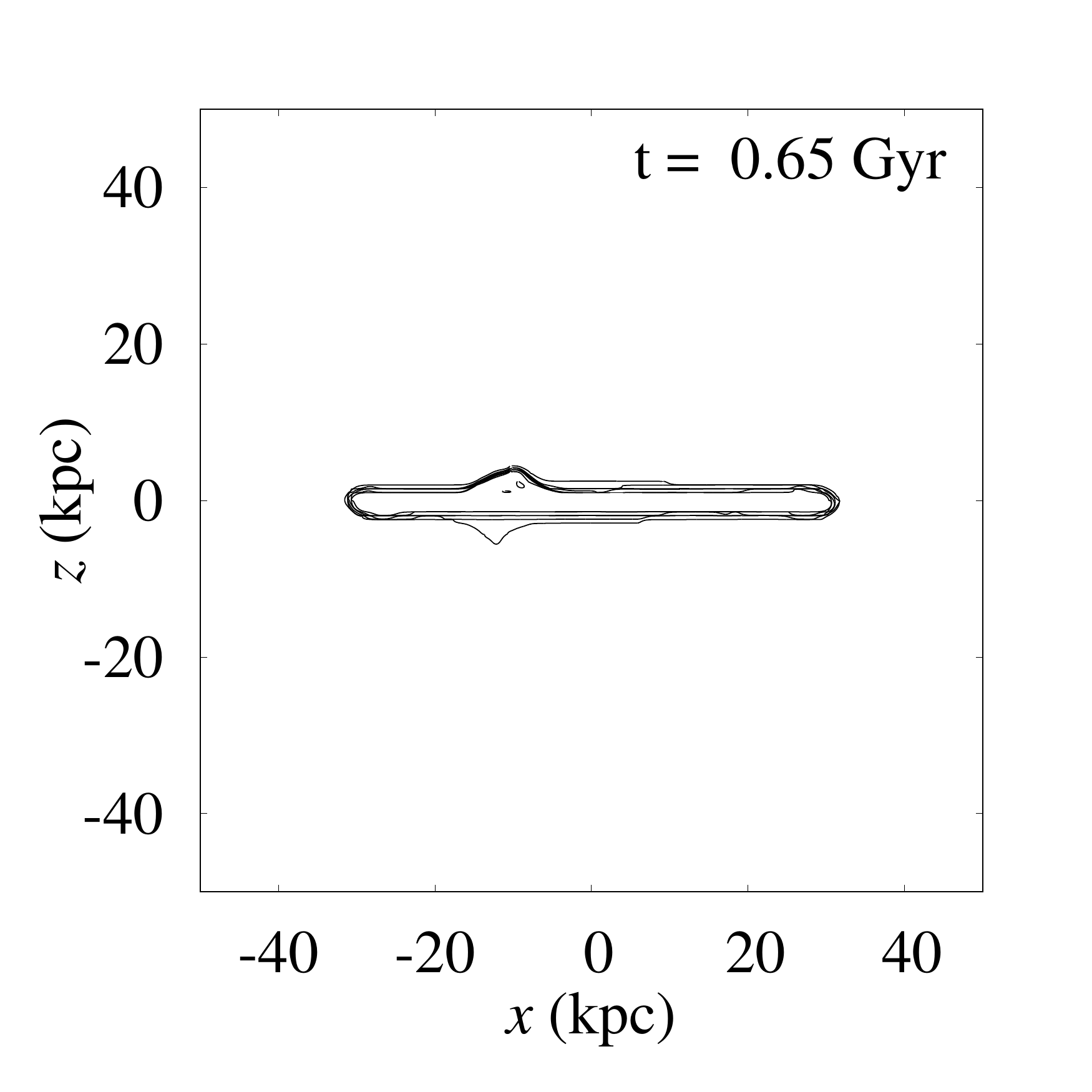}
\includegraphics[width=0.23\textwidth]{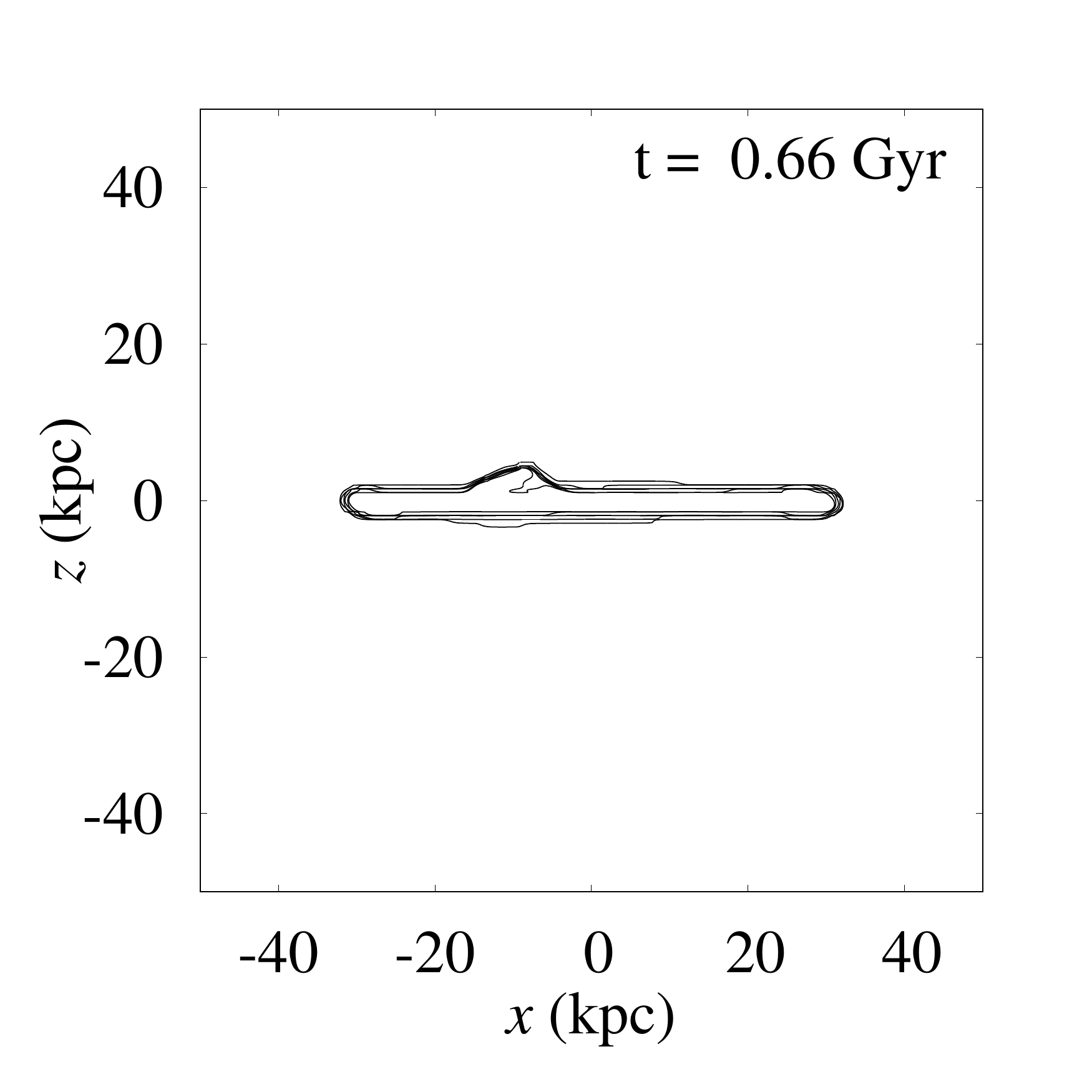}\\
\vspace{-5pt}
\includegraphics[width=0.23\textwidth]{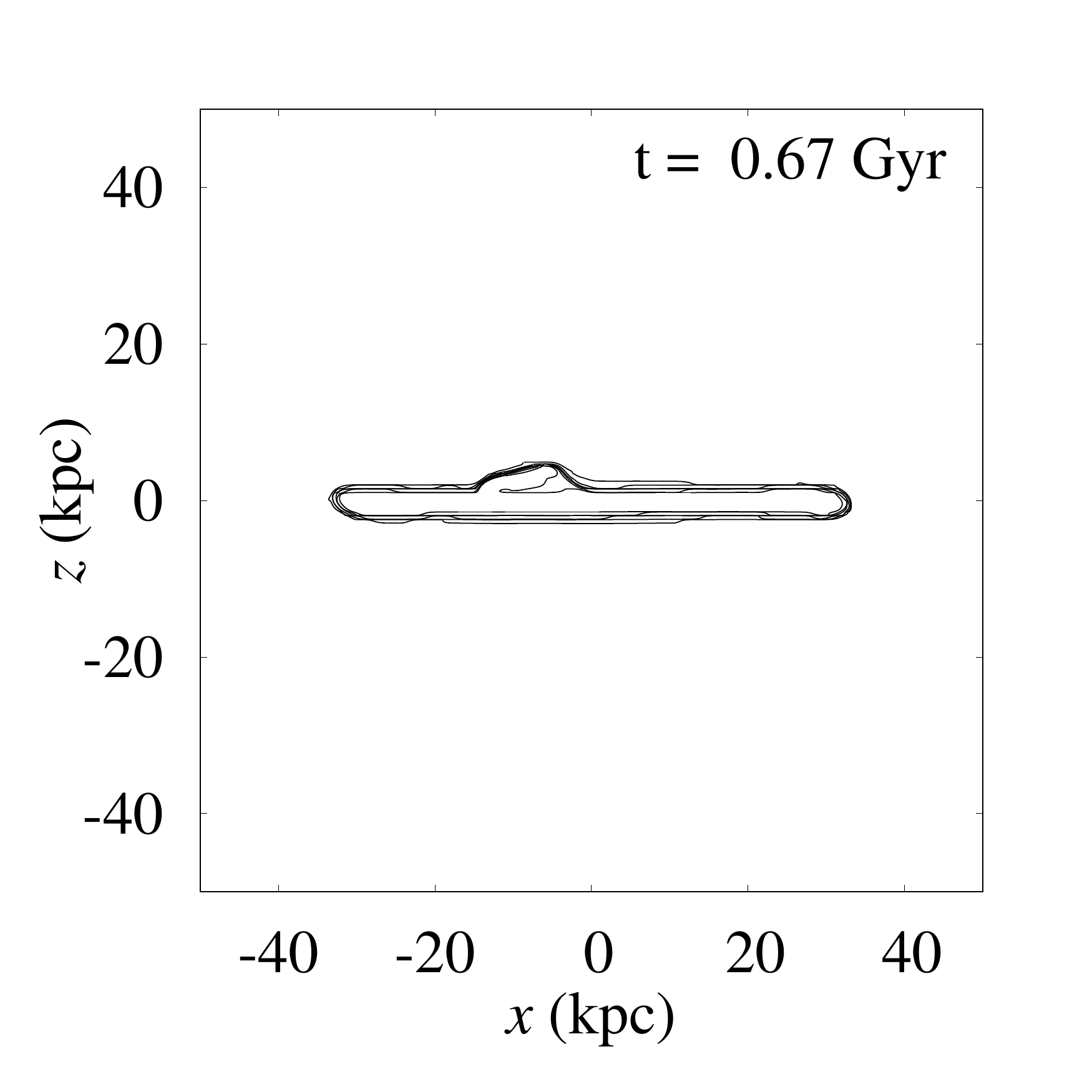}
\includegraphics[width=0.23\textwidth]{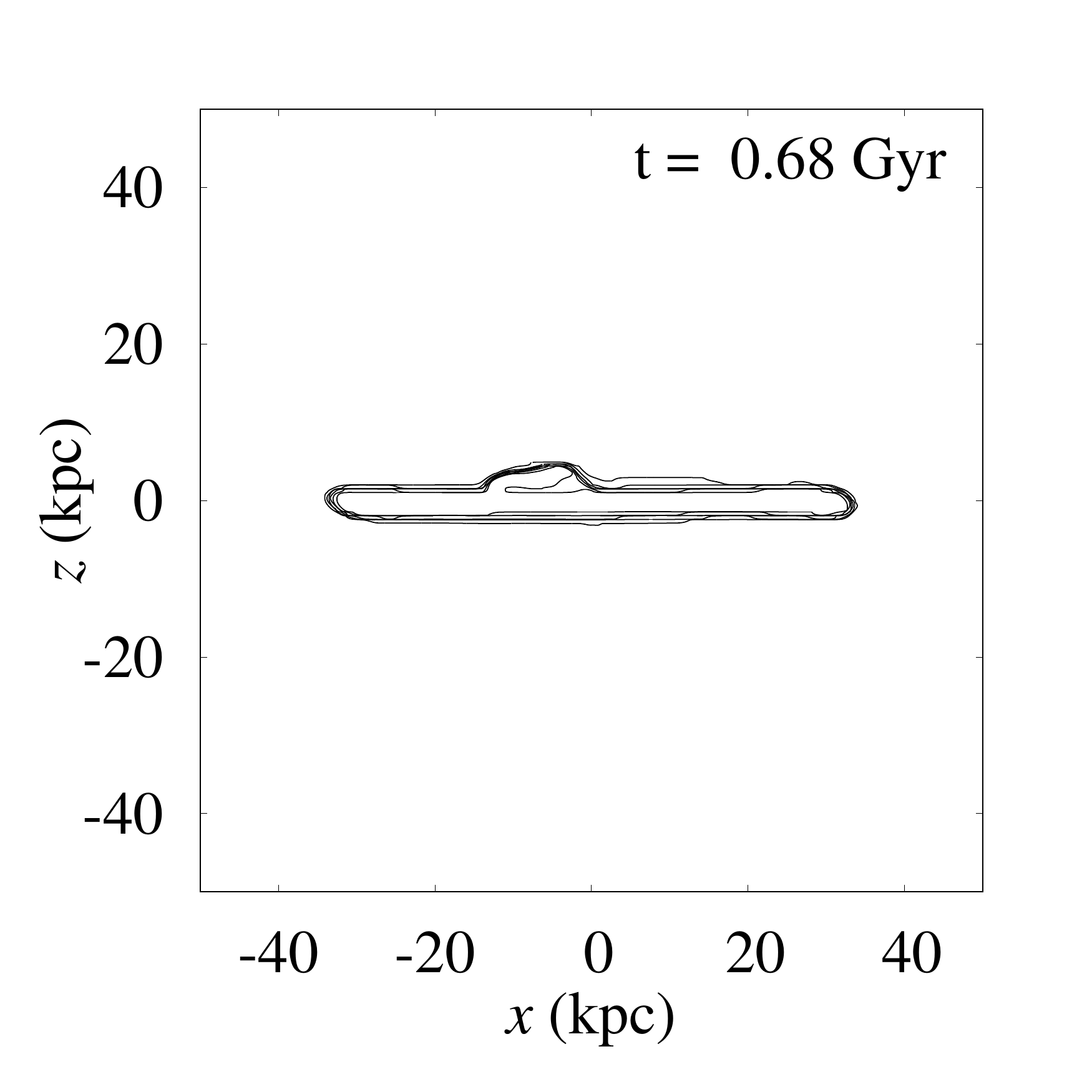}
\includegraphics[width=0.23\textwidth]{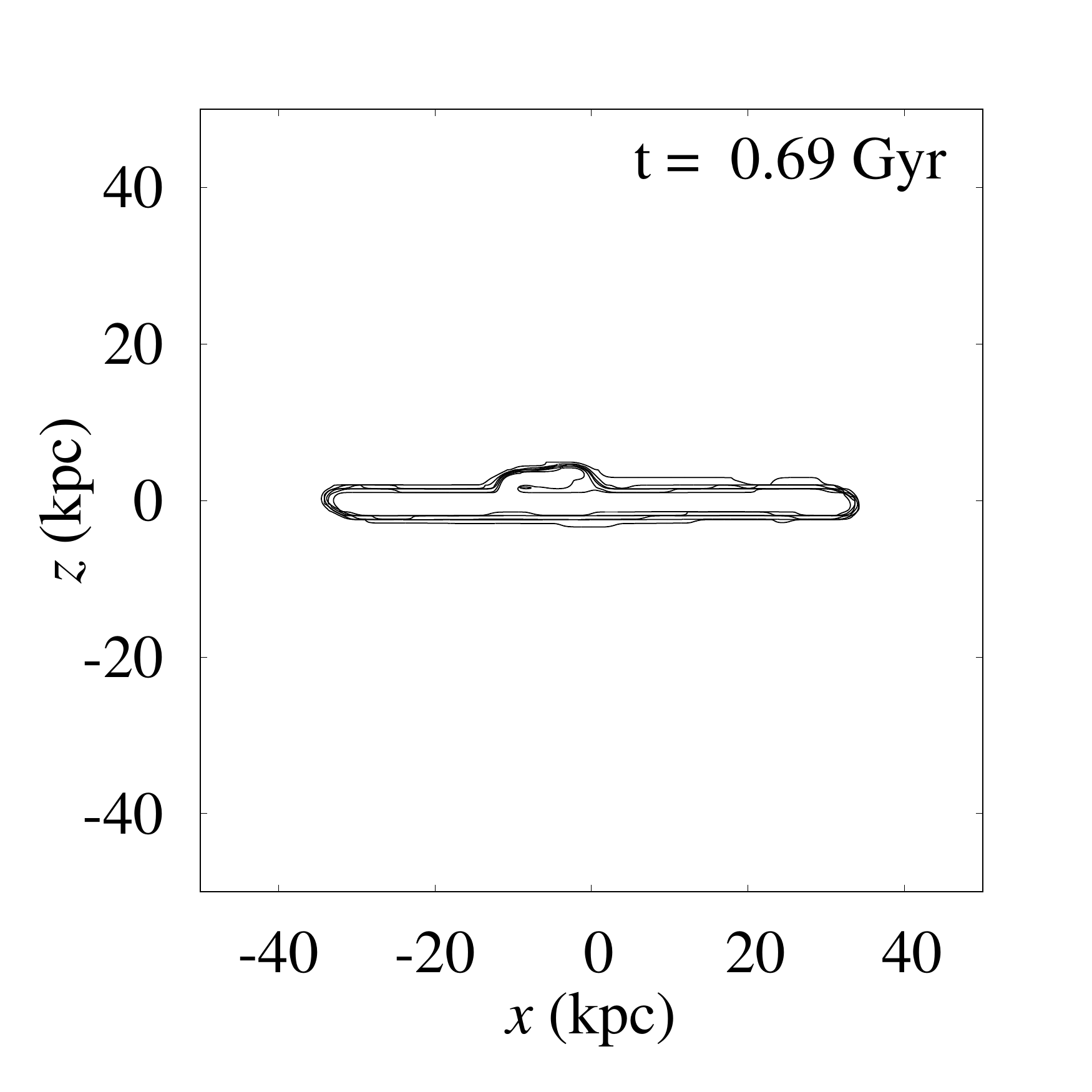}
\includegraphics[width=0.23\textwidth]{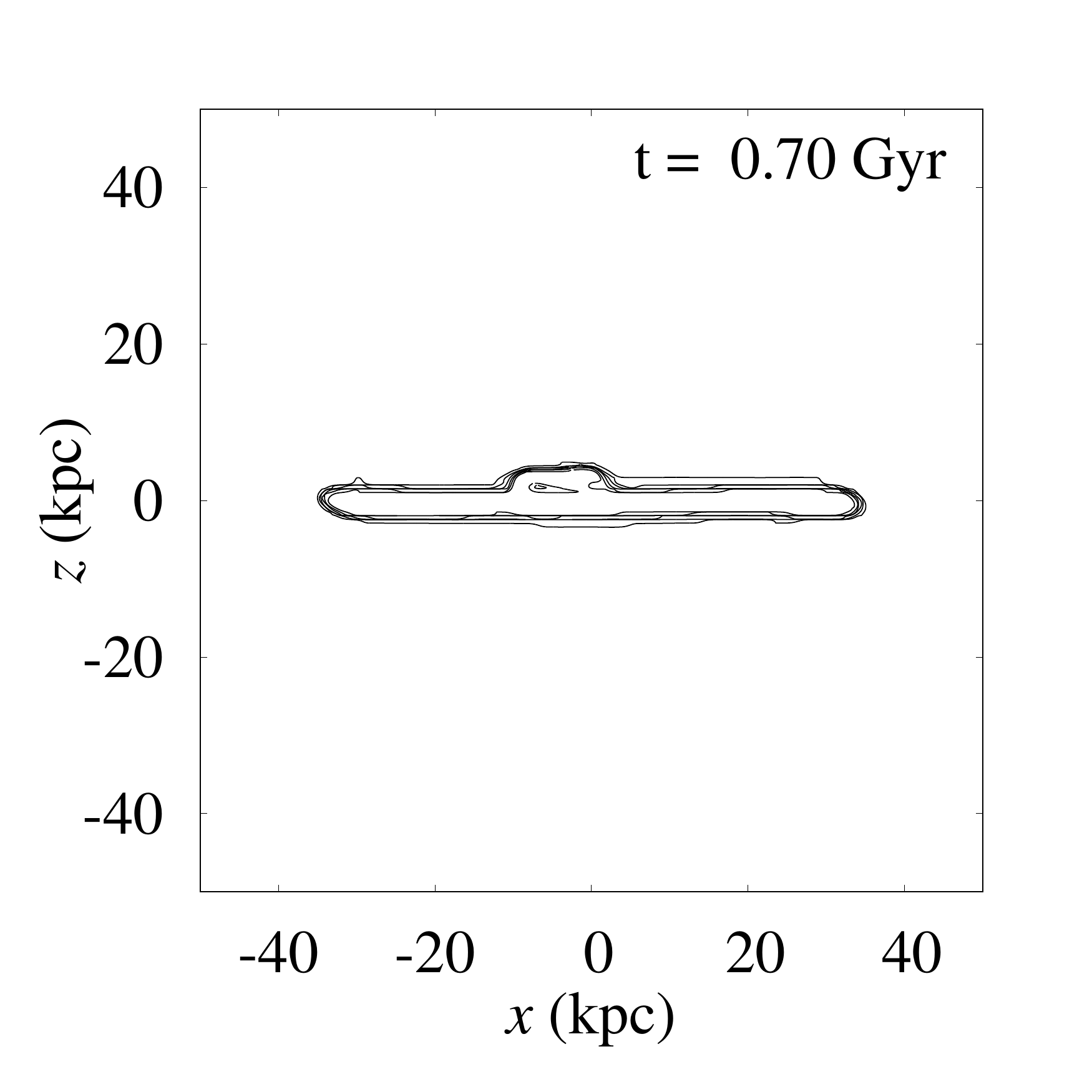}
\caption[ DM-free cloud ]{ Accretion of a DM-free cloud by the Galaxy. The panels show the evolution of a pure gas cloud close to the disc, after having travelled $\sim 100 ~\kpc$ through the hot halo on an initially tangential orbit (with respect to the plane of the disc). Each panel shows an edge-on projection of the system, and contains a series of contours corresponding to total gas column densities in the range $10^{20} ~\psc$ to $10^{21} ~\psc$ in steps of 0.2 dex. The contours corresponding to the gas cloud indicate a total gas column densities, in logarithmic scale, of 20, 20.2 (and 20.4). For an animated version of this figures and additional material follow this \href{http://www.physics.usyd.edu.au/~tepper/proj_smith_paper.html\#snpart_ctr_gc_run4}{link}.}
\label{fig:dmf}
\end{figure*}

It has been argued \citep[][]{nic09a,nic14b,gal16a} that the Smith Cloud, if not confined by DM, is necessarily on its first approach to the disc. To test whether we recover this apparent robust result within our framework, we have run a high-resolution model -- denoted `RUN 4' -- using initial conditions essentially identical to those of RUN 2, with exception of the Smith Cloud progenitor, which in this case consists of a massive ($1.5 \times 10^8 ~\Msun$) cloud {\em without} a supporting DM subhalo around it (see Table \ref{tab:runs}).

In agreement with previous studies, we find that a pure gas cloud does not survive an impact with the disc. Figure \ref{fig:dmf} displays a time sequence of the composite system, starting at roughly 50 Myr before the impact, and ending at $\sim 150 ~\Myr$ after the event. Clearly, the gas cloud is not able to transit the disc, but is rather accreted into it on a short timescale of a few 10 Myr. Perhaps more interesting is the finding that a  massive ($\gtrsim 10^8 ~\Msun$), baryonic gas cloud is able to travel for over 100 kpc at speeds in excess of 100 \kms\ (with respect to the Galaxy's centre) across a hot medium all the way to the disc. This does not contradict earlier work \citep[e.g.][]{hei09b}, since clouds as massive as ours have not been previously considered. It is worth emphasising that the cloud travels on a {\em prograde} orbit with respect to the hot halo (and the disc), which reduces the ram pressure and thus the hydrodynamic interaction. Still, the cloud does not survive its journey intact. Despite its initial high density and relatively high metallicity, cooling is not enough to keep the cloud compact and -- lacking the shielding effect otherwise provide by DM confinement -- it expands as a result of the interaction with the halo gas, thus being subject to stronger hydrodynamic forces.

\begin{figure}
\centering
\hfill\includegraphics[width=0.42\textwidth]{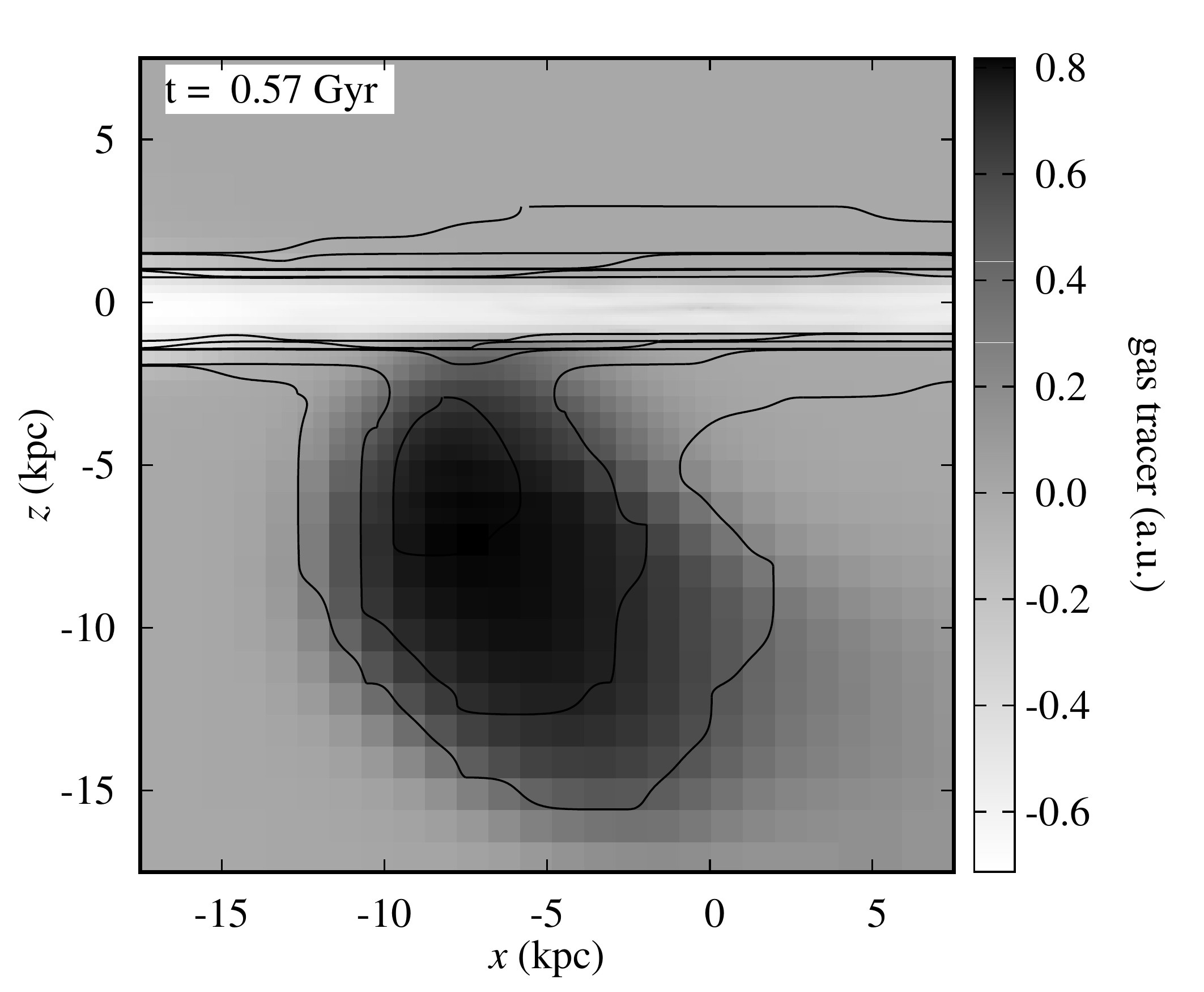}
\includegraphics[width=0.42\textwidth]{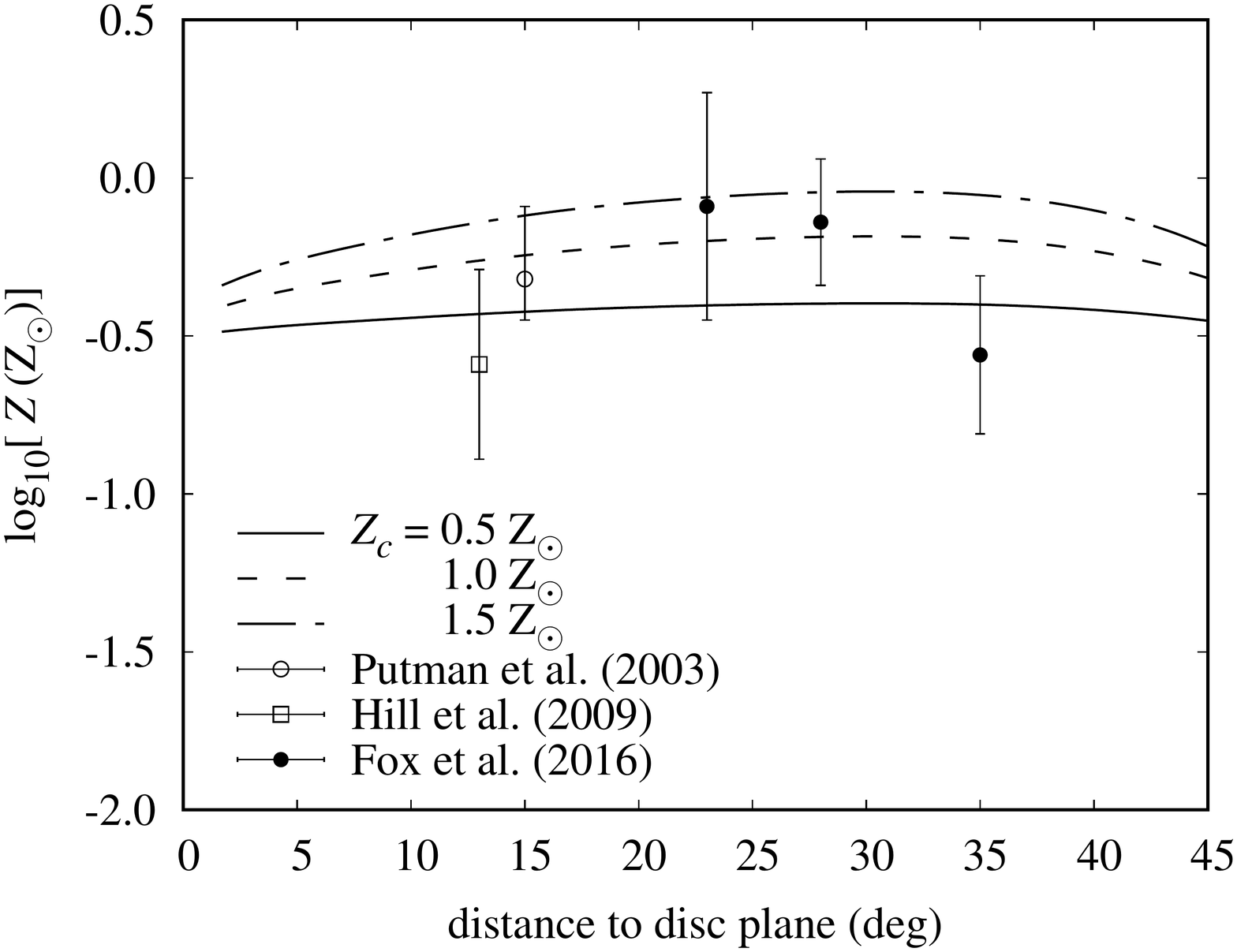}
\caption[ Zoom-in (DM-free) ]{ Top: DM-free cloud during its first approach. The colour scale indicates the value of the gas tracer (see Section \ref{sec:mix}); the contours on top of the gas cloud correspond to total gas column densities, in logarithmic scale, of 20, 20.2, and 20.4 (\psc). Bottom: Gas metallicity along the cloud as shown on top as a function of the angular distance to the disc plane. Each of the model curves correspond to a different {\em initial cloud metallicity} as given in the legend. The data points are identical to those included in Figure \ref{fig:mix}. Clearly, because the cloud has not yet reached the disc, there has been no mixing. In order for this scenario to explain the metallicity trend observed along the Smith Cloud, the DM-free cloud must have an intrinsic high level of heavy elements (cf. Figure \ref{fig:mix}). }
\label{fig:mix2}
\end{figure}

While our DM-free model may tentatively support a first-approach scenario to explain the Smith Cloud, we are inclined to claim differently. The are at least two problems with this interpretation. The first and most important relates to the metallicity measurements along the Smith Cloud \citep[][]{put03b,hil09a,fox16a}. We find that, unless the Smith Cloud has been previously polluted with heavy elements to a relatively high level, the metallicity observed along its main body and tail cannot be explained by a cloud on its first approach to the disc. This mainly a consequence of the lack of mixing with the ISM prior to the transit (see also Section \ref{sec:mix}). Figure \ref{fig:mix2} shows a snapshot of the DM-free cloud just before it reaches the disc (top), and metallicity measurements along the cloud, adopting different values for its intrinsic metallicity. For an adopted (conservative) value for the metallicity of the ambient medium (i.e. hot halo) of 0.3 \Zsun, the cloud's metallicity required to explain the observations is high, $Z \gtrsim 0.5 ~\Zsun$. Halo metallicities below 0.3 \Zsun\ bring the model into better agreement with the data close to the plane, but seem unrealistically low \citep[][]{mil16a}. Virtually the same conclusions are reached by \citet[][]{hen17a}, which is noteworthy in view of the significant differences between their model and ours.

The second, albeit weaker, problem is related to the mass and \HI\ content of the Smith Cloud. The Smith Cloud has a maximum \HI\ column density of $\sim 5 \times 10^{20} ~\psc$ \citep[][]{loc08a}. Our DM-free cloud, despite its mass of $\sim 2 \times 10^7 ~\Msun$ just before reaching the disc, has a maximum {\em total} gas density of $\sim 2 \times 10^{20} ~\psc$. Considering that its temperature is well above $10^4 ~\K$, we conclude that its corresponding \HI\ column density is significantly below $10^{20} ~\psc$. Thus, unless the Smith Cloud progenitor had a gas mass well in excess of $10^8 ~\Msun$ before being accreted by the Galaxy, it seems implausible that it has fallen in devoid of DM. We caution that this last conclusion relies heavily on a range microphysical processes, most notably cooling and heating, which in our current model are very crude or absent.

\section{System stability} \label{sec:sta}

\begin{figure*}
\centering
\includegraphics[width=0.33\textwidth]{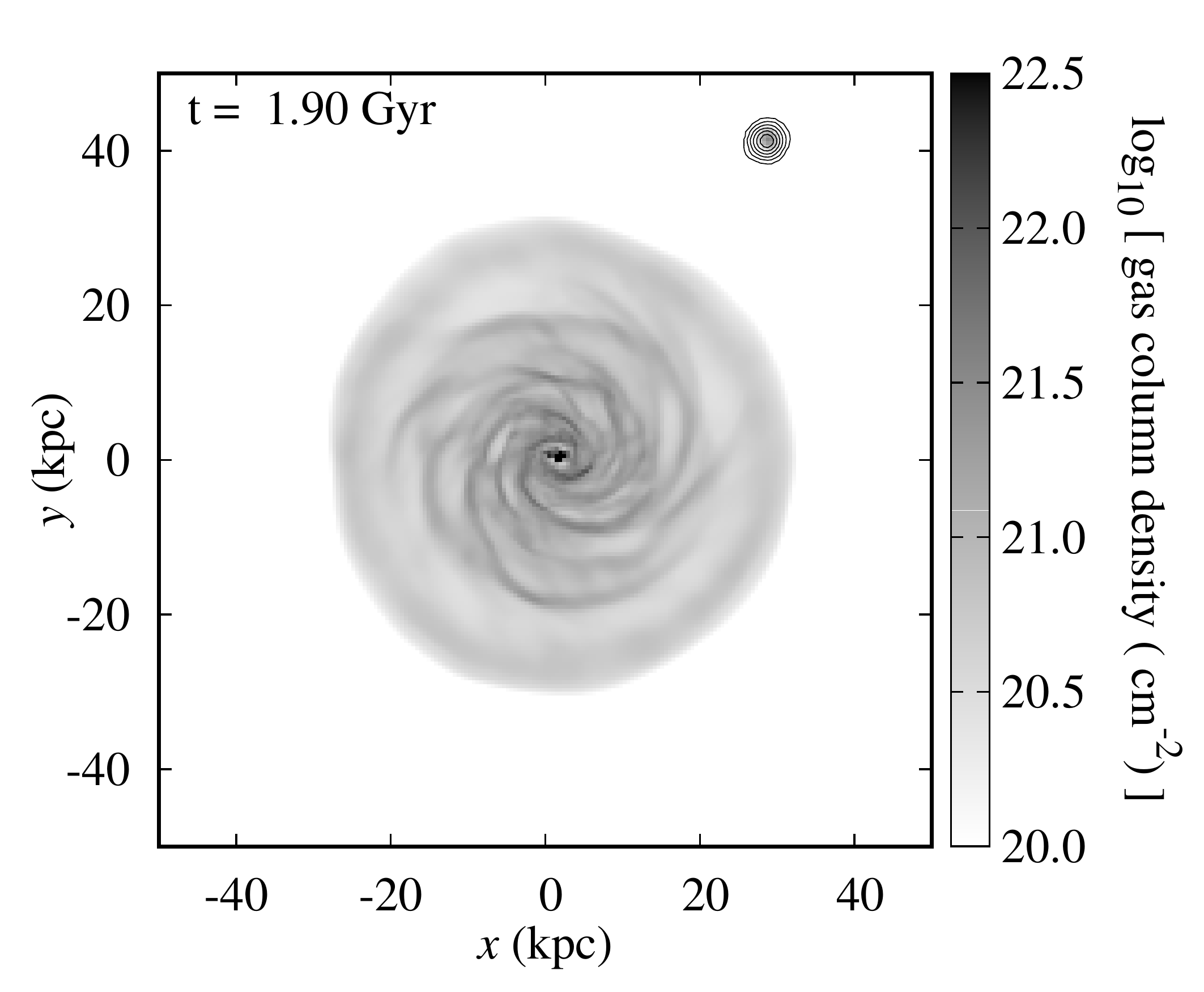}
\hfill
\includegraphics[width=0.33\textwidth]{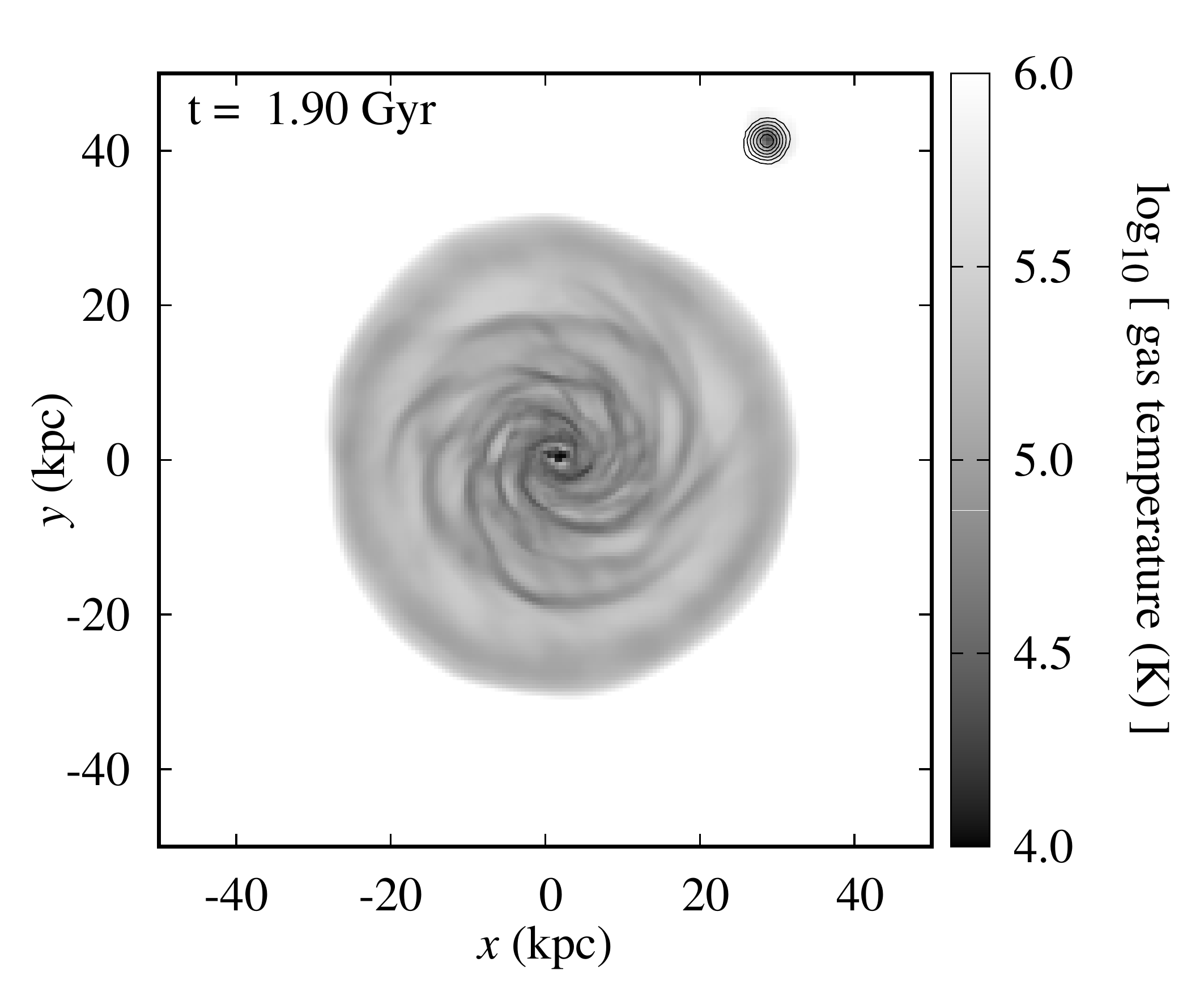}
\hfill
\includegraphics[width=0.33\textwidth]{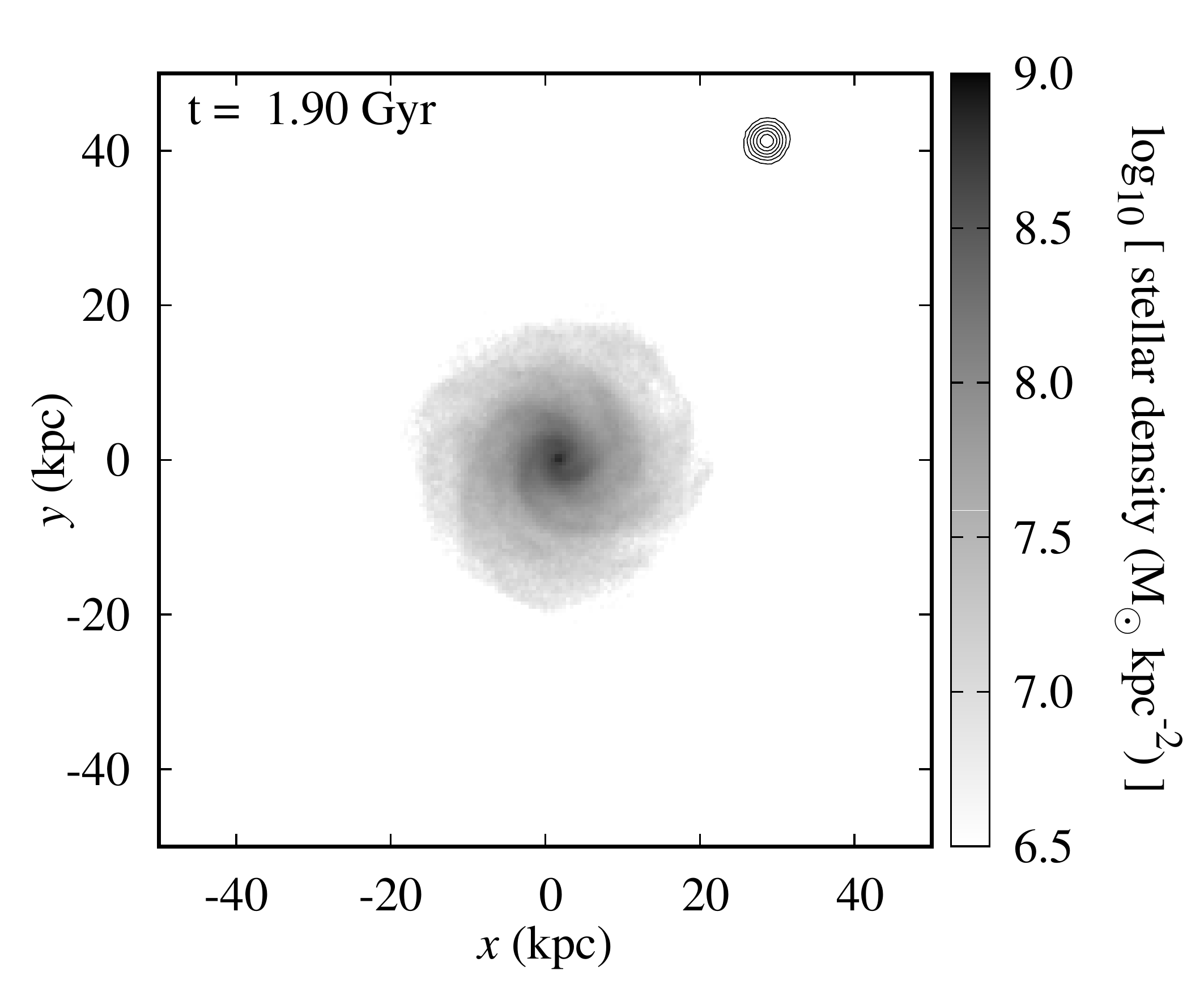}\\
\includegraphics[width=0.33\textwidth]{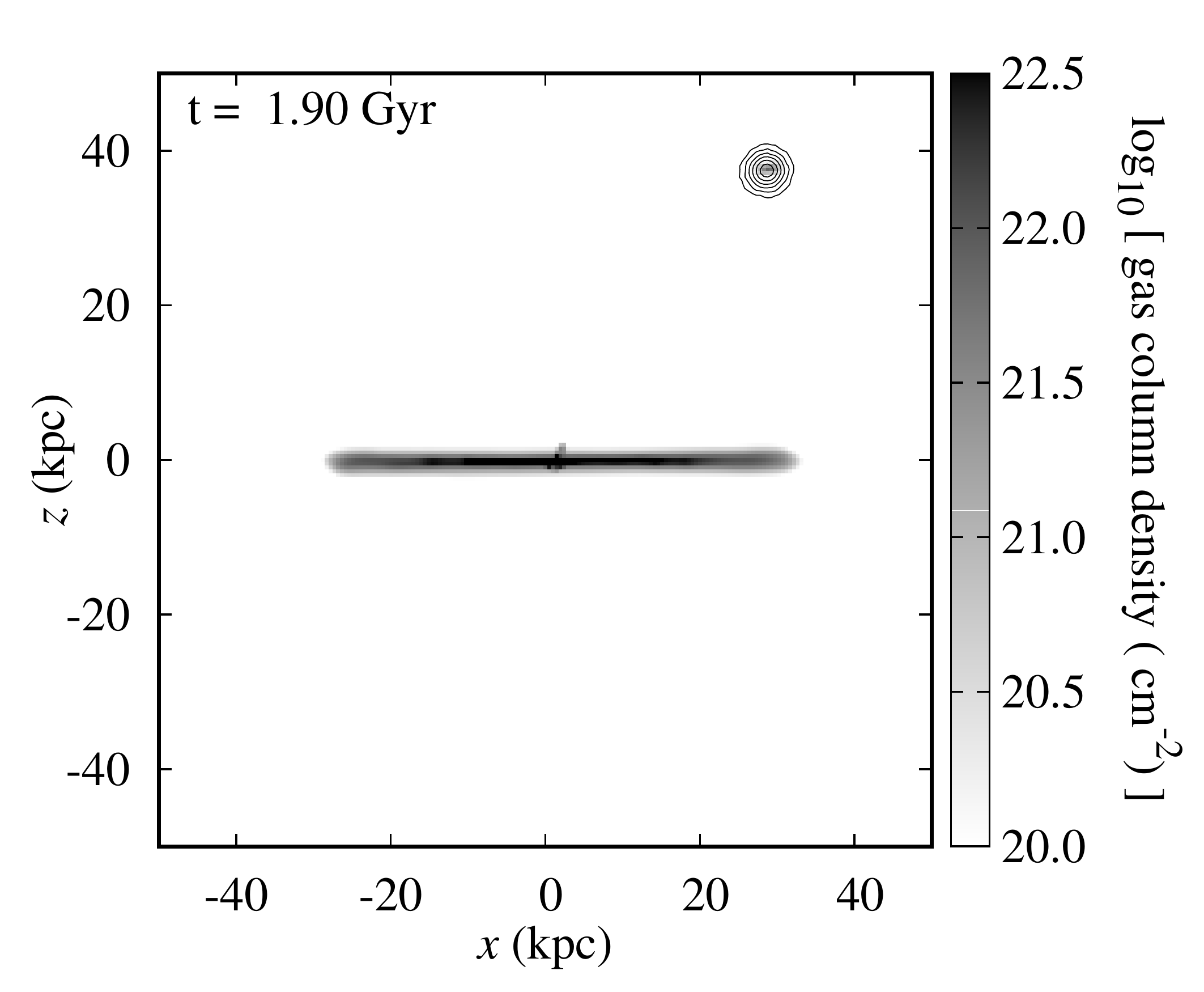}
\hfill
\includegraphics[width=0.33\textwidth]{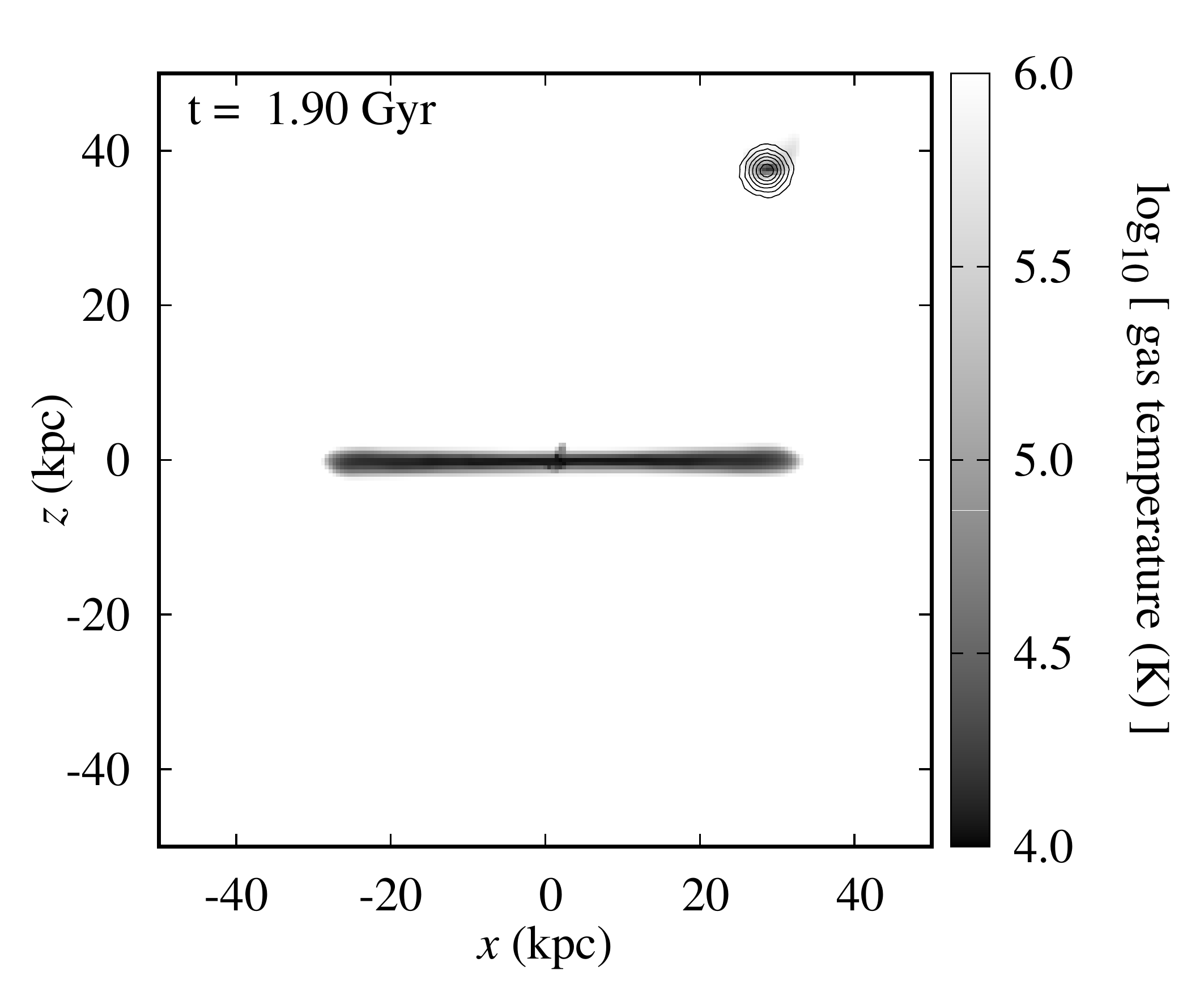}
\hfill
\includegraphics[width=0.33\textwidth]{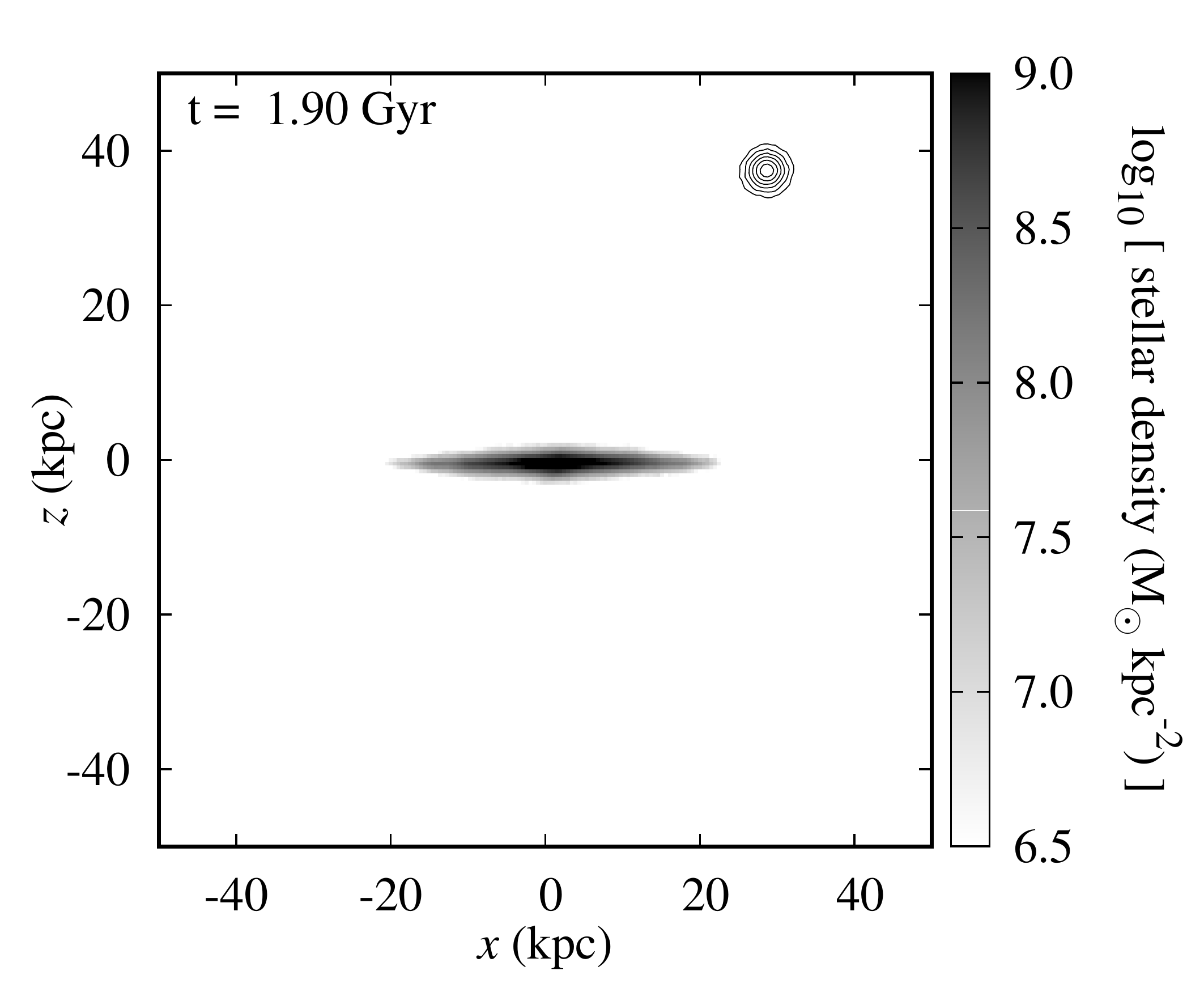}
\caption[ Disc (hi res run) ]{ Gas surface density (left), gas temperature (middle) and stellar surface density (right) in RUN 2 at $t \sim 2 ~\Gyr$. At this point, the DM subhalo is falling back onto the disc (cf. Figure \ref{fig:kin}). Note that the gas disc develops a very dense, warm ($ \sim T = 10^4 ~\K$) core, likely the result of numerical fragmentation. The results are essentially the same in RUN 1 and RUN 2 (with exception of the subhalo's evolution), and are therefore omitted. Supplement material showing the evolution of the system over roughly 3 Gyr can be found following this \href{http://www.physics.usyd.edu.au/~tepper/proj_smith_paper.html}{link}. We refer the reader also to the figure captions in the main text for links to more specific material.}
\label{fig:stable}
\end{figure*}

We claim that our Galaxy model is dynamically stable. In this context, `stability' is regarded as the ability of the composite system as well as each of the individual components to preserve (approximately) its initial state. In this section we thus briefly consider the system's long-term evolution. Figure \ref{fig:stable} shows the state of the system after $\sim 2 ~\Gyr$. Even though it is not itself a proof, comparison of this snapshot with the snapshots shown in Figures \ref{fig:trans1_gas} and \ref{fig:trans1_star} suggests that our model Galaxy appears to be globally stable on the long run. A more rigorous comparison reveals that, although the initial conditions are, by construction, in a strict dynamic equilibrium, they do evolve away from their initial state during the course of our simulations into a slightly different, but equally dynamically stable, state. This is true in particular for the gaseous components.

The main reason behind this behaviour is that, when mapped onto the simulation grid, gas particles falling within the same cell are merged using a CIC interpolation scheme (Section \ref{sec:sys}). As a result, the `bulk' momentum at each cell results from a mass-weighted average of a potentially large number of individual different masses and velocities. The same is true for their internal energy, which in turn sets the pressure (or, equivalently, the temperature). The overall result is generally that the rotation speed of the gas disc at radii where its density is comparable to (or below the) the hot halo's density is lower than required for the gas disc to be in centrifugal equilibrium (cf. Figure \ref{fig:gdisc}). In consequence, the gas disc shrinks during the initial stages, and re-expands later, in an oscillatory fashion until it reaches equilibrium. In doing so, it develops weak sub-structure such as rings and spiral arms (as does the stellar disc). A related effect is a slight loss of angular momentum of the hot halo, although this is mostly a consequence of the infall of zero angular momentum gas through the boundary. None of this is truly a concern for our present work, as we are not interested in the detailed structure of the disc and its evolution; but these issues do need to be addressed in the future. The most problematic issue for the targeted long-term stability is possibly the development of a growing dense core at the centre of the gas disc. This, however, can be avoided by imposing a density dependent temperature floor, as discussed above.

The (perhaps obvious) alternative approach of letting the system evolve towards its true stationary state before dropping the cloud is left for future work. Since the relevant properties of the gas disc, most notably its mean radial surface density, do not significantly change during its evolution, our current approach should be valid.

\bsp	
\label{lastpage}
\end{document}

%% file: defs_mnras.tex
\definecolor{orange}{rgb}{1.0,0.5,0.}

\def\XH{\ifmmode{\>X_{\textnormal{\sc h}}} \else{$X_{\textnormal{\sc h}}$}\fi}
\def\nH{\ifmmode{\>n_{\textnormal{\sc h}}} \else{$n_{\textnormal{\sc h}}$}\fi}

\def\mG{\ifmmode{\>\mu\mathrm{G}}\else{$\mu$G}\fi}
\def\erg{\ifmmode{\> {\rm erg}}\else{erg}\fi}
\def\keV{\ifmmode{\> {\rm keV}}\else{keV}\fi}

\def\deg{\ifmmode{\>^{\circ}}\else{$^{\circ}$}\fi}
\def\onedeg{\ifmmode{\>1^{\circ}}\else{$1^{\circ}$}\fi}

\def\xvir{\ifmmode{\>\!x_{vir}}\else{$x_{vir}$}\fi}
\def\Mvir{\ifmmode{\>\!M_{vir} }\else{$M_{vir} $}\fi}
\def\rvir{\ifmmode{\>\!r_{vir}}\else{$r_{vir}$}\fi}
\def\vvir{\ifmmode{\>\!v_{vir}}\else{$v_{vir}$}\fi}
\def\Vvir{\ifmmode{\>\!V_{vir} }\else{$V_{vir} $}\fi}

\def\tratio{\ifmmode{\>\tau}\else{$\tau$}\fi}

\def\rms{\ifmmode{\>r_{\textnormal{\sc ms}}}\else{$r_{\textnormal{\sc ms}}$}\fi}

\def\Mpc{\ifmmode{\>\!{\rm Mpc}} \else{Mpc}\fi}
\def\kpc{\ifmmode{\>\!{\rm kpc}} \else{kpc}\fi}
\def\pc{\ifmmode{\>\!{\rm pc}} \else{pc}\fi}

\def\Gyr{\ifmmode{\>\!{\rm Gyr}} \else{Gyr}\fi}
\def\Myr{\ifmmode{\>\!{\rm Myr}} \else{Myr}\fi}
\def\yr{\ifmmode{\>\!{\rm yr}} \else{yr}\fi}
\def\pyr{\ifmmode{\>\!{\rm yr}^{-1}}\else{yr $^{-1}$} \fi}
\def\s{\ifmmode{\>\!{\rm s}}\else{s}\fi}
\def\ps{\ifmmode{\>\!{\rm s}^{-1}}\else{s$^{-1}$}\fi}
\def\Hz{\ifmmode{\>\!{\rm Hz}}\else{Hz}\fi}

\def\kms{\ifmmode{\>\!{\rm km\,s}^{-1}}\else{km~s$^{-1}$}\fi}

\def\K{\ifmmode{\>\!{\rm K}}\else{K}\fi}

\def\sr{\ifmmode{\>\!{\rm sr}}\else{sr}\fi}
\def\psr{\ifmmode{\>\!{\rm sr}^{-1}}\else{sr$^{-1}$}\fi}
\def\arcs{\ifmmode{\>\!{\rm arcsec}}\else{arcsec}\fi}
\def\parcs{\ifmmode{\>\!{\rm arcsec}^{-1}}\else{arcsec${-1}$}\fi}
\def\parcss{\ifmmode{\>\!{\rm arcsec}^{-2}}\else{arcsec${-2}$}\fi}

\def\cm{\ifmmode{\>\!{\rm cm}}\else{cm}\fi}
\def\cc{\ifmmode{\>\!{\rm cm}^{3}}\else{cm$^{3}$}\fi}
\def\sqc{\ifmmode{\>\!{\rm cm}^{2}}\else{cm$^{2}$}\fi}
\def\pcc{\ifmmode{\>\!{\rm cm}^{-3}}\else{cm$^{-3}$}\fi}
\def\psc{\ifmmode{\>\!{\rm cm}^{-2}}\else{cm$^{-2}$}\fi}

\def\g{\ifmmode{\>\!{\rm g}}\else{g}\fi}
\def\Msun{\ifmmode{\>\!{\rm M}_{\odot}}\else{M$_{\odot}$}\fi}
\def\hMsun{\ifmmode{\> h^{-1}{\rm M}_{\odot}}\else{$h^{-1}$M$_{\odot}$}\fi}

\def\Zsun{\ifmmode{\>\!{\rm Z}_{\odot}}\else{Z$_{\odot}$}\fi}

\def\rayl{\ifmmode{\>\!{\rm R}}\else{R}\fi}
\def\mR{\ifmmode{\>\!{\rm mR}}\else{mR}\fi}

\renewcommand{\ion}[2]{\hbox{#1\,{\sc #2}}}

\def\lya{\ifmmode{\>\!{\rm Ly}\alpha}\else{Ly$\alpha$}\fi}

\def\Ha{\ifmmode{\>\!{\rm H}\alpha}\else{H$\alpha$}\fi}
\def\Hb{\ifmmode{\>\!{\rm H}\beta}\else{H$\beta$}\fi}

\def\HI{\ifmmode{\> \textnormal{\ion{H}{i}}} \else{\ion{H}{i}}\fi}
\def\HII{\ifmmode{\> \textnormal{\ion{H}{ii}}} \else{\ion{H}{ii}}\fi}
\def\CIV{\ifmmode{\> \textnormal{\ion{C}{iv}}} \else{\ion{C}{iv}}\fi}
\def\SiIV{\ifmmode{\> \textnormal{\ion{S}{iv}}} \else{\ion{Si}{iv}}\fi}

\def\NHI{\ifmmode{\> {\rm N}_{\HI}} \else{N$_{\HI}$}\fi}
\def\MHI{\ifmmode{\> {\rm M}_{ \HI}} \else{M$_{\HI}$}\fi}

\def\mua{\ifmmode{\>\mu_{ \textnormal{\Ha}}}\else{$\mu_{ \textnormal{\Ha}}$}\fi}
\def\alphabha{\ifmmode{\>\alpha_{B}^{(\textnormal{\Ha})}}\else{$\alpha_{B}^{(\textnormal{\Ha})}$}\fi}